\documentclass[12pt]{article}
\usepackage{amsmath}
\usepackage{graphicx,psfrag,epsf}
\usepackage{enumerate}
\usepackage{natbib}
\usepackage{multibib}
\newcites{S}{References}
\usepackage{array}
\usepackage{url} 
\usepackage{algpseudocode}
\usepackage{tabu}
\usepackage{ulem}
\usepackage{bbm}
\usepackage{soul}
\usepackage{verbatim}
\usepackage{amsfonts}
\usepackage[english]{babel}
\usepackage{subcaption}
\usepackage[utf8]{inputenc}
\usepackage{booktabs}
\usepackage[T1]{fontenc}
\usepackage{appendix}
\usepackage{amsthm}
\usepackage{thmtools}
\usepackage{thm-restate}
\usepackage{xcolor} 
\usepackage{mathtools}
\usepackage{bm}
\usepackage{subfiles}
\usepackage[export]{adjustbox}
\usepackage{float}
\usepackage[colorlinks=true, linkcolor=blue, citecolor=blue, urlcolor=blue, allcolors=blue]{hyperref}
\usepackage{tablefootnote}
\usepackage{cleveref}
\usepackage{multirow}

\newif\ifusered
\useredfalse

\ifusered
  \definecolor{customizedColor}{RGB}{255,0,0}
\else
  \definecolor{customizedColor}{RGB}{0,0,0}
\fi

\newif\ifredroundtwo
\redroundtwofalse

\ifredroundtwo
  \definecolor{customizedColorRound2}{RGB}{255,0,0}
\else
  \definecolor{customizedColorRound2}{RGB}{0,0,0}
\fi

\newcommand{\TrainstartDate}{2022-08-01}

\newcommand{\TeststartDate}{2023-06-01}
\newcommand{\TestendDate}{2023-07-31}
\newcommand{\NumObs}{$110,985$}
\newcommand{\significantNum}{60}
\newcommand{\totalNum}{$6,050$}
 
\newcommand{\totalAsset}{$55$}
\newcommand{\significantLevel}{0.01}
\newcommand{\tailQuantile}{0.99}
\newcommand{\tename}{Directional tail dependence} 

\newcommand{\tenameShort}{DTD} 
\newcommand{\tenameLower}{directional tail dependence}

\newtheorem{prop}{Proposition}[section]

\newtheorem*{remark}{Remark}
\newtheorem{lemma}{Lemma}[section] 
\newtheorem{definition}{Definition}[section]

\newenvironment{manualprop1}[1]{%
  \par\noindent
  \textbf{Proposition #1.}\itshape
}{%

  \par
}

\newenvironment{manualprop2}[2]{%
  \par\noindent
  \textbf{Proposition #1{\normalfont~(#2)}.}\itshape
}{%

  \par
}

\begin{document}

\def\spacingset#1{\renewcommand{\baselinestretch}%
{#1}\small\normalsize} \spacingset{1}


\newcommand{\blind}{0}
\if0\blind
{
  \title{\bf The Efficient Tail Hypothesis: An Extreme Value Perspective on Market Efficiency}
  \author{Junshu Jiang$^{\rm a}$\thanks{Corresponding author: junshu.jiang@kaust.edu.sa.\\
    The authors gratefully acknowledge funding from the King Abdullah University of Science and Technology (KAUST) Office of Sponsored Research (OSR) under Award No.\ {OSR-CRG2020-4394}.}\hspace{.2cm}, Jordan Richards$^{\rm b}$, Rapha\"el Huser$^{\rm a}$, David Bolin$^{\rm a}$  \\
    $\;^{\rm a}$ Statistics Program, CEMSE Division, \\ King Abdullah University of Science and Technology, Saudi Arabia. \\
    $\;^{\rm b}$ School of Mathematics, University of Edinburgh, Edinburgh, UK.
    }
  \maketitle
} \fi

\if1\blind
{
  \bigskip
  \bigskip
  \bigskip
  \begin{center}
    {\LARGE\bf The Efficient Tail Hypothesis: An Extreme Value Perspective on Market Efficiency}
\end{center}
  \medskip
} \fi

\bigskip
\begin{abstract}
In econometrics, the Efficient Market Hypothesis posits that asset prices reflect all available information in the market. Several empirical investigations show that market efficiency drops when it undergoes extreme events. Many models for multivariate extremes focus on positive dependence, making them unsuitable for studying extremal dependence in financial markets where data often exhibit both positive and negative extremal dependence. To this end, we construct regular variation models on the entirety of $\mathbb{R}^d$ and develop a bivariate measure for asymmetry in the strength of extremal dependence between adjacent orthants. Our directional tail dependence (DTD) measure allows us to define the Efficient Tail Hypothesis (ETH)---an analogue of the Efficient Market Hypothesis---for the extremal behaviour of the market. Asymptotic results for estimators of DTD are described, and we discuss testing of the ETH via permutation-based methods and present novel tools for visualization. An empirical study of China's futures market leads to a rejection of the ETH and we identify potential profitable investment opportunities \textcolor{customizedColor}{during extreme episodes}. To promote the research of microstructure in China's derivatives market, we open-source our high-frequency data, which are being collected continuously from multiple derivative exchanges.
\end{abstract}

\noindent%
{\it Keywords:}  asymmetric dependence, China's derivatives market, extremal dependence, market efficiency, regular variation
\vfill

\newpage
\spacingset{1.8} 

\section{Introduction}\label{sec:intro}

Market efficiency is a fundamental concept in \textcolor{customizedColorRound2}{finance}, which indicates how fast the market can absorb new information and how swiftly market prices change to reflect the fair value of an asset. The Efficient Market Hypothesis, first introduced by~\cite{fama1970efficient}, states that, in an ideal situation, the market price of an asset reflects all available information in the market and must be equal to the asset's fair value. In other words, when the market is fully efficient, the price returns are entirely unpredictable. The debate on the use of the Efficient Market Hypothesis for statistical analysis has been ongoing for decades, and the Efficient Market Hypothesis has been challenged by many empirical studies; see, e.g.,~\cite{lo2004adaptive,lo2005reconciling} and~\cite{malkiel2003efficient}. Various statistical methods have been proposed to test the Efficient Market Hypothesis~\citep{lo1988stock,choi1999testing,escanciano2009automatic,charles2011small,wang2024generalized}. However, these methods are often based on the assumption that log-returns are normally distributed, while, in reality, the returns typically have much heavier tails and non-Gaussian dependencies, especially in a high-frequency setting \citep{cont2001empirical,de2023modeling}. Moreover, increasing evidence shows that the market is not always efficient, especially when it experiences extreme events during financial turmoil \citep{choi2021analysis}; this kind of inefficiency in tails is often ignored by traditional Efficient Market Hypothesis tests.

Many extremal dependence models have been developed within the last 20 years \citep[see, e.g.,][]{hartmann2004asset,poon2004extreme,castro2018time,huang2019new,engelke2020graphical,chan2022extreme,gong2022asymmetric,gong2024partial}, and such models are widely used in risk management and insurance. For example,\textcolor{customizedColor}{~\cite{poon2004extreme} study extremal dependence between returns of risky assets in major financial markets, and \cite{hartmann2004asset} study cross-asset extremal dependence between stocks and government bonds and find that stock crises tend to be linked with booms in government bonds; this phenomenon is known as the ``flight to quality'', and serves as an example of cross-directional extremal dependence. Other more recent studies include:} \cite{gong2024partial}, who study extremal dependence between the depreciation of different currencies;~\cite{bouaddi2022systematic}, who study the asymmetry in contagion effects between the US stock market and other international markets by comparing extreme losses and gains;~\cite{chan2022extreme}, who examine the extremal dependence between trading volumes and returns in cryptomarkets; in the same vein,~\cite{van2019estimating} propose an estimator for linear coefficients of a bivariate regular variation model under extreme adverse market conditions;~\cite{lu2024extreme} study the extreme co-movement between three decomposed oil prices by estimating a tail correlation coefficient.  For a review of many more applications of extremal dependence models in finance, see~\cite{nolde2021extreme}.

\textcolor{customizedColor}{Empirical studies show that the efficiency of a market drops when it undergoes extreme events~\citep{choi2021analysis}, and extreme value theory is an effective framework to analyze extremal dependence between market assets~\citep{poon2004extreme}}. However, traditional extremal dependence models may not be adequate for studying market efficiency as they are often founded on an underlying assumption that data are regularly varying in the positive orthant of $\mathbb{R}^d$ only, i.e., when all variables are positive and jointly heavy-tailed; for details on regular variation, see, e.g.,~\citet[Chapter~6]{resnick2007heavy}. Thus, such models can assess only positive extremal dependence, i.e., the dependence between variables as they jointly grow large. While joint losses may be studied by flipping the sign of the data, extremal dependence in all orthants (of $\mathbb{R}^d$) is important in the study of market efficiency, as we seek to model the joint occurrence of combinations of both losses and gains of different assets. \textcolor{customizedColor}{In this work, we develop a bivariate measure based on extreme value theory to quantify the difference in the strength of tail dependence in two adjacent quadrants; this allows us to investigate the direction of dependence during extremal events}. 

We consider random vectors, $\bm{X}:=(X_1,\dots,X_p)^\top \in \mathbb{R}^p$, that are regularly-varying on all orthants of $\mathbb{R}^p$. \textcolor{customizedColor}{Focusing on dependence modeling, we standardize the marginal distributions and} derive mass balance constraints for these vectors to construct, without loss of generality \textcolor{customizedColor}{(see Section~\ref{sec:meth:margin})}, \textit{balanced} regularly-varying random vectors in which both upper and lower tails are heavy-tailed and of unit scale. Furthermore, we define a new pairwise measure of \tenameLower\ (DTD), denoted by $\lambda(\cdot,\cdot)$, which quantifies the strength of asymmetry in the extremal dependence exhibited in adjacent orthants of $\mathbb{R}^d$ for these balanced regularly-varying random vectors; this measure extends the popular pairwise Extremal Dependence Measure (EDM), proposed by~\cite{resnick2004extremal}. \textcolor{customizedColor}{The terminology of ``asymmetric'' tail dependence has been broadly used in the literature to describe asymmetry between the joint upper-tail and joint lower-tail regions \citep[see, e.g.,][]{manner2010testing,jondeau2016asymmetry,echaust2021asymmetric,gong2022asymmetric}.} To avoid confusion, we use here the terminology of ``\textit{directional (a)symmetry}'' for describing the (a)symmetry of extremal dependence in two adjacent quadrants, as one variable plays a leading role in this framework for the occurrence of extreme events in the other variable. We propose two estimators for $\lambda(\cdot,\cdot)$ and study their asymptotic properties, as well as construct a non-parametric permutation test for the symmetric case where $\lambda(\cdot,\cdot)=0$.


In financial markets, let $\bm{Y}\in \mathbb{R}^{p_2}$ represent the asset returns at a given time, and $\bm{X}\in \mathbb{R}^{p_1}$ represent the information that is measurable before the return of the assets. We then define the Efficient Tail Hypothesis (ETH), a new concept analogous to the Efficient Market Hypothesis, which corresponds to the situation where $\lambda(X_i,Y_j)=0$ for $i=1,\dots,p_1, j=1,\dots,p_2,$ where $p_2$ is the number of assets and $p_1$ is the number of indicators. We also provide a means of visualizing directional tail asymmetry for high-dimensional random vectors by projecting values of $\lambda$ to the surface of a sphere. From an economic standpoint, the ability to detect directional tail dependence, between extreme gains and losses, provides an informative view of the shock-propagation mechanism that occurs in financial markets during extreme periods. Moreover, a rejection of the ETH suggests the existence of assets that, when extreme, lead to extremes of other assets in the following time step; this can serve as an early-warning system for risk management or can be used to develop trading strategies. Therefore, our proposed methodology contributes to the literature on risk spillover analysis \citep[see, e.g.,][]{billio2012econometric,bollerslev2018risk,jian2018asymmetric} and lead-lag detection \citep[see, e.g.,][]{hou2007industry,tolikas2018lead,buccheri2021high} by incorporating directional tail dependence.




We apply our method to study the market-wide tail-efficiency of China's futures market, using high-frequency data containing second-level market information that we mined for use in statistical analyses. We find that most of the time-leading tail dependencies are symmetric, i.e., $\lambda(X_i,Y_j)=0,$ but some inefficiencies persist. To capitalize on this finding, we construct an artificial dynamic portfolio consisting of assets that do not satisfy the conditions of the ETH and show that it can generate profits using out-of-sample data. 

China's futures market has emerged as one of the most important markets globally, with transactions amounting to 80 trillion United States (U.S.) dollars annually and 70 types of commodity futures. An increasing body of research indicates that China's derivatives market exerts a significant pricing influence on the global market \citep{arslanalp2016china}. Given its magnitude and influence, there is considerable potential for empirical studies of this market in order to validate theoretical results. However, despite China's growing impact, there are limited accessible and user-friendly datasets available for researchers specializing in China's derivatives market. While there are commercial data providers, such as Wind, Bloomberg, and Choice, their data are not open-source. To promote research on China's derivatives market, we have gathered millisecond-level, high-frequency data since 2022 from multiple exchanges, totaling more than 1.3 Terabytes of data, and we provide it as open-source for academic purposes. A subset of these data have been considered previously by \ifnum\blind=0\cite{jiang2022market} \else \textcolor{blue}{Anonym et al. }(\textcolor{blue}{2022}) \fi in market-making research.


The rest of the paper is organized as follows: Section \ref{sec:bg} reviews the literature on regular variation and the EDM. Section \ref{sec:meth} introduces balanced regular variation, our \tenameLower\ measure, the ETH, and our hypothesis test. Section~\ref{sec:simulation} provides a simulation study that gives empirical evidence for the power of our permutation test. Section~\ref{sec:app} presents our application to asset returns from China's futures market; the dataset and its construction are described in the Appendix~\textcolor{blue}{B}. Relevant proofs are included in the Appendix~\textcolor{blue}{A}. Supporting code and the dataset used in our analysis are available at the Github repository, \ifnum\blind=0\href{https://github.com/junshujiang/tailEfficientTest}{https://github.com/junshujiang/tailEfficientTest}\else\href{https://github.com/anonymous/anonymous}{https://github.com/anonymous/anonymous}\fi. The full dataset can be obtained upon request.

\section{Multivariate extreme-value background}
\label{sec:bg}

Multivariate regular variation is a common assumption used to describe the tail behaviour of random variables. Under the setting of regular variation, the probability of extreme events decays according to a power law of the extent of extremeness. This property makes it suitable to study dependence between jointly heavy-tailed random variables.

\begin{definition}[Multivariate regular variation (\citealp{resnick2007heavy}; Chapter~6)]\label{def:rv}
    A $p$-dimensional random vector $\bm{X}\in \mathbb{R}^p_+$ is regularly varying (RV) with tail index $\alpha>0$, denoted by $\bm{X}\in {\rm RV}^p_+(\alpha)$, if there exists a sequence $b_n\to \infty$ such that $n\mathbb{P}\{b_n^{-1}\bm{X} \in\cdot\}\xrightarrow{v} v_{\bm{X}}(\cdot)$ as $n\to\infty$, where $v_{\bm{X}}(\cdot)$ is a Radon measure on the cone $\mathbb{E}^p_+:=[0,\infty]^p\setminus{\{\bm{0}\}}$ and $\xrightarrow{v}$ denotes vague convergence.

\end{definition}

The measure $v_{\bm{X}}(\cdot)$ is referred to as the limit measure and satisfies the homogeneity property, $v_{\bm{X}}(rB)=r^{-\alpha}v_{\bm{X}}(B)$ for any $r>0$ and any Borel subset $B\subset \mathbb{E}^p_+$. Courtesy of its homogeneity property, we can decompose $v_{\bm{X}}(\cdot)$ into radial and angular mass measures. Define by $H_{\bm{X}}(\cdot)$ the angular mass measure for $\bm{X}$ on the positive part of the unit $(p-1)$-sphere $\mathbb{S}^{p-1}_+:=\{\bm{x}\in \mathbb{R}_+^p: ||\bm{x}||_2=1\},$ where $||\cdot||_2$ denotes the $l_2$-norm. Then, for $r>0$, we have ${v_{\bm{X}}(\{\bm{x}\in \mathbb{E}^p_+: ||\bm{x}||_2\geq r, \bm{x}/||\bm{x}||_2\in B_H \})}=r^{-\alpha}H_{\bm{X}}(B_H)$, where $B_H \subset\mathbb{S}^{p-1}_+ $ is a Borel subset. The angular mass measure can be normalized to give a valid probability measure, which we denote by $N_{\bm{X}}(\cdot):=H_{\bm{X}}(\cdot)/m$ where $m=\int_{\mathbb{S}^{p-1}_+ }{\rm d}H_{\bm{X}}(\bm{w})$ is the total mass of $H_{\bm{X}}(\cdot)$. Note that $H_{\bm{X}}(\cdot)$ is not invariant to changes in the normalization sequence $b_n$: it is only unique up to a multiplicative constant. For example, let $b_n'=cb_n$ for some $c>0$ be a new normalization sequence; then the corresponding limit and angular measures are $c^{-\alpha}v_{\bm{X}}(\cdot)$ and $c^{-\alpha}H_{\bm{X}}(\cdot)$, respectively, while $N_{\bm{X}}(\cdot)$ is unique.

To construct an inner product space for regularly-varying random variables, we follow \cite{lee2021transformed} and take $\alpha=2$ throughout and consider $\bm{X}\in {\rm RV}^p_+(2)$. Such a choice is not overly restrictive as one can always transform margins to any specific scale; we defer the reader to \cite{lee2021transformed} for full details. Consider a $q$-dimensional random vector  $\bm{Z}:=(Z_1,\dots,Z_q)^\top$ with independent components $Z_i\in {\rm RV}^1_+(2)$ for $i=1,\dots,q$. Then, we can construct the space
\[\mathcal{V}^q_+=\{X\in {\rm RV}^1_+(2):X=\bm{a}^\top\circ \bm{Z}=(a_{1}\circ Z_1)\oplus \dots \oplus (a_{q}\circ Z_q), \bm{a}\in \mathbb{R}^q_+\},\]
where $\oplus$ and $\circ$ are transformed-linear operators \citep[see][for details]{cooley2019decompositions} that satisfy $X_1\oplus X_2=t\{t^{-1}(X_1)+t^{-1}(X_2)\}$ and $a\circ X=t\{at^{-1}(X)\}$, for constant $ a\in \mathbb{R}$ and ${t(\cdot)=\log\{1+\exp(\cdot)\}}$. The space $\mathcal{V}^q_+$ is a subspace of ${\rm RV}^1_+(2)$, and contains elements that can be spanned (using the transformed-linear operations) by a vector of independent regularly varying ${\rm RV}^{1}_+(2)$ random variables with non-negative coefficients.  This space has two important properties, which make it conducive to studying the tails of random variables (see, e.g., the transformed linear model of \cite{mhatre2022transformed} and the partial tail correlation coefficients of \cite{lee2022partial} and \cite{gong2024partial}): first, the corresponding angular measure $H_{\mathbf{X}}(\cdot)$ for any $\bm{X}\in {\rm RV}^p_+(2)$ can be approximated using a sequence of transformed-linear combinations of independent regularly varying random variables \cite[Proposition 4; ][]{cooley2019decompositions} with increasing $q$. This implies that $\mathcal{V}^q_+$ is dense in ${\rm RV}^1_+(2)$ as $q\to\infty$. Second, the inner product in $\mathcal{V}^q_+$, i.e., $\langle X,Y\rangle:= \bm{a}_x^T\bm{a}_y$ where $\bm{a}_x$ and $\bm{a}_y$ are the coefficient vectors of $X$ and $Y$, respectively, in the transformed-linear space $\mathcal{V}^q_+$, can be related to the Extremal Dependence Measure (EDM). This measure was first introduced by \cite{resnick2004extremal} and later studied by \cite{larsson2012extremal} and \cite{cooley2019decompositions}.

The EDM for a bivariate random vector ${(X,Y)^\top\in {\rm RV}^2_+(2)}$ measures extremal dependence with associated angular mass and probability measures, $H_{(X,Y)}(\cdot)$ and $N_{(X,Y)}(\cdot)$, respectively. The EDM, denoted by $\sigma(X,Y)$, is defined as 
\begin{equation}
    \label{eq:edm}
    \sigma(X,Y):=\int_{\mathbb{S}^{1}_+ }\omega_x\omega_y{\rm d}N_{(X,Y)}(\bm{w})=\lim_{r\to \infty} \mathbb{E}\left[\frac{XY}{R^2}\mid R>r\right],
\end{equation}

where $R=||(X,Y)||_2$ and $\bm{\omega}=(\omega_x,\omega_y)^\top \in \mathbb{S}^{1}_+$. \cite{lee2022partial} showed that $m\sigma(X,Y)=\langle X,Y\rangle,$ where $m=\int_{\mathbb{S}^{1}_+ }{\rm d}H_{(X,Y)}(\bm{w})$ and $\langle \cdot,\cdot \rangle$ is the inner product for $\mathcal{V}^q_+$. The value of $\sigma(X,Y)$ ranges between 0 and 0.5, with $\sigma(X,Y)=0$ if and only if the angular measure puts all mass onto the axes; i.e., the support of $H_{(X,Y)}(\cdot)$ is $\{(0,1),(1,0)\}$. This implies asymptotic independence between $X$ and $Y$. The value of $\sigma(X,Y)$ increases with the strength of asymptotic dependence in $(X,Y)^\top$, with its maximal value, $\sigma(X,Y)=0.5$, attained if and only if the angular measure puts all mass along the diagonal, which corresponds to perfect dependence in the tail. We can estimate $\sigma(X,Y)$ empirically using $n$ data samples, $\{(x_i,y_i)^\top\}_{i=1}^{n}$. The empirical estimator is 
\begin{equation}
    \label{eq:estimationSigma}
    \hat{\sigma}(X,Y)=N^{-1}\sum_{i=1}^{n}\frac{x_{i}y_i}{r^2_i}\mathbbm{1}(r_i>r_0),
\end{equation}
where $r_i=||(x_i,y_i)^\top||_2$, $r_0>0$ is a suitably-chosen high threshold, and $N=\sum_{i=1}^n \mathbbm{1}(r_i>r_0)$ is the number of threshold exceedances, where $\mathbbm{1}(\cdot)$ is the indicator function.

\section{Methodology}
\label{sec:meth}
In Section~\ref{sec:meth:RegularVarying}, we construct balanced regularly varying random vectors on $\mathbb{R}^p$ which have both upper and lower tails of unit scale and allow extremal dependence to be modeled simultaneously for both tails. Such a setting appeals to fields where extremal events can happen in both directions, such as the financial sector. \textcolor{customizedColor}{In Section~\ref{sec:meth:margin}, we demonstrate various marginal transformations that can be applied to data to produce balanced regularly varying random variables.} In Section~\ref{sec:meth:visual}, we discuss a method for visualizing directional asymmetric extremal dependence in such random vectors. In Section~\ref{sec:meth:asymmetric}, we propose our measure of \tenameLower. In Section~\ref{sec:meth:eth}, we introduce the Efficient Tail Hypothesis (ETH) and propose methods for testing it.

\subsection{Balanced regular variation}

\label{sec:meth:RegularVarying}

We begin by considering regular variation on the entirety of $\mathbb{R}^p$. This concept is briefly discussed in \S6.5.5 of \cite{resnick2007heavy} and by \cite{cooley2019decompositions}.

\begin{definition}[Regular variation on $\mathbb{R}^p$]\label{def:rv_whole}
    A $p$-dimensional random vector $\bm{X}\in \mathbb{R}^p$ is regularly varying with tail index $\alpha>0$, denoted by $\bm{X}\in {\rm RV}^p(\alpha)$, if $|\bm{X}|\in {\rm RV}_+^p(\alpha)$ with the normalizing sequence $b_n\to\infty$ and for all ${\bm{z}\in [0,\infty)^p\setminus\{\bm{0}\}}$ and $\bm{s} \in \{-1, 1\}^p$,
    \[\lim_{n\rightarrow \infty}\frac{\mathbb{P}\{b_n^{-1}\bm{s}\odot\bm{X}\in [\bm{0},\bm{z}]^c\}}{\mathbb{P}\{b_n^{-1}|\bm{X}|\in [\bm{0},\bm{z}]^c\}}\in (0,1),\]
    where $\odot$ is the element-wise (Hadamard) product and $|\cdot|$ denotes the element-wise absolute value. 
\end{definition}

\textcolor{customizedColor}{The properties of $\bm{X} \in {\rm RV}^p(\alpha)$ similarly involve a radial-angular decomposition, as previously described in Section~\ref{sec:bg} for ${\rm RV}^p_+(\alpha)$ random variables. The difference here is that the limit and angular mass measures are now defined on the punctured real space, $\mathbb{E}^p := \mathbb{R}^p \setminus \{\bm{0}\},$ and the full unit $(p-1)$-sphere, respectively. Similarly (to ${\rm RV}^p_+(\alpha)$ random variables), the angular mass measure can be normalized to a unique probability measure $N_{\bm{X}}(\cdot) := H_{\bm{X}}(\cdot)/m$, with the normalizing constant $m = H_{\bm{X}}(\mathbb{S}^{p-1})$ representing the total angular mass on the unit sphere.}

To focus on describing extremal dependence structure, we \textcolor{customizedColor}{apply a marginal scaling of $\bm{X}$ to ensure that the scale of all of its marginal upper and lower tails are  equal,} and consider a ``balanced'' version of $\bm{X}\in{\rm RV}^p(\alpha)$. We define scaling factors for upper and lower tails by ${c^+_i:=\nu_{\bm{X}}\left(\{\bm{x}\in \mathbb{E}^p:x_i>1\}\right)}$ and  ${c^-_i:=\nu_{\bm{X}}(\{\bm{x}\in \mathbb{E}^p:x_i<-1\})},$ respectively, and consider $\lim_{n\rightarrow \infty}{\mathbb{P}\{X_i<-xb_n\} \over \mathbb{P}\{X_i>xb_n\}}={c^-_i \over c^+_i}$ for all $i=1,\dots,p$. It is possible that the tails are unbalanced in the sense that $c^+_i$ and $c^-_i$ are unequal for at least one of $i\in\{1,\dots,p\}$. To ensure that $c_i^+=c_i^-$ for all $i=1,\dots,p$, we use the transformation defined in Proposition~\ref{prop:Rebalance} below and denote the resulting random vector as $\bm{X}\in {\rm BRV}^p(\alpha)$; see Definition~\ref{def:brv}.

\begin{definition}[Balanced regular variation]\label{def:brv}
    A $p$-dimensional regularly varying random vector $\bm{X}\in {\rm RV}^p(\alpha)$ is balanced, denoted by $\bm{X}\in {\rm BRV}^p(\alpha)$, if 
    ${v_{\bm{X}}(\{\bm{x} \in \mathbb{E}^p:x_i>1\})}={v_{\bm{X}}\left(\{\bm{x} \in \mathbb{E}^p:x_i<-1\}\right)=1}$ for all $i=1\dots,p$.
\end{definition}

\begin{prop}\label{prop:Rebalance}
    Let $\bm{X}^*=(X^*_1,\dots,X^*_p)^\top\in {\rm RV}^p(\alpha)$ with limit measure $v_{\bm{X}^*}(\cdot)$ satisfying $v_{\bm{X}^*}\left(\{\bm{x} \in \mathbb{E}^p:x_i>1\}\right)=c_i^+>0$ and $v_{\bm{X}^*}\left(\{\bm{x} \in \mathbb{E}^p:x_i<-1\}\right)=c_i^->0$, for ${i=1,\dots,p}$. Then $\bm{X}=(X_1,\dots,X_p)^\top$, with $X_i=\max\{(c_i^+)^{-1/\alpha}X^*_i,0\}+\min\{(c_i^-)^{-1/\alpha}X_i^*,0\}$ for all ${i=1,\dots,p}$, satisfies $\bm{X}\in{\rm BRV}^p(\alpha)$. 
\end{prop}

Recall that we consider $\alpha=2$ for constructing the inner product space. Whilst ${\rm BRV}^p(2)$ random vectors present a natural modeling framework for studying multivariate extremal dependence, it is not trivial to immediately quantify their associated extremal dependence. This is because the total mass of the angular measure $H_{\bm{X}}(\cdot)$ is not split evenly between the orthants of $\mathbb{R}^p$, and many of the standard tools available for multivariate extremes have been developed for ${\rm RV}^p_+(\alpha)$; see, e.g., \cite{cooley2019decompositions,lee2021transformed}. Hence, we now consider a transformation of $\bm{X}\in {\rm BRV}^p(2)$ to the positive orthant which preserves information on the degree of asymmetry in the upper and lower tails. Such a representation is briefly mentioned in \S6.5.5 of \cite{resnick2007heavy}.

Let upper tail $X_i^+:=\max\{X_i,0\}$ and lower tail $X_i^-:=-\min\{X_i,0\}$ for $i=1,\dots,p$. Then ${\bm{X}^+:=(X_1^+,\dots,X_p^+)^\top}$ and ${\bm{X}^-:=(X_1^-,\dots,X_p^-)^\top}$ contain information on the upper and lower tails, respectively, of ${\bm{X}\in {\rm BRV}^p(2)}$. We concatenate $\bm{X}^+$ and $\bm{X}^-$ into the $2p$-dimensional vector  $\bar{\bm{X}}:=\left((\bm{X}^+)^\top,(\bm{X}^-)^\top\right)^\top$.  Properties of $\bar{\bm{X}}$ are described in Proposition~\ref{prop:flatten:rv} (see the Appendix~\textcolor{blue}{A.2} for the proof).

\begin{prop}[Properties of $\bar{\bm{X}}$]\label{prop:flatten:rv}
    Let $\bm{X}\in {\rm BRV}^p(2)$ and let $\bar{\bm{X}}:=\left((\bm{X}^+)^\top,(\bm{X}^-)^\top\right)^\top$. Then $\bar{\bm{X}} \in {\rm RV}_+^{2p}(2)$ with limit measure $\nu_{\bar{\bm{X}}}(\cdot)$ that satisfies $\nu_{\bar{\bm{X}}}(\{\bm{x}\in \mathbb{E}^{2p}_+:x_i>1\})=1$ for $i=1,\dots,2p$, angular measure $H_{\bar{\bm{X}}}(\cdot)$ satisfying ${H_{\bar{\bm{X}}}(\mathbb{S}^{2p-1}_+)}=2p$, and with $\sigma(X_i^+,X_i^-)=0$ for $i=1,\dots,p$ (orthogonality property).
\end{prop}

This representation of $\bar{\bm{X}}$ has many advantages. First, the extremal dependence structure information in $\bm{X}$ is preserved in $\bar{\bm{X}}$, and any subvector of $\bar{\bm{X}}$ is still regularly varying. Second, the total angular mass is altered in a deterministic way, which is helpful for proving theoretical properties of $\bar{\bm{X}}$. Third, the orthogonality property simplifies the analysis and visualization of extremal dependence (see Section \ref{sec:meth:visual}).

\subsection{Marginal transformation to balanced regular variation}
\label{sec:meth:margin}

\textcolor{customizedColor}{In practice, the upper and lower tails of a random variable may exhibit different tail heaviness, and the margins of our asset return data, denoted here by $\mathbf{R}\in\mathbb{R}^p,$ may not be balanced, i.e., they may have different scales or tail-heaviness. However, the methodology in this paper remains applicable after standardizing the data margins to ensure balanced regular variation is satisfied. The pre-standardization of data to common margins is common practice for modeling of multivariate extremes \citep[see discussion by, e.g., ][]{naveau2024multivariate}. We describe here two marginal transformations of $\mathbf{R}$ to a balanced regularly varying random vector ${\rm BRV}^p(2)$: the first assumes that the upper and lower tails of all components of $\mathbf{R}$ are marginally regularly varying (we find this to be the case for the data in our application; see Section~\ref{sec:app} and the Appendix~\textcolor{blue}{D.2}), whilst the second is applicable for general data margins. }

\textcolor{customizedColor}{
The first transformation follows from Proposition~\ref{prop:Rebalance}: let $\lim_{r\rightarrow \infty}\mathbb{P}\{R_i>r\}/r^{-\alpha_i^+}=c_i^+$ and $\lim_{r\rightarrow \infty}\mathbb{P}\{R_i<-r\}/r^{-\alpha_i^-}=c_i^-$, where $c^{+}_i,c^{-i}_i,\alpha^{+}_i,\alpha^-_i>0$ for $i=1,\dots,p$, i.e., the upper and lower tails of each component of $\mathbf{R}$ are regularly varying, albeit with different tail index and scale. We can transform $\mathbf{R}$ to $\bm{X}$ satisfying the balanced regular variation condition, by setting
\begin{equation}
    \label{eq:transformation}
    X_i=\begin{cases} 
        (c_i^+)^{-\frac{1}{2}}\times R_i^{{\alpha}_i^+/2}, \quad& \text{for } R_i\geq 0,\\
        -(c_i^-)^{-\frac{1}{2}}\times|R_i|^{{\alpha}_i^-/2}, \quad& \text{for } R_i< 0,
       \end{cases} 
\end{equation}
for all $i=1,\dots,p$. In practice, we need to estimate the tail indices, $\alpha_i^+$ and $\alpha_i^-$, e.g., using the \cite{hill1975simple} estimator, and the scaling coefficients, $c_i^+$ and $c_i^-$, using {$q(r^{(1-q)}_i)^{\alpha_i^+}$} and $-q(r^{(q)}_i)^{\alpha_i^-},$ respectively, where $r^{(q)}_i$ denotes the $q$-quantile of $R_i$ for $q>0$ close to zero.}

\textcolor{customizedColor}{If all data margins cannot reasonably be assumed to be regularly varying, one can instead apply an empirical rank-based transformation of the data onto uniform margins before transforming them back to symmetric Pareto with unit scale and shape 2; this has distribution and quantile functions given by
\begin{equation}\label{eq:transformation:dist}
    \begin{aligned}
    F(x) & = 
    \begin{cases} 
    \displaystyle
    (\sqrt{2}-x)^{-2}, & x \leq 0, \\[10pt]
    \displaystyle
    1 - (x+\sqrt{2})^{-2}, & x > 0,
    \end{cases}
    \quad
    F^{-1}(q) & =
    \begin{cases}
    \displaystyle
    \sqrt{2} - q^{-1/2}, 
    & 0 < q \le \tfrac{1}{2}, \\[10pt]
    \displaystyle
    (1-q)^{-1/2} - \sqrt{2},
    & \tfrac{1}{2} < q < 1,
    \end{cases}
    \end{aligned}
    \end{equation}
respectively. Note that the distribution function (and its corresponding inverse) in~\eqref{eq:transformation:dist} is asymptotically equivalent to the distribution (inverse) function of $X_i$ in~\eqref{eq:transformation}.}

\textcolor{customizedColor}{The parametric tail-index-based transformation in Equation~\eqref{eq:transformation} requires that the original data have regularly varying tails, whilst the rank-based transformation in Equation~\eqref{eq:transformation:dist} has no such requirement (and is suitable for general data margins); however, it might lack efficiency when estimating the tail behavior due to the scarcity of extreme events. Sensitivity of the dependence analysis to the choice of marginal transformation should be investigated; see Section~\ref{sec:app}.}



\subsection{Visualizing directional tail asymmetry}
\label{sec:meth:visual}

To simplify the notation, we now constrain our focus to the case where $p=2$ and consider the vector $(X,Y)^\top\in {\rm BRV}^2(2)$. We introduce this new notation also to stress that $X$ is an ``explanatory variable'' for the target variable $Y$. The dependence structure in high-dimensional vectors $\bm{X}$ can later be studied in a pairwise fashion. We assume hereon that $X^+,Y^+,Y^-$ are elements of the vector space $\mathcal{V}^q_+$.

By exploiting the orthogonality property in Proposition~\ref{prop:flatten:rv}, we can visualize asymmetry in the tail dependence structure of $(X,Y)^\top$ by considering the triplet $(X^+,Y^+,Y^-)^\top$ where ${X^+=\max\{X,0\}}$, ${Y^+=\max\{Y,0\}}$, and ${Y^-=-\min\{Y,0\}}$. We consider both EDMs $\sigma(X^+,Y^+)$ and $\sigma(X^+,Y^-)$ as defined in Equation~\eqref{eq:edm}, and project these values onto the surface of $\mathbb{S}^2_+$.~
We denote this projected unit ball as the \textit{extremal ball} and define its coordinate system in Definition~\ref{def:ebCoordinate}. We can perform a similar analysis for $(-X,Y)^\top$ by instead considering the triplet $(X^-,Y^+,Y^-)^\top$.

\begin{definition}[Extremal ball]
    For $(X, Y)^\top\in {\rm BRV}^2(2)$ with $X^+,Y^+,Y^-\in \mathcal{V}^q_+$, the dependence structure in the tails can be visualized on the extremal ball with coordinates $(\theta_{X^+,Y^+},\theta_{X^+,Y^-})$, where
    \begin{equation}
        \label{eq:ebCoordinate}
    \theta_{X,Y}=\cos^{-1}\frac{\sigma({X,Y})}{\sqrt{\sigma({X,X})\sigma(Y,Y})},
    \end{equation}
    denotes the angle between $X$ and $Y$.
    \label{def:ebCoordinate}
\end{definition}

\begin{figure}[ht!]
    \centering
    \includegraphics[width=0.55\linewidth]{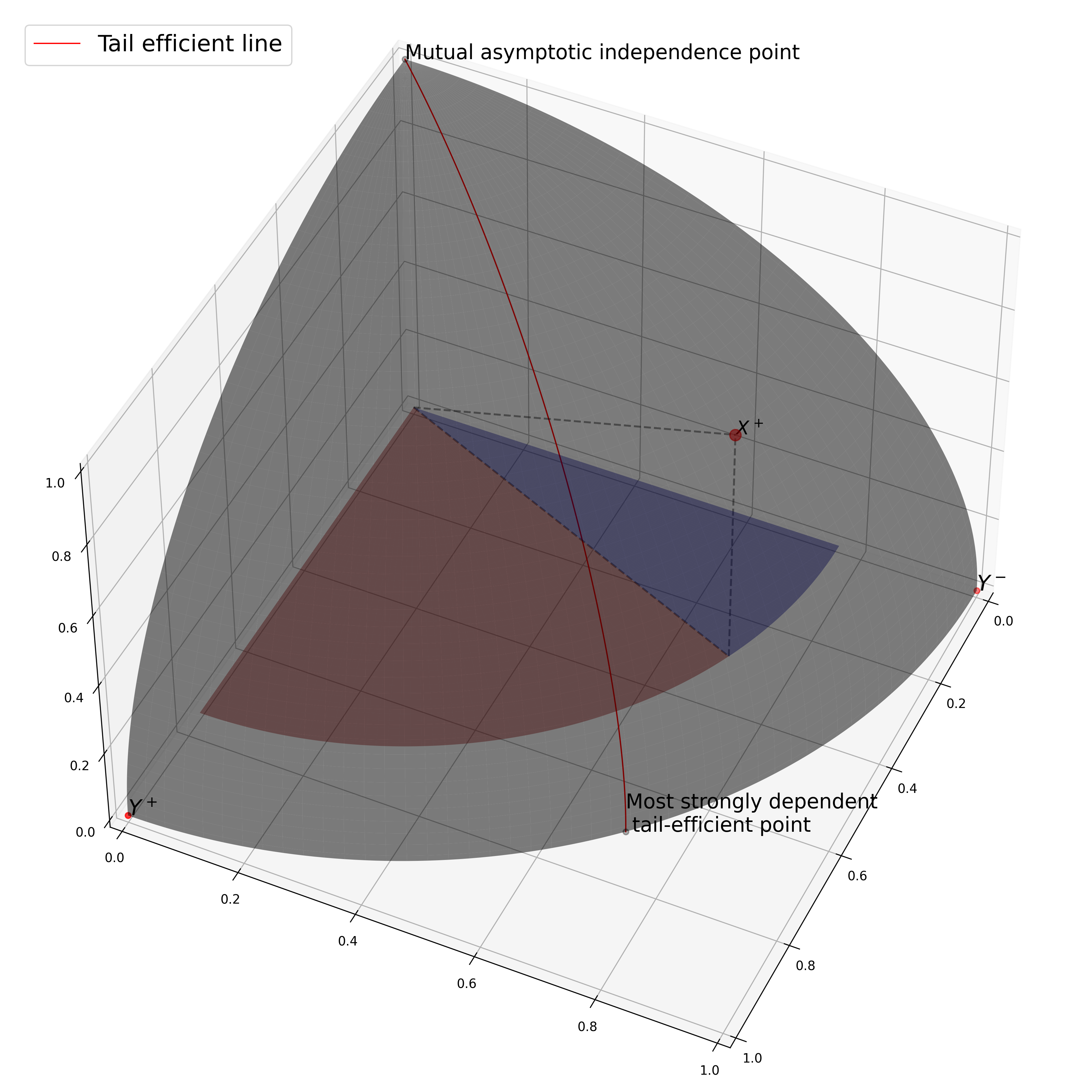}
    \caption{Illustration of the extremal ball for visualising asymmetric tail dependence.}
    \label{fig:ebAreaRatio}
\end{figure}

Figure \ref{fig:ebAreaRatio} illustrates the extremal ball for visualisation of asymmetric tail dependence. Here $Y^+$ and $Y^-$ are represented by coordinates $(1,0,0)$ and $(0,1,0)$, respectively. The angle between $Y^+$ and $Y^-$ is $90^\circ$ as $\sigma(Y^+,Y^-)=0$ and they are orthogonal in the space $\mathcal{V}^q_+$. The position of $X^+$ on the extremal ball can be uniquely determined by $\theta_{X^+,Y^+}$ and $\theta_{X^+,Y^-}$. Along the center line (\textit{tail efficient line}), the tail dependence structure is symmetric in the sense that $\sigma(X^+,Y^+)=\sigma(X^+,Y^-)$; as the coordinates deviate from the tail efficient line, they indicate stronger directional asymmetry in $(X^+,Y)^\top$, i.e., $\sigma(X^+,Y^+)\neq\sigma(X^+,Y^-)$.

The two boundary cases of the tail efficient line,  where ${\sigma(X^+,Y^+)=\sigma(X^+,Y^-)=\frac{\sqrt{2}}{4}}$ and ${\sigma(X^+,Y^+)=\sigma(X^+,Y^-)=0}$, are achieved when $X^+$ is at the \textit{most strongly dependent tail efficient point} $(\frac{\sqrt{2}}{2},\frac{\sqrt{2}}{2},0)$ and \textit{mutual asymptotic independence point} $(0,0,1)$, respectively. The most strongly dependent tail efficient point indicates the upper limit of achievable extremal dependence between $X^+$ and $Y^+$ (or $X^+$ and $Y^-$), under directional symmetry. The mutual asymptotic independence point reflects another special case where $X^+$, $Y^+$, and $Y^-$ are all mutually asymptotically independent. The extremal dependence for $(X^+,Y^+)^\top$ and $(X^+,Y^-)^\top$ are entangled, as stated in Proposition~\ref{prop:tailInequality} (see the Appendix~\textcolor{blue}{A.3} for the proof).

\begin{prop}[Extremal ball constraint]\label{prop:tailInequality}
    For $(X, Y)^\top\in {\rm BRV}^2(2)$ with $X^+,Y^+,Y^-\in \mathcal{V}^q_+$, the angles $\theta_{X^+,Y^+}$ and $\theta_{X^+,Y^-}$ satisfy the constraint $\frac{\pi}{2}-\theta_{X^+,Y^+} \leq \theta_{X^+,Y^-} \leq \frac{\pi}{2}$.
\end{prop}

This proposition shows that knowledge of extremal dependence in $(X^+,Y^+)^\top$ provides an upper bound on the extremal dependence between $X^+$ and $Y^-$. This theoretical result suggests the existence of interaction between the upper and lower tails of ${\rm BRV}^2(2)$ random vectors and rules out some possible extremal dependence structures for $(X^+,Y^+,Y^-)^\top$.

\subsection{\tename\ measure}
\label{sec:meth:asymmetric}

We now define the concept of \tenameLower\ (\tenameShort) in Definition~\ref{def:TE}.

\begin{definition}[\tename\ measure]\label{def:TE}
    For $(X,Y)^\top\in {\rm BRV}^2(2)$, the \tenameLower\ measure, $\lambda(X,Y)$, is defined by
    \begin{equation}
        \lambda(X,Y)=\int_{\{(\omega_x,\omega_y)^\top\in \mathbb{S}^1:\omega_x\geq 0\}} \omega_x\omega_y {\rm d}H_{(X,Y)}(\bm{\omega}),
    \end{equation}
    where $\bm{\omega}=(\omega_x,\omega_y)^\top \in \mathbb{S}^1$.
\end{definition}

The \tenameShort\ measure can be used to describe the dependence from target $Y$ on the explanatory variable $X$, particularly when focusing on extremal situations. The case $\lambda(X,Y)=0$ implies that the dependence structure for $(X^+,Y^+,Y^-)^\top$ is directionally symmetric. We design this measure to satisfy certain desirable properties; see, e.g., \cite{embrechts2002correlation}. Specifically, the \tenameShort\ measure satisfies the following two properties as stated in Proposition~\ref{prop:propertiesOflambda} (see the Appendix~\textcolor{blue}{A.4} for the proof).

\begin{prop}[\textcolor{customizedColor}{Properties of \tenameLower\ $\lambda$}]
    \label{prop:propertiesOflambda}
    \textcolor{customizedColor}{The directional tail dependence $\lambda(X,Y)$ satisfies ${\lambda(X,Y)\in[-1,1]}$, where ${\lambda(X,Y)=-1}$ and ${\lambda(X,Y)=1}$ are achieved when ${\lim_{a\rightarrow \infty}\mathbb{P}[X>a \mid Y<-a]=1}$ and ${\lim_{a\rightarrow \infty}\mathbb{P}[X>a\mid Y>a]=1}$, respectively. Furthermore, $\lambda(X,Y)$ is an odd function of $Y$, i.e., $\lambda(X,-Y)=-\lambda(X,Y)$ for all $Y$ (this property does not hold for $X$).}
\end{prop}

The boundary cases ${\lim_{a\rightarrow \infty}\mathbb{P}[X>a \mid Y<-a]=1}$ and ${\lim_{a\rightarrow \infty}\mathbb{P}[X>a\mid Y>a]=1}$ correspond to perfect asymptotic dependence between the upper tail of $X$ and, respectively, the lower and upper tails of $Y$. Moreover, $\lambda(X,Y)$ is an odd function of $Y$ only. Thus, $\lambda(X,Y)$ and $\lambda(-X,Y)$ describe the asymmetry in the influence on $Y$ from the upper tail and lower tail, respectively, of $X$. In summary, the \tenameLower\ measure is a flexible description of directional asymmetry in the tails of $(X,Y)^\top$, and it is directional, and so has potential usage in causal-effect analyses. 

Estimation of $\lambda$ does not follow directly from the limit form of the conditional expectation in Equation \eqref{eq:edm}, as the angular mass on the right half circle, $H_{(X,Y)}(\{(\omega_x,\omega_y)^\top\in \mathbb{S}^1: \omega_x\geq 0 \})$, is, in practice, unknown and difficult to estimate empirically. We can instead exploit two equivalent forms for $\lambda(X,Y)$, given in Proposition~\ref{prop:linkWithEDM}.

\begin{prop}
    \label{prop:linkWithEDM}
    \quad\vspace{-3mm}
    Consider the \tenameLower\ $\lambda(X,Y)$ as in Definition~\ref{def:TE}. Then
    \begin{equation}
        \label{eq:linkWithEDM}
        \lambda(X,Y)=3\int_{\{(\omega_x,\omega_y)^\top\in \mathbb{S}^1:\omega_x\geq 0\}}\omega_x\omega_y{\rm d}N_{(X^+,Y)}(\bm{\omega})=2\{\sigma(X^+,Y^+)-\sigma(X^+,Y^-)\},
    \end{equation}
    where $N_{(X^+,Y)}(\cdot)$ is the normalized angular measure of $(X^+,Y)^\top \in {\rm BRV}^2(2)$.
\end{prop}

We can view the first form of $\lambda$ in Equation \eqref{eq:linkWithEDM} as an extension of the EDM for the vector $(X^+,Y)^\top$, which is regularly varying on the right half space of $\mathbb{R}^2$. The second form of $\lambda$ is the difference between the quadrant-specific measures $\sigma(X^+,Y^+)$ and $\sigma(X^+,Y^-)$. These two forms are equivalent, but permit two different estimators; for the proof of Proposition~\ref{prop:linkWithEDM}, see the Appendix~\textcolor{blue}{A.5}.

\begin{definition}\label{def:TEestimator}
    Let $\{(x_i,y_i)^\top\}_{i=1}^{n}$ be $n$ independent replicates of $(X,Y)^\top$. Then the \tenameShort\ measure, $\lambda(X,Y)$, may be estimated using the following two estimators:
    
    1) $\widehat{\lambda}^1_n=3\frac{1}{N}\sum_{i=1}^{n}\frac{x^+_{i}y_{i}}{r^2_i}\mathbbm{1}(r_i\geq r_0)$, where $r_i=||(x^+,y)^\top||_2$, $N=\sum_{i=1}^n \mathbbm{1}(r_i>r_0)$, and $r_0$ is a suitably-chosen high threshold;
    
    2) $\widehat{\lambda}^2_n=2\left(\frac{1}{N^+}\sum_{i=1}^{n}\frac{x^+_{i}y^+_{i}}{(r_i^+)^2}\mathbbm{1}(r_i^+\geq r^+_0)-\frac{1}{N^-}\sum_{i=1}^{n}\frac{x^+_{i}y^-_{i}}{(r_i^-)^2}\mathbbm{1}(r_i^-\geq r^-_0)\right) $, where $r_i^+$ and $r_i^-$ equals to $||(x^+_i,y^+_i)^\top||_2$ and $||(x_i^+,y_i^-)^\top||_2$ respectively; and $N^+=\sum_{i=1}^n \mathbbm{1}(r^+_i>r^+_0)$, {$N^-=\sum_{i=1}^n \mathbbm{1}(r^-_i>r^-_0)$}, where $r^+_0$ and $r^-_0$ are suitably-chosen high thresholds. 
\end{definition}

The estimator $\widehat{\lambda}^1_n(X,Y)$ is asymptotically normal under similar assumptions as the EDM; see \cite{larsson2012extremal}, Theorem 1, and Proposition~\ref{prop:asymptoticNormality} below (proved in the Appendix~\textcolor{blue}{A.6}).


\begin{prop}[Asymptotic normality of $\hat{\lambda}^1_n$]
    \label{prop:asymptoticNormality}
        Let $(X,Y)^\top\in {\rm BRV}^2(2)$ with ${\omega^+_{x}=\frac{X^+}{R}}$, ${\omega_{y}=\frac{Y}{R}}$, where $X^+=\max\{X,0\}$ and $R=||(X^+,Y)^\top||_2$. Let $a(n)$ be a function satisfying $a(n)\rightarrow \infty$ and $n\mathbb{P}\{R>a(n)\}\rightarrow 1$ as $n\to \infty$. Define the ``bias'' process by
        \[B_n(t)=\frac{n}{\sigma_n\sqrt{k}}(\mathbb{E}\left[\omega^+_{x}\omega_{y}\mathbbm{1}\{a(n/k)^{-1}R\geq t\}\right]-\frac{\lambda}{3}\mathbb{P}[a(n/k)^{-1}R\geq t]),\]
    
        where $\sigma^2_{n}={\rm Var}[\omega^+_{x}\omega_{y}\mid a(n/k)^{-1}R\geq t]$. Assume that $n,k\to\infty$, $n/k\to \infty$, and that ${\lim_{n\to \infty}B_n(1)\xrightarrow{P} 0}$. Then,
        \begin{equation}
            \label{eq:clt}
            \frac{\sqrt{k}}{3}(\hat{\lambda}^1_n-\lambda)\xrightarrow{d}N(0,\sigma^2),
        \end{equation} 
        where $\sigma^2={\rm Var}[\tilde{\omega}^+_{x}\tilde{\omega}_{y}]$, $\lambda=3\mathbb{E}[\tilde{\omega}^+_{x}\tilde{\omega}_{y}]$, and $(\tilde{\omega}^+_{x},\tilde{\omega}_{y})^\top$ has distribution $N_{(X^{+},Y)}(\cdot)$.
\end{prop}

\textcolor{customizedColorRound2}{The assumptions $k \rightarrow \infty$ and $n/k \rightarrow \infty$ ensure that the number of samples with radius $R$ exceeding a high threshold grows to infinity, which guarantees the convergence of the angular component $(\omega^+_{x}, \omega_{y})$ to the limiting distribution $N_{(X^{+},Y)}(\cdot)$. The second assumption, ${\lim_{n \to \infty} B_n(1) \xrightarrow{P} 0}$, implies that as $R \rightarrow \infty$, the dependence between the angular component $(\omega^+_{x}, \omega_{y})$ and the radius $R$ diminishes, and that the first term in $B_n(t)$ can be decomposed.} Asymptotic properties for $\widehat{\lambda}^2_n$ are more difficult to derive and, instead, we construct a permutation test based on this statistic to check the Efficient Tail Hypothesis; see Section~\ref{sec:meth:eth}. 

\subsection{Efficient Tail Hypothesis (ETH)}
\label{sec:meth:eth}

The situation where $\lambda(X,Y)=0$ is of particular interest in financial markets; see, e.g., \cite{bouaddi2022systematic}. For $(X,Y)^\top \in {\rm BRV}^2(2)$, we say that $Y$ is \textit{tail-efficient} with respect to $X$ if $\lambda(X,Y)=0$. We can extend this concept to a setting where $\bm{X}$ and $\bm{Y}$ are vectors of interest.

For a $p_1$-dimensional random vector $\bm{X}$ and a $p_2$-dimensional random vector $\bm{Y}$ satisfying ${\left(\bm{X}^\top,\bm{Y}^\top\right)^\top\in {\rm BRV}^{p_1+p_2}(2)}$, we posit the pairwise ETH in Definition \ref{hyp:vector}.

\begin{definition}[Pairwise Efficient Tail Hypothesis]\label{hyp:vector}
    For $\left(\bm{X}^\top,\bm{Y}^\top\right)^\top\in {\rm BRV}^{p_1+p_2}(2)$, the Efficient Tail Hypothesis states that all pairs $(X_i,Y_j)^\top$, $i=1,\dots,p_1$, $j=1,\dots,p_2$, are tail-efficient, that is, $\lambda(X_i,Y_j)=0$.
\end{definition}

\textcolor{customizedColorRound2}{In the application to financial markets (Section~\ref{sec:app}), we let the target random variables, contained within $\bm{Y},$ denote financial asset returns, and the explanatory variables, in $\bm{X}$, represent available prior information available. The null hypothesis of the ETH states that the explanatory variables exhibit symmetric tail dependence with respect to both the upper and lower tails of asset returns. When the ETH is rejected, it suggests that the explanatory variables can help to predict the direction of large price movements in the target assets, indicating market inefficiency. However, when the ETH cannot be rejected, one cannot conclude that the market is efficient, not only because statistical tests can never be used to formally confirm a hypothesis when it is true---they can only provide evidence against it when it is false---but more importantly because the statistic $\lambda$ focuses only on extreme events.}

For a bivariate vector $(X,Y)^\top$, either of the estimators introduced in Definition~\ref{def:TEestimator} can be used to test for tail efficiency. For the first estimator $\widehat{\lambda}^1_n$, a standard $t$-test based on its asymptotic normality can be used. However, asymptotic normality holds under the strict assumption that $k,n\rightarrow\infty$ and $k/n\rightarrow 0$. In practice, careful selection of the threshold $r_0$ is crucial to meet the asymptotic assumption. We instead propose a finite-sample permutation test for tail efficiency which is a non-parametric test and requires fewer assumptions. Thanks to the property ${\lambda(X,Y)=-\lambda(X,-Y)}$, we have $\lambda(X,-Y)=\lambda(X,Y)$ under the null hypothesis; we can then construct a permutation test for tail efficiency by considering reflections of $Y$ about zero.

Let $\widehat{\lambda}_n$ be an estimator of $\lambda(X,Y)$ using sample $\{(x_i,y_i)^\top\}_{i=1}^{n}$ and $\widehat{\lambda}_n^{(l)}$ be an estimator based on its $l$-th permutation $\{(x_i,y^{(l)}_i)^\top\}_{i=1}^{n}$ where $y^{(l)}_i=y_i$ with probability $0.5$ and ${y^{(l)}_i=-y_i}$, otherwise. Note that we can use either the first or second estimator of Definition~\ref{def:TEestimator}. In practice, considering all possible permutations of $Y$ is infeasible with large $n$; we instead use a random subset of permutations with a size $P$ to approximate the p-value. Then, the (approximate) p-value of the permutation test with $P$ permutations is
\begin{equation}
    \frac{1}{P}\sum_{l=1}^{P}\mathbbm{1}(|\widehat{\lambda}^{(l)}_n|>|\widehat{\lambda}_n|).
    \label{eq:pvalue}    
\end{equation}

Testing of the ETH for vectors $\bm{X}$ and $\bm{Y}$ requires multiple tests, which increases the risk of obtaining false positive results, i.e., incorrectly rejecting a null hypothesis. Many methods have been proposed to correct the multiple testing problem \citep[see, e.g.,][]{bonferroni1936teoria,holm1979simple, benjamini1995controlling}. In Section~\ref{sec:app}, we use the Benjamini--Hochberg correction to test for tail efficiency of China's futures market.

\section{Simulation study}
\label{sec:simulation}

Here we investigate the power of the tail efficiency tests for bivariate random vectors. Section \ref{sec:sim:design} introduces the simulation design and Section \ref{sec:sim:results} presents the results.

\subsection{Simulation design}
\label{sec:sim:design}

To simulate bivariate random vectors ${(X,Y)}^\top\in {\rm BRV}^2(2)$, we first simulate random vectors $(U',V')^\top$ with uniform margins using a copula which exhibits extremal dependence. Then, we use a mixture to simulate vectors ${(U,V)}^\top$ with different extents of tail asymmetry, before marginally transforming ${(U,V)}^\top\mapsto {(X,Y)}^\top\in {\rm BRV}^2(2)$ to the Symmetric Pareto scale; see Equation \eqref{eq:transformation:dist}.

Specifically, we construct $(U,V)^\top$ as 
\begin{equation} 
    \label{eq:mixtureCopula}
    U = U', \quad V = \begin{cases} V' & \text{with probability }\phi , \\ 1-V' & \text{otherwise,}\end{cases}    
\end{equation}
where $\phi\in (0,1)$ is  the mixing probability. For $(U,V)^\top$, we consider the Student's $t$-copula \citep{demarta2005t}, with degrees of freedom $\nu=4$ and correlation $\rho > 0$, and the Gumbel copula \citep{gumbel1960distribution}, with dependence parameter $\theta \in [1,\infty)$. For the Student's $t$-copula, the tail dependence strength increases with $\rho$; for the Gumbel copula, the tail dependence strength increases with $\theta$. When $\phi=0.5$, $(X,Y)^\top$ exhibits tail symmetry, i.e., $\lambda(X,Y)=0$, and  $\lambda(X,Y)\neq 0$, otherwise, with increasing asymmetry as $|\phi-0.5|$ increases.

\begin{figure}[htp!]
    \centering
    \begin{subfigure}[b]{0.44\textwidth}
        \caption*{Symmetric}
        \includegraphics[width=\linewidth]{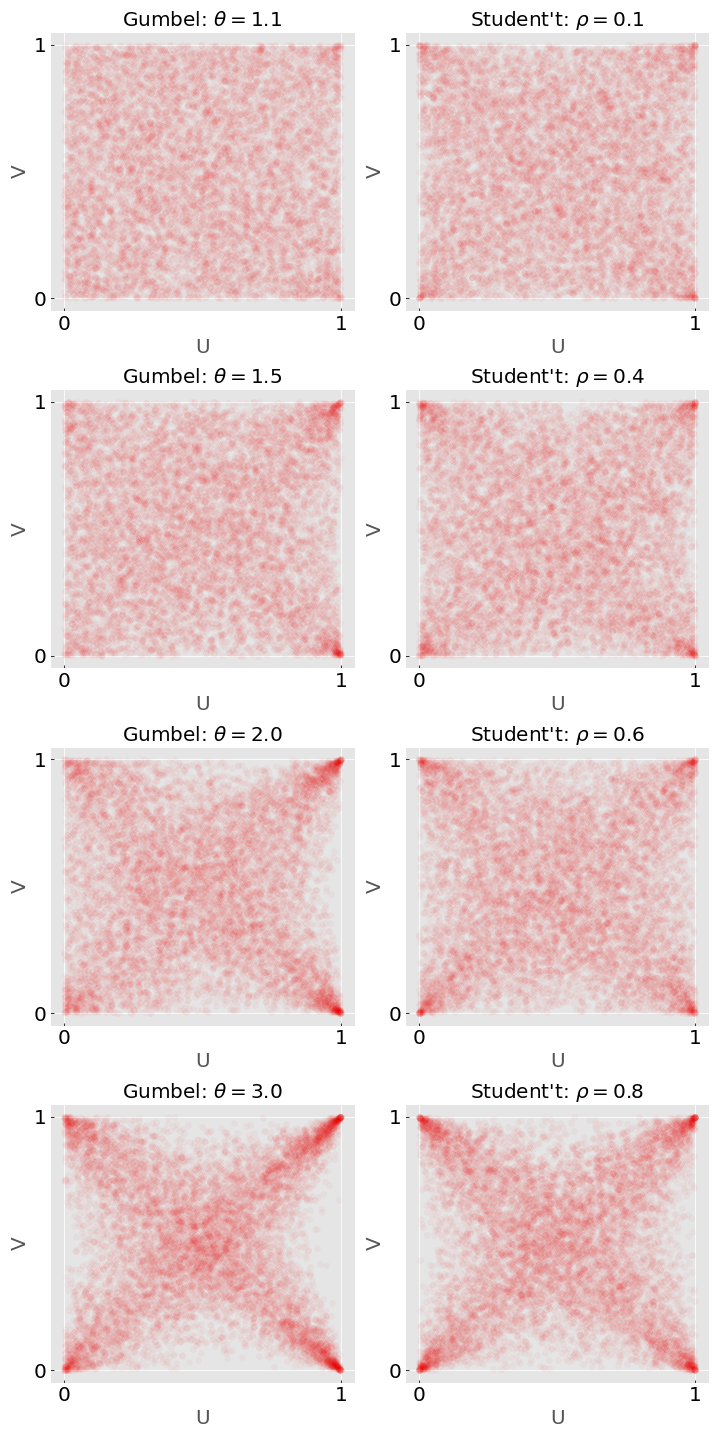}
        \label{fig:sim:sym:uniform}
    \end{subfigure}
    \hspace{5mm}
    \begin{subfigure}[b]{0.44\textwidth}
        \caption*{Asymmetric}
        \includegraphics[width=\linewidth]{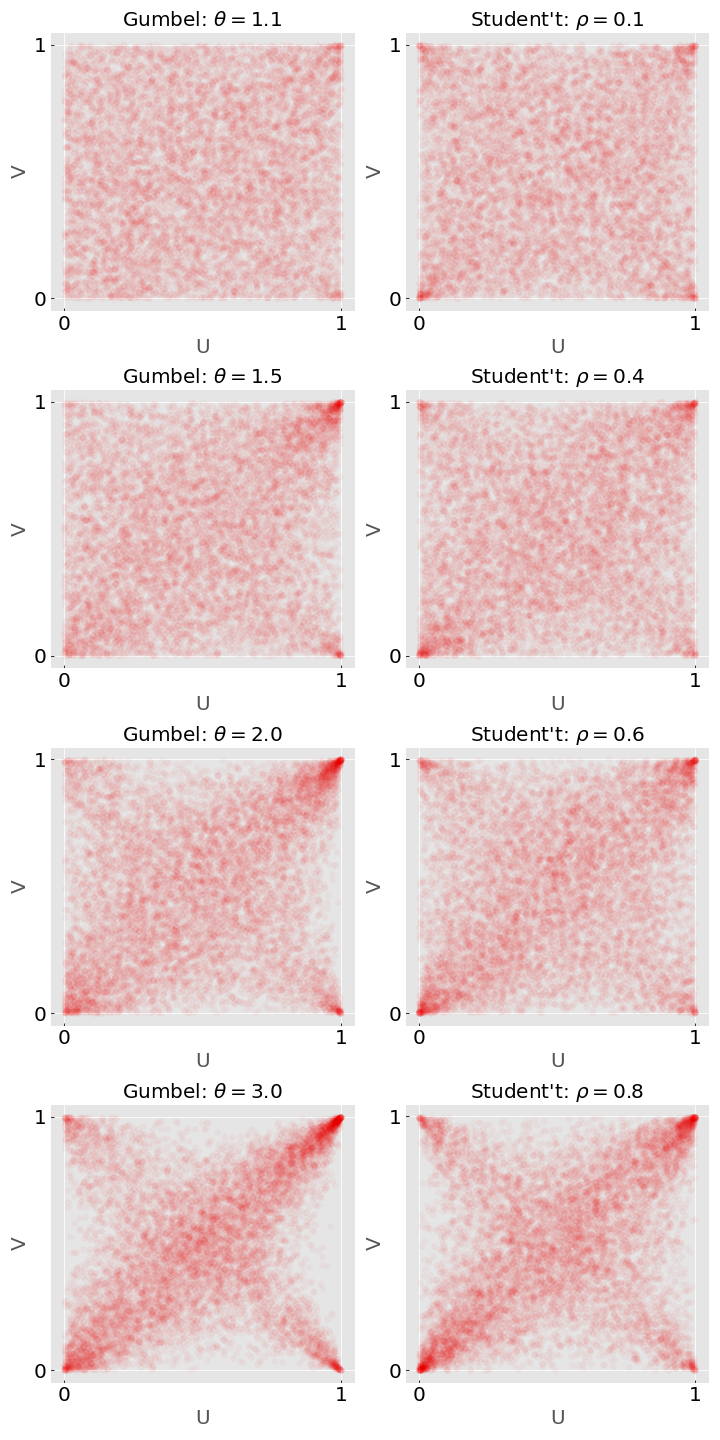}
        \label{fig:sim:assym:uniform}
    \end{subfigure}
    \caption{Scatter plots for simulated $(U,V)^\top$ under the null hypothesis with $\phi=0.5$ (Left two columns) and alternative hypothesis with $\phi=0.7$ (Right two columns). The underlying copula differs with the columns: the first and third columns are scatter points from the Gumbel copula and the second and fourth columns are scatter points from the Student's $t$-copula. Rows correspond to different parameters for the copulas: for the Gumbel copula, $\theta\in\{1.1,1.5,2.5,3.0\}$; for the Student's $t$-copula, $\rho\in\{0.1,0.4,0.6,0.8\}$ with degrees of freedom $ \nu$ fixed to be $4$.}
    \label{fig:sim:uniform}
\end{figure}

Scatter plots for $(U,V)^\top$ are shown in Figure \ref{fig:sim:uniform}. We use four different settings for both copula families: for the Student's $t$-copula, we take $\rho \in \{0.1, 0.4, 0.6, 0.8\}$ and for the Gumbel copula, we take $\theta\in\{1,1.2,1.5,2\}$. For the mixing probability $\phi$, we consider $\phi= 0.5$ for sampling under the null hypothesis and $\phi= 0.7$ for the alternative. Figure \ref{fig:sim:uniform} illustrates asymmetry in the quadrants when $\phi=0.7$. 

\subsection{Results}
\label{sec:sim:results}

\begin{figure}[htp!]
    \centering
    \begin{subfigure}[b]{0.44\textwidth}
        \caption*{Symmetric}
        \includegraphics[width=\linewidth]{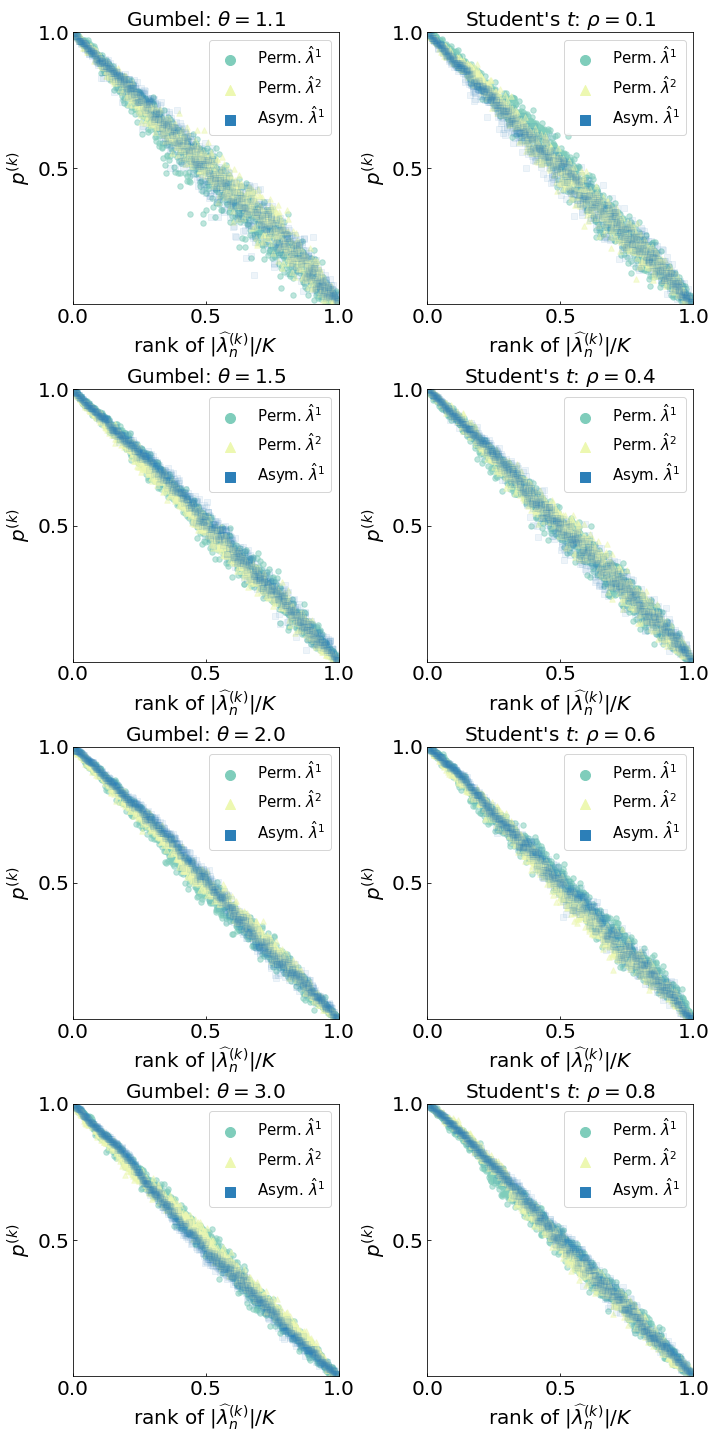}
    \end{subfigure}
    \vspace{5mm}
    \begin{subfigure}[b]{0.44\textwidth}
        \caption*{Asymmetric}
        \includegraphics[width=\linewidth]{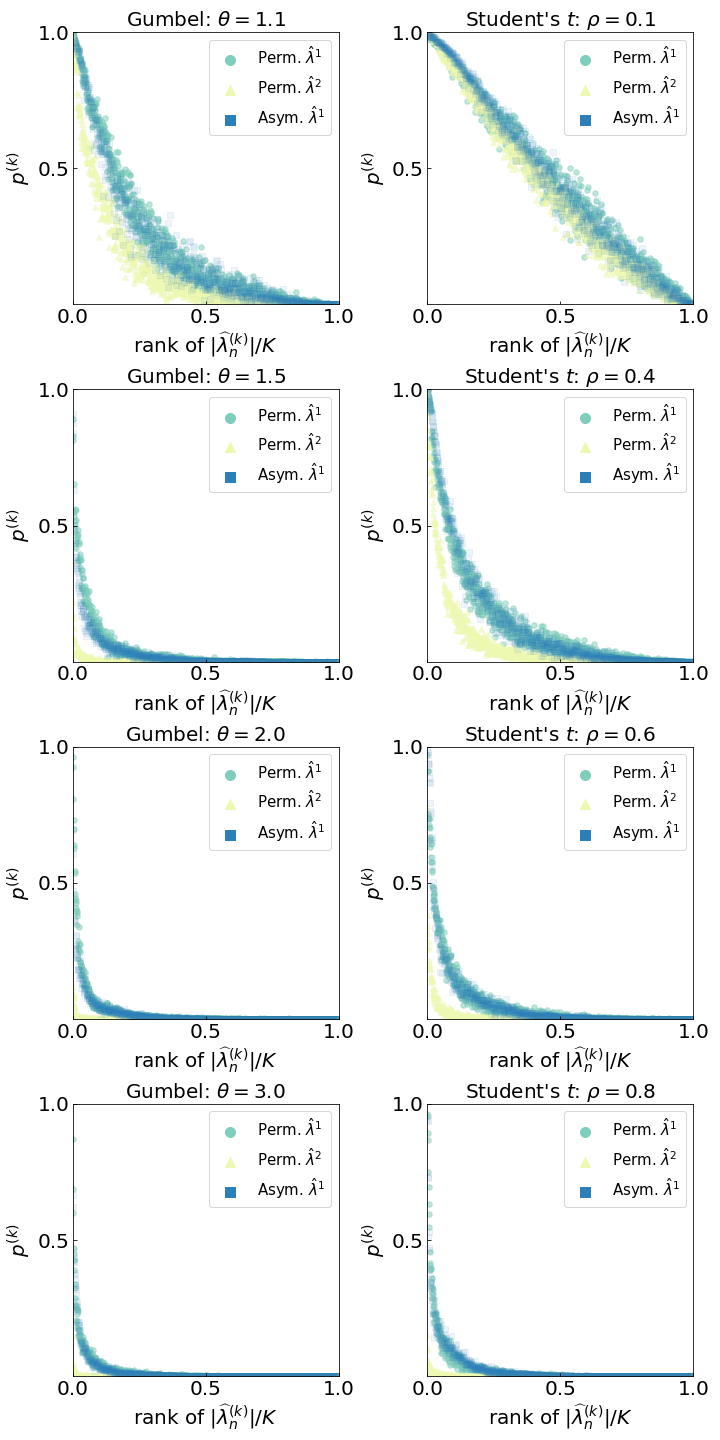}
    \end{subfigure}
    \caption{Plot of p-values ($y$-axis) against empirical rank of test statistic $|\widehat{\lambda}_n^{(k)}(X,Y)|/K$ ($x$-axis); case $\lambda(X,Y)=0$ (Left two columns) and case $\lambda(X,Y)\neq0$ (Right two columns); Circular (green) and triangular (yellow) points correspond to permutation tests using $\widehat{\lambda}_n^1$ and $\widehat{\lambda}^2_n$, respectively; Square (blue) points correspond to $t$-tests using $\widehat{\lambda}_n^1$. The underlying copula differs with the columns: the first and third columns are results for the Gumbel copula and the second and fourth columns are results for the Student's $t$-copula. Rows correspond to different parameters for the copulas: for the Gumbel copula, $\theta\in\{1.1,1.5,2.5,3.0\}$; for the Student's $t$-copula, $\rho\in\{0.1,0.4,0.6,0.8\}$ with degrees of freedom $ \nu$ fixed to be $4$.}
    \label{fig:sim:test}
    \end{figure}

For each setting, we test the null $\mathcal{H}_0: \lambda(X,Y)=0$ for $K=1,000$ repetitions. For each test, we simulate independent samples $\{(x_i,y_i)^\top\}_{i=1}^{n}$ with sample size $n=10,000$ and calculate the test statistic with threshold $r_0$ taken to be the empirical $q=0.99$-quantile of $\{r_i=||(x_i,y_i)^\top||_2\}_{i=1}^n$ (and similarly for $r_0^+$ and $r_0^-$). We compare the permutation test (with $P=1,000$) for both estimators $\widehat{\lambda}^1_n$ and $\widehat{\lambda}^2_n$ in Definition~\ref{def:TEestimator}, with the $t$-test based on the asymptotic normality of $\widehat{\lambda}^1_n$. In the end, for each of the three test types, we have $K$ independent pairs of test statistics $|\widehat{\lambda}_n^{(k)}|$ and their corresponding p-value, $p^{(k)}$, for $k=1,\dots,K$. Figure~\ref{fig:sim:test} visualises the relationship between the test statistics and their p-values. We also investigate the effect of the sample size $n$ and lower exceedance thresholds (see the Appendix~\textcolor{blue}{C}); the results are similar, and so are not discussed further here.

    Under the null hypothesis, $p^{(k)}$ should be roughly equal to $1-{\rm rank}(|\widehat{\lambda}^{(k)}_n)|/ K$. In Figure~\ref{fig:sim:test}, left column-block, we see that all points are clustered around the second diagonal under symmetry, which means that the rejection percentages for each significance level are approximately equal to the corresponding nominal value (see the Appendix~\textcolor{blue}{C} for further analysis on the empirical sizes of the three tests).  From the right column-block of Figure~\ref{fig:sim:test}, which displays the results when data are simulated under the alternative hypothesis (with $\phi=0.7$), all points are below the diagonal line. Points closer to the $x$-axis indicate tests with larger power of rejection. 

    The power of the tests, estimated as the percentage of rejections of the null hypothesis, when data are generated under the alternative hypothesis is investigated in Table~\ref{tab:power:more} and, as expected, the statistical power for all three testing methods becomes higher with increasing tail asymmetry (across descending rows in Figure~\ref{fig:sim:test}). The permutation test using the second estimator $\widehat{\lambda}^2_n$ have the strongest power among the three testing methods. This may be attributed to its data efficiency, as $\widehat{\lambda}^2_n$ uses more unique data than $\widehat{\lambda}^1_n$. Recall from Definition~\ref{def:TEestimator} that $\widehat{\lambda}^2_n$ is an estimator for ${\lambda(X,Y)=2\{\sigma(X^+,Y^+)-\sigma(X^+,Y^-)\}}$, which involves estimating the EDM for $\sigma(X^+,Y^+)$ and $\sigma(X^+,Y^-)$, separately. For each permuted samples, the set of indices for which the $l_2$-norm exceeds the threshold $r^+_0$ and $r_0^-$ will vary; for estimator $\widehat{\lambda}^1_n$, this set remains consistent across all $P$ permutations. The advantage of using $\widehat{\lambda}^2_n$ comes with a higher computational cost as estimation requires re-selecting the exceedances for each permutation. Another advantage of using $\widehat{\lambda}^2_n$ is that it has the flexibility of choosing two thresholds, $r_0^+$ and $r_0^-$, instead of a single threshold $r_0$. In our application in Section~\ref{sec:app}, we use the permutation-based test with estimator $\widehat{\lambda}^2_n$.

    \begin{table}[t!]
        \centering
        \caption{Power of the three proposed statistical tests (i.e., percentage of null hypotheses rejected) at significance levels $\alpha^*=0.01$ and $\alpha^*=0.05$, across different copulas and sample sizes.}
        \footnotesize
        \renewcommand{\arraystretch}{0.40}
        \begin{tabular}{cccc|ccc|ccc}
        \toprule
        \multirow{2}{*}{$n$} & \multirow{2}{*}{copula} & \multirow{2}{*}{$\theta$ or $\rho$} & \multicolumn{1}{c|}{} & \multicolumn{3}{c|}{$\alpha^*=0.01$} & \multicolumn{3}{c}{$\alpha^*=0.05$} \\
        \cmidrule(lr){5-7} \cmidrule(lr){8-10}
        & & & & Perm. $\hat{\lambda}_n^1$ & Perm. $\hat{\lambda}_n^2$ & Asym. $\hat{\lambda}_n^1$ & Perm. $\hat{\lambda}_n^1$ & Perm. $\hat{\lambda}_n^2$ & Asym. $\hat{\lambda}_n^1$ \\
        \midrule
        \multirow{8}{*}{1000} 
        & \multirow{4}{*}{Gumbel} 
        & $\theta=1.1$ & & 5 & 10 & 7 & 18 & 26 & 19 \\
        & & $\theta=1.5$ & & 51 & 83 & 51 & 74 & 95 & 77 \\
        & & $\theta=2.0$ & & 66 & 96 & 67 & 86 & 99 & 87 \\
        & & $\theta=3.0$ & & 75 & 98 & 74 & 92 & 100 & 91 \\
        \cmidrule(lr){2-10}
        & \multirow{4}{*}{Student's $t$} 
        & $\rho=0.1$ & & 2 & 3 & 2 & 9 & 10 & 10 \\
        & & $\rho=0.4$ & & 22 & 40 & 22 & 44 & 66 & 45 \\
        & & $\rho=0.6$ & & 46 & 77 & 49 & 71 & 93 & 74 \\
        & & $\rho=0.8$ & & 67 & 94 & 67 & 86 & 98 & 88 \\
        \midrule
        \multirow{8}{*}{10000} 
        & \multirow{4}{*}{Gumbel} 
        & $\theta=1.1$ & & 18 & 31 & 18 & 38 & 57 & 37 \\
        & & $\theta=1.5$ & & 62 & 93 & 64 & 84 & 99 & 84 \\
        & & $\theta=2.0$ & & 74 & 98 & 73 & 90 & 99 & 90 \\
        & & $\theta=3.0$ & & 76 & 98 & 78 & 91 & 100 & 92 \\
        \cmidrule(lr){2-10}  
        & \multirow{4}{*}{Student's $t$} 
        & $\rho=0.1$ & & 2 & 4 & 3 & 9 & 13 & 10 \\
        & & $\rho=0.4$ & & 24 & 51 & 28 & 49 & 74 & 51 \\
        & & $\rho=0.6$ & & 52 & 87 & 53 & 76 & 96 & 75 \\
        & & $\rho=0.8$ & & 70 & 96 & 68 & 88 & 100 & 85 \\
        \bottomrule
        \end{tabular}
        \label{tab:power:more}
    \end{table}

\section{Application}
\label{sec:app}

\subsection{Overview and basic data analysis}
\label{sec:app:overview}
We apply the framework developed in this paper to analyze the market-wide extremal dependence of China's futures market. The data are represented as a multivariate time series of percentage price changes of futures for commodities with high liquidity. The return is calculated at 30-second intervals and the detailed procedure of data generation and access are given in the Appendix~\textcolor{blue}{B}. We consider a subset of the dataset containing 55 assets, which covers a wide range of commodities, including agricultural products, base metals, and ferrous metals. We select a one-year period of data from \TrainstartDate\ to \TestendDate, which contains $T=\;$\NumObs\ observations.

\textcolor{customizedColor}{For each of the $55$ assets, we study bulk and tail temporal dependence by estimating their autocorrelation and extremogram \citep{davis2012towards} functions, respectively. The estimated (extremal) dependence decays rapidly, dropping below values of $0.05$ and $0.1$ (for the auto-correlation and extremogram, respectively) after $10$ time lags. Therefore, to remove temporal dependence from the data, we subsample every tenth observation. A detailed description of this dependence analysis is deferred to the Appendix~\textcolor{blue}{D.1}. After sub-sampling, we transform the data to be balanced regularly varying. In the Appendix~\textcolor{blue}{D.2}, we investigate the marginal tail behaviour of the data and find reasonable evidence to suggest that the upper and lower tails of each asset are regularly varying; we thus employ the first marginal transformation as described in Section~\ref{sec:meth:margin}. Unreported experiments found this choice to have negligible effect on the final results. We denote the transformed data by $\bm{Z}_t=(Z_{1,t},\dots,Z_{55,t})^\top$ for $t=1,\dots, T$.}

\textcolor{customizedColor}{The rest of this section is organized as follows: In Section~\ref{sec:app:visualization}, we visualize extremal dependence in the market using the extremal ball. In Section~\ref{sec:app:tail_efficiency}, we test the ETH for China's futures market and interpret our findings. In Section~\ref{sec:app:portfolio}, we capitalize on our test results to construct an artificial dynamic portfolio consisting of tail-inefficient assets and backtest an investment strategy on out-of-sample data. We also study the uncertainty arising from the marginal transformation via a bootstrap experiment, which is deferred to the Appendix~\textcolor{blue}{D.4}.}



\subsection{Visualization of market-wide extremal dependence}
\label{sec:app:visualization}

\begin{figure}[t!]
    \centering
    \includegraphics[width=0.55\linewidth]{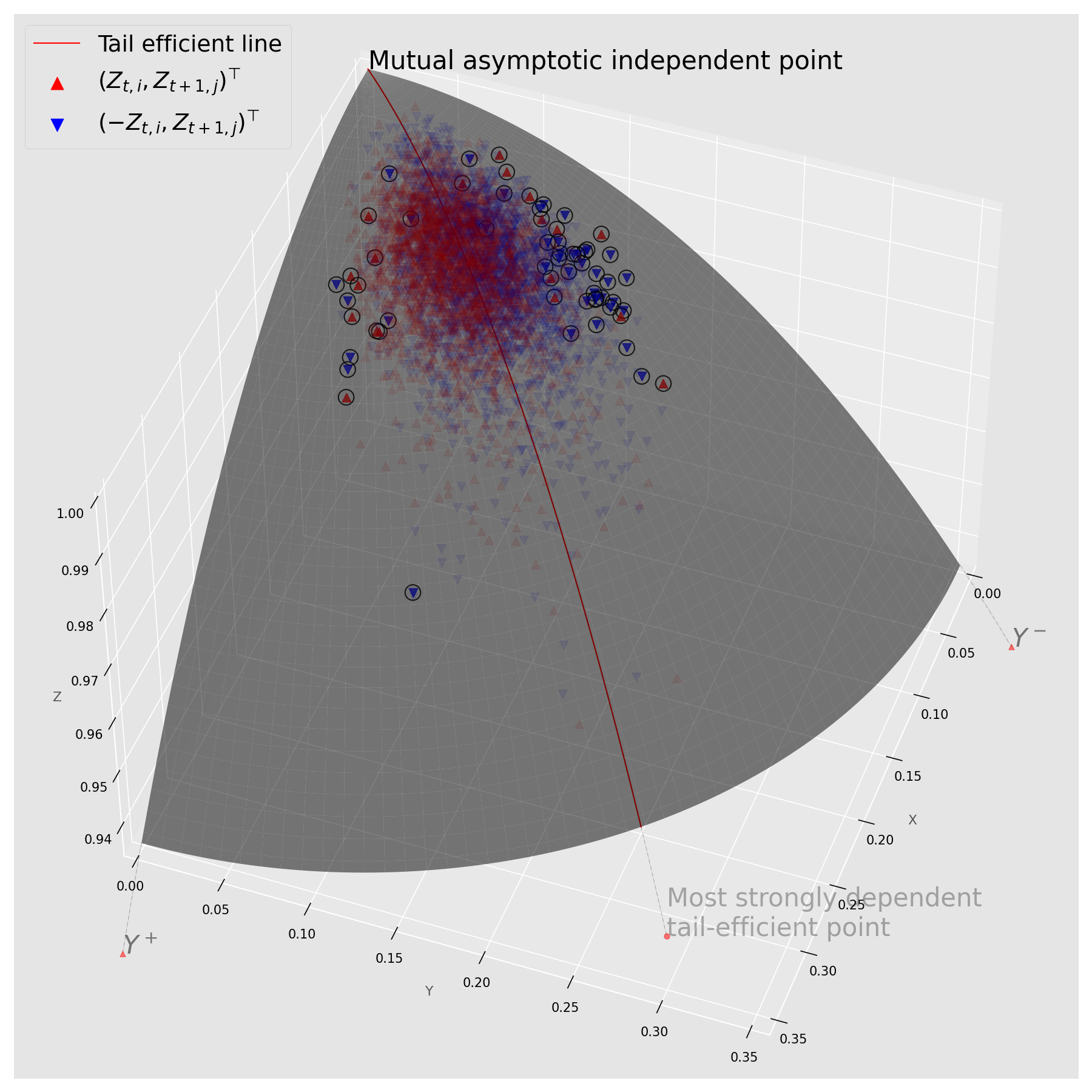}
    \caption{Extremal ball for one-step forward dependencies in China's futures market. \textcolor{customizedColor}{Significant tail-inefficient pairs (p-values < \significantLevel) are circled} based on the permutation test using $\hat{\lambda}^2_n$. Red: $(Z_{i,t},Z_{j,t+1})^\top$; Blue: $(-Z_{i,t},Z_{j,t+1})^\top$.}
    \label{fig:ExtremeBall}
\end{figure}

\textcolor{customizedColor}{We are specifically interested in one-step forward extremal dependence, i.e., the extremal dependence in the bivariate vectors $(-Z_{i,t},Z_{j,t+1})^\top$ and $(Z_{i,t},Z_{j,t+1})^\top$ where $i,j=1,\dots,55$.} Figure~\ref{fig:ExtremeBall} visualises directional tail asymmetry in the market-wide extremal dependence for all pairs $(-Z_{i,t},Z_{j,t+1})^\top$ and $(Z_{i,t},Z_{j,t+1})^\top$ where $i,j=1,\dots,55$. The former group corresponds to those where the driving event is extremal loss, whilst the latter corresponds to pairs where the driving asset is the extremal gain.

We observe that most of the points concentrate on the tail efficient line, covering a wide range of strengths of extremal dependence. This suggests that most one-step forward extremal dependencies are directionally tail symmetric. \textcolor{customizedColor}{Although some extreme events appear to be closely linked to subsequent large movements in asset prices, this connection does not necessarily yield a clear directional signal. At best, it serves as an indicator of heightened volatility, implying that the market remains relatively efficient even under extreme conditions.} A similar visualization is conducted for the contemporaneous extremal dependence structure which reveals the pattern of concurrence of extremal events across assets; we defer this to the Appendix~\textcolor{blue}{D.3} for brevity.

\subsection{Testing the ETH for China's futures market}
\label{sec:app:tail_efficiency}

Although the extremal dependence structure mainly concentrates on the tail efficient line of the extremal ball, one can still notice some deviations. To verify the significance of those deviations, we test the ETH posited in Section~\ref{sec:meth:eth}, excluding the last two months (\TeststartDate\ to \TestendDate) for an out-of-sample analysis in following Section~\ref{sec:app:portfolio}. We let $\bm{X}=(Z_{1,t},\dots,Z_{55,t}, -Z_{1,t},\dots,-Z_{55,t})^\top$ be the explanatory vector and asset return after one period $\bm{Y}=(Z_{1,t+1},\dots,Z_{55,t+1})^\top$ be the target vector. There are a total of $M=$ \totalNum\ pairs $(X_{i},Y_{j})^\top$ for $i=1,\dots,110$ and $j=1,\dots,55$. For each pair, we estimate $\lambda(X_{i},Y_{j})$ and its p-value with $P=10,000$ permutations using the second proposed estimator $\hat{\lambda}^2_n$ (see Section~\ref{sec:sim:results}). The quantile for thresholds $r_0^+, r_0^-$ are derived using the empirical \tailQuantile-quantiles. The significance level is set to $\alpha^*=$\significantLevel\ and we use the~\cite{benjamini1995controlling} multiple testing correction. The p-values for each pair are sorted as \( p_{(1)} < \dots < p_{(M)} \). The Benjamini--Hochberg rejection threshold is $p_{(H)}$ with order $H=\max\{ i : p_{(i)} < l_i\}$, where $l_i= \frac{i\alpha^*}{C_M M}$ for $ C_M = \sum_{i=1}^{M}i^{-1}$, which accounts for dependence between the tests. \textcolor{customizedColor}{The Efficient Tail Hypothesis (ETH) for China's futures market, that is, that the pairwise ETH is rejected for at least one pair of assets (see Definition~\ref{hyp:vector}), is rejected under the significance level of \significantLevel. This result is consistent across the 100 bootstrap replicates, where the market-level ETH is rejected for all replicates.} 

By studying pairs with p-values smaller than \significantLevel, we find that \textcolor{customizedColor}{\significantNum}\ pairs are tail-inefficient. These pairs are marked in Figure~\ref{fig:ExtremeBall}. Here, we focus on analyzing the significant pairs with the top 10 largest test statistic, $|\widehat{\lambda}^2_n(X_i,Y_j)|$; see Table~\ref{tab:effects}. \textcolor{customizedColor}{Table~\ref{tab:effects} also shows the bootstrap mean and standard deviation of $\widehat{\lambda}^2_n(X_i,Y_j)$ across $100$ bootstrap replicates. The largest value corresponds to a measure of directional asymmetric dependence between the eggs' futures and itself in the next time step, with a p-value of $0.0051$. This suggests that extreme depreciation in futures of eggs tends to occur with an extreme appreciation in the next period.}

\begin{table}[t!]
    \caption{Asset pairs (and their acronyms) with the top 10 largest asymmetry test statistics for pairs with p-values smaller than \significantLevel. \textcolor{customizedColor}{The bootstrap mean and standard deviation of $\hat{\lambda}^2_n$ are calculated using $100$ bootstrap replicates. Each of the estimated $\hat{\lambda}^2_n$ for the observed data are within one standard deviation of the bootstrap mean, and omitted.}}
    \centering
    \renewcommand{\arraystretch}{0.60}
    \resizebox{\textwidth}{!}{
    \begin{tabular}{llllll}
        \hline
        Explanatory asset $X_i$      & Target asset $Y_j$  & Sign of $X_i$ &  \textcolor{customizedColor}{p-values} & \textcolor{customizedColor}{Boot. mean $\widehat{\lambda}^2_n$ (std)}\\ \hline
        eggs (jd)       & eggs (jd) & negative                   & 0.0051	 & 0.145 (0.054)          \\
        zinc (zn)   & fuel oil (fu)         & negative                 &    0.0012     &-0.130 (0.044)         \\
        live hog (lh)   & pulp (sp)     & negative          & 0.0001    & -0.127 (0.036)                   \\
        Polypropylene (pp)       & Polyvinyl chloride (v)    & negative          & 0.0018  &   -0.120 (0.036)               \\
        corn starch (cs)     & aluminum (al)   & positive &   0.0002  &     -0.113 (0.052)       \\
        pulp (sp) & red dates (CJ)    & negative          &    0.0006   & 0.111 (0.035)           \\
        Ethylene Glycol (eg)& Soda Ash (SA) & negative            &   0.0004   &    -0.108 (0.036)       \\
        corn starch (cs)       & lead (pb)    & negative          &0.0011    &    -0.105 (0.055)             \\ 
        eggs (jd)           & lead (pb) & positive        &    0.0018    & -0.092 (0.063)            \\
        natural rubber (ru)    & natural rubber (ru)& positive           &0.0063	  &  -0.074 (0.064) 	                  \\\hline
        \end{tabular}
        }
    \label{tab:effects}
    \end{table}

We find that seven out of the ten aforementioned pairs are driven by negative extreme events. This suggests that the extremal depreciation of assets are more influential than the extremal appreciation of assets. Similar conclusions are found by~\cite{bouaddi2022systematic}, who show that negative extreme movements are more influential than positive extreme movements in the U.S. stock market. 

\textcolor{customizedColorRound2}{By analyzing whether $X_i$ and $Y_j$ belong to the same asset, we classify the significant pairs into two types: \textit{in-product} pairs (i.e., $X_i$ and $Y_j$ belong to the same asset) and \textit{cross-product} pairs (i.e., they belong to different assets). For the in-product significant pairs, all of them exhibit \textit{contrarian} effects, i.e., ${\rm Sgn}(X_i) \neq {\rm Sgn}(\lambda(X_i, Y_j))$. Contrarian (momentum) effects are well-known phenomena in financial markets, reflecting a tendency for price changes to follow opposite (same) directional movements relative to recent returns of the same asset. The phenomenon of \textit{contrarian in extremes} observed here may suggest that the market tends to overreact during extreme events and subsequently reverses in the following periods. For the cross-product pairs, the sign of the explanatory variable $X_i$ and the sign of $\lambda(X_i, Y_j)$ tend to be the same, which we refer to as \textit{co-directional lead-lag effects} across different assets.}

Statistics on all \significantNum\ significant pairs exhibit similar patterns: 41 pairs are driven by negative extreme events;  There are only three in-product significant pairs: \textcolor{customizedColor}{egg futures (driven by an extremal loss), natural rubber (driven by an extremal gain), and stainless steel (driven by an extremal gain)}. These three pairs are all contrarian in extremes. And for the remaining 57 cross-product pairs, 41 out of them are co-directional lead-lag effects across different assets.



Overall, the findings suggest that negative extreme movements are more influential than positive ones; contrarian effects in extreme dominates in in-product pairs; and co-directional lead-lag effects dominates in cross-product pairs.

Such statistics show that, in the context of extremes of China's futures market, an asset's own price often exhibits an initial overreaction, followed by a subsequent correction, which may form the basis of contrarian effects in extremes for tail-inefficient in-product pairs. As for the \textcolor{customizedColorRound2}{co-directional lead-lag effects} in extremes for the cross-product pairs, it can be explained by  the ``slow information diffusion hypothesis''~\citep{hou2007industry}. One plausible explanation is that certain assets swiftly incorporate relevant news, leading to initial price adjustments in explanatory assets, while others reflect the same information only after a delay. \textcolor{customizedColorRound2}{Moreover, the significant cross-product pairs may be interpreted as potential channels of tail risk transmission, thus connecting our analysis to the literature on spillover effects and lead-lad detection under market stress.}



\subsection{Exploiting the Efficient Tail Hypothesis analysis}
\label{sec:app:portfolio}

To motivate the practical application of the \tenameShort\ measure, we construct an artificial dynamic portfolio based on the significant tail-inefficient pairs found in Section~\ref{sec:app:tail_efficiency}. \textcolor{customizedColor}{Each inefficient pair falls into one of the following four cases: i) \(\hat{\lambda}_n^2(Z_{i,t},Z_{j,t+1})>0\): an extremal appreciation event for asset \( i \) is associated with an extremal appreciation event for asset \( j \) in the next timestep; ii) \(\hat{\lambda}_n^2(Z_{i,t},Z_{j,t+1})<0\): an extremal appreciation event for asset \( i \) is associated with an extremal depreciation event for asset \( j \) in the next timestep; iii) \(\hat{\lambda}_n^2(-Z_{i,t},Z_{j,t+1})>0\): an extremal depreciation event for asset \( i \) is associated with an extremal appreciation event for asset \( j \) in the next timestep; iv) \(\hat{\lambda}_n^2(-Z_{i,t},Z_{j,t+1})<0\): an extremal depreciation event for asset \( i \) is associated with an extremal depreciation event for asset \( j \) in the next timestep.} 

\begin{figure}[t!]
    \centering
    \includegraphics[width=\textwidth]{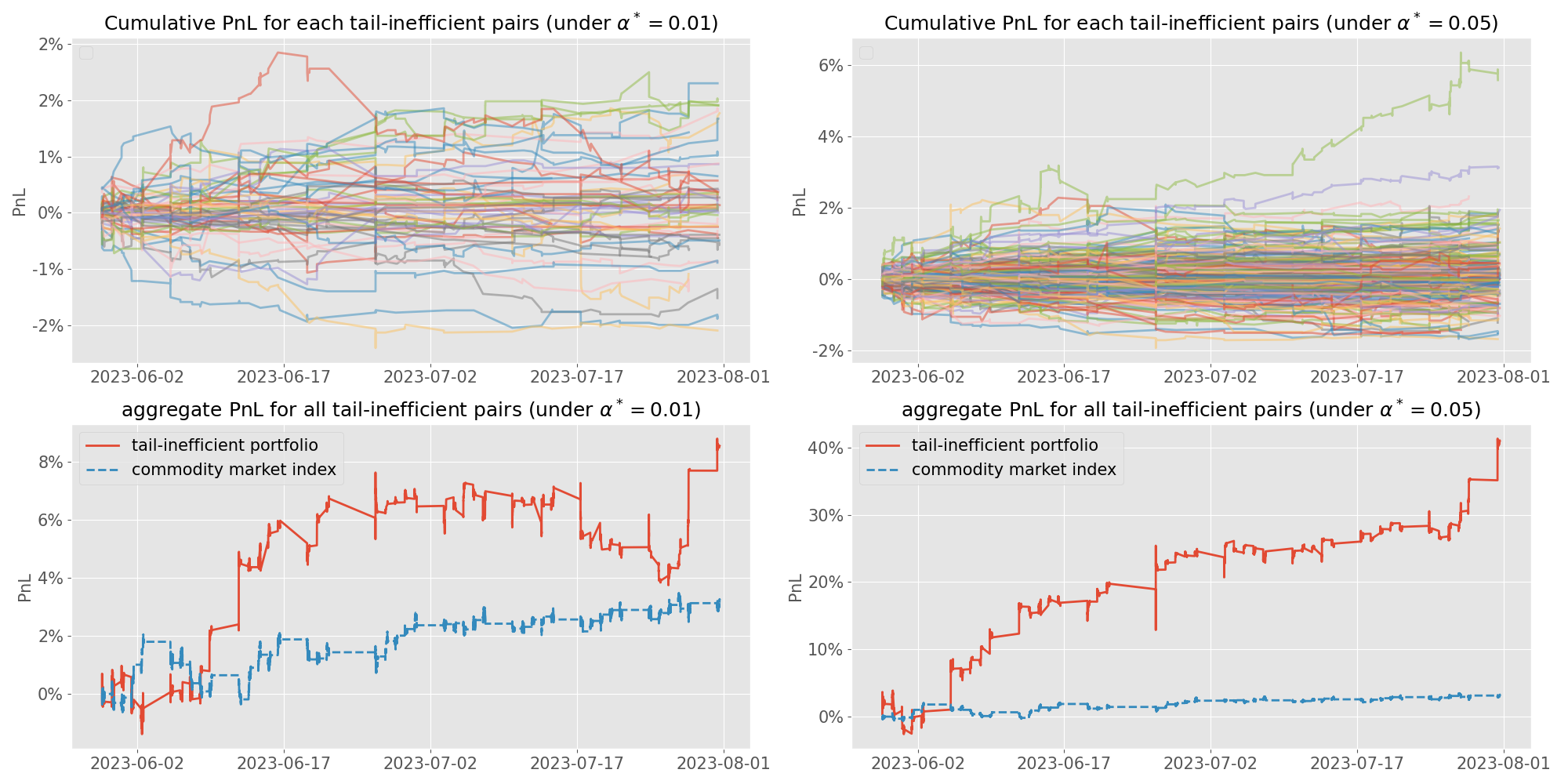} 
    \caption{\textcolor{customizedColor}{Profit and Loss of a dynamic portfolio consisting of the \significantNum\ or 294 tail-inefficient pairs under significance level $\alpha^*=0.01$ (left) or $\alpha^*=0.05$ (right), respectively. Upper panels show the Profit and Loss for each pair and the lower panels show the running cumulative Profit and Loss for the portfolio. The $y$-axis is the cumulative Profit and Loss (\%) and the $x$-axis is the time index for the two-month test period. \textcolor{customizedColorRound2}{The dashed blue curve represents the Profit and Loss of an equally-weighted portfolio of the 55 assets.} }}
    \label{fig:backtest} 
\end{figure}

For each pair satisfying $|\hat{\lambda}_n^2(X_i,Y_j)|> 0$, we suppose that we enter into the market for asset $Y_j$ (long position if $\hat{\lambda}_n^2(X_i,Y_j)>0$, otherwise short position) when we observe an extreme value for asset $X_i$ (given it exceeds its 99.5\% historical quantile for positive extremes and falls below its 0.5\% quantile when considering negative extremes). We hold the asset for one period and then exit the market by closing the position. With the trading cost ignored, we backtest this strategy on the out-of-sample data from \TeststartDate\ to \TestendDate. The cumulative Profit and Loss for each of the \textcolor{customizedColor}{\significantNum}\ significant pairs is given in Figure~\ref{fig:backtest}, alongside the overall Profit and Loss for the portfolio. We also study the case when set significance level to $0.05$ under which we find \textcolor{customizedColor}{294} significant pairs. Both results show a positive Profit and Loss throughout, suggesting that a trading strategy based on tail inefficiencies has the potential to make excess returns \textcolor{customizedColorRound2}{relative to a baseline buy-and-hold portfolio consisting of the same 55 assets held with equal weights}. This result is consistent with the bootstrap experiment, which is deferred to the Appendix~\textcolor{blue}{D.4}.

\textcolor{customizedColorRound2}{As stated by the Efficient Market Hypothesis, past information cannot be used to predict future price movements. However, our backtesting on out-of-sample data suggests that our trading strategy using tail-inefficient assets is profitable (under the ideal situation where the trading cost is not considered), which gives evidence against the Efficient Market Hypothesis. Therefore, the Efficient Tail Hypothesis, which is testable, can be viewed as an tail analogue of the Efficient Market Hypothesis with a stronger focus on extreme events.}

\section{Conclusion}
\label{sec:conclusion}

In this paper, we studied the pairwise directional tail dependence asymmetry for components of a random vector, $\bm{X}$. \textcolor{customizedColorRound2}{In comparison with  traditional tail asymmetry, this new type of asymmetry focuses on describing asymmetry of tail dependence in two adjacent quadrants.} Under balanced regular variation of $\bm{X}$ over the entire $\mathbb{R}^p$, we study the theoretical properties characterizing the dependence structure of $\bm{X}$ in all lower/upper tail quadrants, and we then create a novel tool, the extremal ball, to visualize the pairwise directional asymmetry of tail dependence within high-dimensional random vectors. Furthermore, we proved constraints on the tail dependence in two adjacent quadrants. Such a theoretical result suggests that knowledge of extremal dependence can aid in understanding the extremal dependence in adjacent quadrants of $\mathbb{R}^p$.

We proposed a new measure, the \tenameLower\ $\lambda(X,Y)$, to quantify the extent of directional tail symmetry. Two estimators are proposed for $\lambda(X,Y)$. We proved asymptotic normality for the first estimator; deriving the asymptotic properties for the second estimator remains future work. We further developed a permutation test for the null hypothesis $\mathcal{H}_0: \lambda(X,Y)=0$ and verified its effectiveness and power through simulation studies. We showcased that the permutation test with the second estimator $\widehat{\lambda}^2_n$ is more efficient and can achieve better statistical power than using the first estimator $\widehat{\lambda}^1_n$ with the standard $t$-test based on asymptotic theory. We introduce the concept of the Efficient Tail Hypothesis (ETH), an testable analogue of the Efficient Market Hypothesis, to investigate the directional symmetry behavior of the market in its tails. \textcolor{customizedColorRound2}{The statistical essence of ETH is directional tail symmetry between explanatory variables and future asset returns.}

In our data application, we investigate the market-wide tail efficiency of China's futures market. Our empirical results suggest that the Efficient Tail Hypothesis (ETH) for China's futures market is rejected under a significance level of $\alpha^*=0.01$. A deeper analysis of those tail-inefficient assets shows that negative extreme depreciation is more influential than positive extreme events. Moreover, we find evidence \textcolor{customizedColorRound2}{of co-directional lead-lag effects across different assets} in extremes, which suggests that the market tends to be under-active in extreme situations. To capitalize on our findings, we construct an artificial dynamic portfolio consisting of those exceptional assets. The strategy is back-tested on out-of-sample data and shows a significant overall excess return. \textcolor{customizedColor}{Those results suggest that most one-step forward extremal dependencies are directionally tail symmetric, while significant deviations indicate potential trading opportunities. Therefore}, the analysis of directional tail dependence is useful for risk management and for constructing trading strategies in the financial market.

Several directions can be explored in future research. One avenue is to extend the concept of directional tail dependence to higher dimensions. Additionally, current causal inference methods in extremes can be expanded to consider both upper and lower tails, using our new directional tail dependence measure. \textcolor{customizedColorRound2}{It is also of interest to study directional tail dependence during high-volatility periods, such as financial crises or at market opening and closing times.} Last but not least, further investigation into simulation methods for time series data with specific directional tail dependence structures could enable the generation of time series data with varying degrees of directional asymmetry in financial applications.

\section*{Acknowledgments}
We thank the Editor, the Associate Editor, and the two anonymous referees for their insightful comments, which have significantly improved the paper.

\section*{Disclosure statement}
The authors report there are no competing interests to declare.

\clearpage
\appendix

\section{Proofs}
\label{sec:Proof}

\subsection{Proof of Proposition~\textcolor{blue}{3.1}}
\begin{manualprop1}{3.1}
    Let $\bm{X}^*=(X^*_1,\dots,X^*_p)^\top\in {\rm RV}^p(\alpha)$ with limit measure $v_{\bm{X}^*}(\cdot)$ satisfying $v_{\bm{X}^*}\left(\{\bm{x} \in \mathbb{E}^p:x_i>1\}\right)=c_i^+>0$ and $v_{\bm{X}^*}\left(\{\bm{x} \in \mathbb{E}^p:x_i<-1\}\right)=c_i^->0$, for ${i=1,\dots,p}$. Then $\bm{X}=(X_1,\dots,X_p)^\top$, with $X_i=\max\{(c_i^+)^{-1/\alpha}X^*_i,0\}+\min\{(c_i^-)^{-1/\alpha}X_i^*,0\}$ for all ${i=1,\dots,p}$, satisfies $\bm{X}\in{\rm BRV}^p(\alpha)$. 
\end{manualprop1}

\begin{proof}[Proof of Proposition~\textcolor{blue}{3.1}]\label{proof:Rebalance}
  We require three conditions to hold: i) $|\bm{X}|\in {\rm RV^p_+(\alpha)}$; ii) $\bm{X}$ is regularly varying in every quadrant; iii) $\bm{X}$ is balanced, i.e., $\nu_{\bm{X}}(\{\bm{x}\in \mathbb{E}^p: x_i>1\})=\nu_{\bm{X}}(\{\bm{x}\in \mathbb{E}^p: x_i<-1\})=1$ for all $i=1,\dots,p$.

    Condition i) follows by Lemma 6.1 of \cite{resnick2007heavy}, which is a criterion for vague convergence to measures; It states that to show $\lim_{n\to\infty}n\mathbb{P}\{b_n^{-1}|\bm{X}|\in \cdot\}=\nu_{|\bm{X}|}(\cdot)$ on ${\mathbb{E}^p_+=\mathbb{R}_+\setminus\{\boldsymbol{0}\}}$, it is enough to show that
$\lim_{n\to\infty}n\mathbb{P}\{b_n^{-1}|\bm{X}|\in [\bm{0},\bm{z}]^c\}=\nu_{|\bm{X}|}([\bm{0},\bm{z}]^c)$ for all $\bm{z}:=(z_1,\dots,z_p)^\top \in [{0},\infty)^p\setminus \{\bm{0}\}$. We show that $\nu_{|\bm{X}|}(\cdot)$ can be expressed in terms of $\nu_{\bm{X}^{*}}(\cdot),$ which is the limit measure for $\bm{X}^*$. Consider
    \begin{equation*}
        \begin{aligned}
          \mathbb{P}\{b_n^{-1}|\bm{X}|&\in [\bm{0},\bm{z}]^c\}=\mathbb{P}\left\{\bm{X}\in \mathbb{R}^p:\textcolor{customizedColor}{\bigcup}_{i=1}^p \left[b_n^{-1}\left|X_i\right|>z_i\right]\right\}\\
          &=\mathbb{P}\left\{\bm{X}^*\in \mathbb{R}^p:\textcolor{customizedColor}{\bigcup}_{i=1}^p \left[b_n^{-1}\left|\max\{(c_i^+)^{-1/\alpha}X^{*}_i,0\}+\min\{(c_i^-)^{-1/\alpha}X_i^{*},0\}\right|>z_i\right]\right\}\\
            &=\mathbb{P}\left\{\bm{X}^*\in \mathbb{R}^p:\textcolor{customizedColor}{\bigcup}_{i=1}^p \left[ \left(b_n^{-1}(c_i^+)^{-1/\alpha}X^{*}_i>z_i\right) \textcolor{customizedColor}{\cup} \left(b_n^{-1}(c_i^-)^{-1/\alpha}X_i^{*}<-z_i\right)  \right]\right\}\\
            &=\mathbb{P}\left\{\bm{X}^*\in \mathbb{R}^p:\textcolor{customizedColor}{\bigcup}_{i=1}^p \left[ \left(b_n^{-1}X^{*}_i>z_i(c^+_i)^{1/\alpha}\right) \textcolor{customizedColor}{\cup} \left(b_n^{-1}X^{*}_i<-z_i(c_i^-)^{1/\alpha}\right)  \right]\right\}.
        \end{aligned}
    \end{equation*}
Then, for all $\bm{z} \in [{0},\infty)^p \setminus \{\bm{0}\} $, we have that
\begin{align*}
    &\lim_{n\to\infty}n\mathbb{P}\{b_n^{-1}|\bm{X}|\in [\bm{0},\bm{z}]^c\}=\lim_{n\to\infty}n\mathbb{P}\left\{\bm{X}\in \mathbb{R}^p:\textcolor{customizedColor}{\bigcup}_{i=1}^p \left[b_n^{-1}\left|X_i\right|>z_i\right]\right\}\\
    =&  \lim_{n\to\infty}n\mathbb{P}\left\{\bm{X}^*\in \mathbb{R}^p:\textcolor{customizedColor}{\bigcup}_{i=1}^p \left[ \left(b_n^{-1}X^{*}_i>z_i(c^+_i)^{1/\alpha}\right) \textcolor{customizedColor}{\cup} \left(b_n^{-1}X^{*}_i<-z_i(c_i^-)^{1/\alpha}\right)  \right]\right\}\\
    =&\nu_{\bm{X}^*}(\{\bm{x}\in \mathbb{E}^p:\textcolor{customizedColor}{\bigcup}_{i=1}^px_i\in \{[-\infty, - z_i(c_i^-)^{1/\alpha})\cup (z_i(c_i^+)^{1/\alpha},\infty]\}\})=: \nu_{|\bm{X}|}( [0,\bm{z}]^c)
\end{align*} as needed.

For Condition ii), we show that $\lim_{n\rightarrow \infty}\frac{n\mathbb{P}\{b_n^{-1}\bm{s}\odot\bm{X}\in [\bm{0},\bm{z}]^c\}}{n\mathbb{P}\{b_n^{-1}|\bm{X}|\in [\bm{0},\bm{z}]^c\}} \in (0,1)$ for all $\bm{z}\in [0,\infty)^p\setminus\{\bm{0}\}$ and any $\bm{s} \in \{-1, 1\}^p$. Consider the numerator; we first show that this is strictly greater than zero. We have that  $\lim_{n\rightarrow \infty}n\mathbb{P}\{\bm{s}\odot\bm{X}\in b_n[\bm{0},\bm{z}]^c\}\geq  \lim_{n\rightarrow \infty}\textcolor{customizedColor}{n}\mathbb{P}\{s_1b_n^{-1} X_1>z_1\}$, as the region $\{s_1b_n^{-1} X_1>z_1\}$ is a subspace of the region $\{\bm{s}\odot\bm{X}\in b_n[\bm{0},\bm{z}]^c\}$,  and 
\begin{equation*}
        \begin{aligned}
 \lim_{n\rightarrow \infty}\textcolor{customizedColor}{n}\mathbb{P}\{s_1b_n^{-1} X_1>z_1\}&=  \begin{cases}
                \lim_{n\rightarrow \infty}n\mathbb{P}\{b_n^{-1} X^{*}_1>z_1(c^+_1)^{1/\alpha}\}, & \text{if } s_1 = 1, \\
                \lim_{n\rightarrow \infty}n\mathbb{P}\{b_n^{-1} X^{*}_1<-z_1(c^-_1)^{1/\alpha}\}, & \text{if } s_1 =-1 ,
              \end{cases} \\
            &=  \begin{cases}
                \nu_{\bm{X}^{*}}(\{\bm{x}\in \mathbb{E}^p:x_1>z_1(c^{+}_1)^{1/\alpha}\}), & \text{if } s_1 = 1, \\
                \nu_{\bm{X}^{*}}(\{\bm{x}\in \mathbb{E}^p:x_1<-z_1(c^{-}_1)^{1/\alpha}\}), & \text{if } s_1 =-1 ,
                \end{cases} \\
            &= \begin{cases}
                z_1^{-\alpha}(c^{+}_1)^{-1}\nu_{\bm{X}^{*}}(\{\bm{x}\in \mathbb{E}^p:x_1>1\}), & \text{if } s_1 = 1 ,\\
                z_1^{-\alpha}(c^{-}_1)^{-1}\nu_{\bm{X}^{*}}(\{\bm{x}\in \mathbb{E}^p:x_1<-1\}), & \text{if } s_1 =-1 ,
            \end{cases} \\
            &= z_1^{-\alpha}>0,\\
        \end{aligned}
    \end{equation*}
    where the third line follows due to the homogeneity property of $\nu_{\bm{X}^{*}}(\cdot)$. Also, we have that the numerator smaller the denominator, i.e., $\mathbb{P}\{\bm{s}\odot\bm{X}\in b_n[\bm{0},\bm{z}]^c\}< \mathbb{P}\{|\bm{X}|\in b_n[\bm{0},\bm{z}]^c\}$. Therefore, $0<{\lim_{n\rightarrow \infty}\mathbb{P}\{b_n^{-1}\bm{s}\odot\bm{X}\in [\bm{0},\bm{z}]^c\}}<{\lim_{n\rightarrow \infty}\mathbb{P}\{b_n^{-1}|\bm{X}|
\in [\bm{0},\bm{z}]^c\}}$ and Condition ii) holds.
    
    For Condition iii), we show that $\nu_{\bm{X}}(\{\bm{x}\in \mathbb{E}^p: x_i>1\})=\nu_{\bm{X}}(\{\bm{x}\in \mathbb{E}^p: x_i<-1\})=1$ for all $i=1,\dots,p$. This follows as, for any $i=1,\dots,p$,
    \begin{equation*}
        \begin{aligned}
            \nu_{\bm{X}}(\{\bm{x}\in \mathbb{E}^p: x_i>1\})&=\lim_{n\rightarrow \infty}n\mathbb{P}\{b_n^{-1} \left(\max\{(c_i^+)^{-1/\alpha}X^{*}_i,0\}+\min\{(c_i^-)^{-1/\alpha}X_i^{*},0\}\right)>1\}\\
            &=\lim_{n\rightarrow \infty}n\mathbb{P}\{b_n^{-1}(c_i^+)^{-1/\alpha}X^{*}_i>1\}\\
            &=(c_i^+)^{-1}\lim_{n\rightarrow \infty}n\mathbb{P}\{b_n^{-1}X^{*}_i>1\}=   (c_i^+)^{-1}\nu_{\bm{X}^*}(\{\bm{x}\in \mathbb{E}^p: x_i>1\})=1,
        \end{aligned}
    \end{equation*}
    where the last line follows due to the homogeneity property of the limit measure. A similar argument can be used to show that  $\nu_{\bm{X}}(\{\bm{x}\in \mathbb{E}^p: x_i<-1\})=1$ for any $i=1,\dots,p$. Thus, all three conditions are satisfied and this completes the proof.
\end{proof}

\subsection{Proof of Proposition~\textcolor{blue}{3.2}}

\begin{manualprop2}{3.2}{Properties of $\bar{\bm{X}}$}
    Let $\bm{X}\in {\rm BRV}^p(2)$ and let $\bar{\bm{X}}:=\left((\bm{X}^+)^\top,(\bm{X}^-)^\top\right)^\top$. Then $\bar{\bm{X}} \in {\rm RV}_+^{2p}(2)$ with limit measure $\nu_{\bar{\bm{X}}}(\cdot)$ that satisfies $\nu_{\bar{\bm{X}}}(\{\bm{x}\in \mathbb{E}^{2p}_+:x_i>1\})=1$ for $i=1,\dots,2p$, angular measure $H_{\bar{\bm{X}}}(\cdot)$ satisfying ${H_{\bar{\bm{X}}}(\mathbb{S}^{2p-1}_+)}=2p$, and with $\sigma(X_i^+,X_i^-)=0$ for $i=1,\dots,p$ (orthogonality property).
\end{manualprop2}

\begin{proof}[Proof of Proposition~\textcolor{blue}{3.2}]
We first show that $\bar{\bm{X}}:=(\bar{X}_1,\dots,\bar{X}_{2p})^\top\in {\rm RV}^{2p}_+(2)$ with limit measure $\nu_{\bar{\bm{X}}}(\cdot)$, which we express in terms of $\nu_{{\bm{X}}}(\cdot)$. By Lemma 6.1 of \cite{resnick2007heavy}, we have, for all $\bm{\bar{z}}:=((\bm{z^+})^\top;(\bm{z^-})^\top)^\top \in [0,\infty)^{2p}\setminus \{\bm{0}\}$, that
        \begin{equation*}
            \begin{aligned}
                \lim_{n\to\infty}n\mathbb{P}\{b_n^{-1}\bar{\bm{X}}\in [\bm{0},\bm{\bar{z}}]^c\}&=\lim_{n\to\infty}n\mathbb{P}\{\textcolor{customizedColor}{\bigcup}_{i=1}^p[X^+_i> z^+_i \textcolor{customizedColor}{\cup} X^-_i > z^-_i]\}\\
                &=\lim_{n\to\infty}n\mathbb{P}\{\textcolor{customizedColor}{\bigcup}_{i=1}^p[\max\{X_i,0\}> z^+_i\textcolor{customizedColor}{\cup}-\min\{X_i,0\}> z^-_i]\}\\
                &=\lim_{n\to\infty}n\mathbb{P}\{\textcolor{customizedColor}{\bigcup}_{i=1}^p[X_i \in (-\infty,-z^-_i)\cup(z^+_i,\infty)\}\\
                &=\nu_{\bm{X}}(\{\textcolor{customizedColor}{\bigcup}_{i=1}^p[(-\infty,-z^-_i)\cup(z^+_i,\infty)]\} )=:\nu_{\bar{\bm{X}}}([\bm{0},\bm{\bar{z}}]^c).
            \end{aligned}
        \end{equation*}
    Thus, $\bar{\bm{X}}\in {\rm RV}^{2p}_+(2)$ with limit measure $\nu_{\bar{\bm{X}}}([\bm{0},\bm{\bar{z}}]^c)=\nu_{\bm{X}}(\{\textcolor{customizedColor}{\bigcup}_{i=1}^p[(-\infty,-z^-_i)\cup(z^+_i,\infty)]\} )$.

We now show that the angular measure, $H_{\bar{\bm{X}}}$, for $\bar{\bm{X}}$ satisfies $H_{\bar{\bm{X}}}(\mathbb{S}^{2p-1}_+)=2p$. Consider $i\leq p$; then ${\bar{X}}_i=\max\{X_i,0\}$ and ${{\bar{X}}_{i+p}=-\min\{X_i,0\}}$. Further, we have that
        \begin{equation*}
            \begin{aligned}
                \nu_{\bar{\bm{X}}}(\{\bm{x}\in \mathbb{E}^{2p}_+:x_i>1\})&=\lim_{n\rightarrow\infty}\textcolor{customizedColor}{n}\mathbb{P}\{b_n^{-1}\bar{X}_i>1\}=\lim_{n\rightarrow\infty}\textcolor{customizedColor}{n}\mathbb{P}\{b_n^{-1}\max\{X_i,0\}>1\}\\
                =\lim_{n\rightarrow\infty}\textcolor{customizedColor}{n}\mathbb{P}\{&(b_n^{-1}X_i>1) \textcolor{customizedColor}{\cap} (X_i>0)\}+ \lim_{n\rightarrow\infty}\textcolor{customizedColor}{n}\mathbb{P}\{(b_n^{-1}\times 0>1) \textcolor{customizedColor}{\cap} (X_i<=0)\}\\
                &=\lim_{n\rightarrow\infty}\textcolor{customizedColor}{n}\mathbb{P}\{b_n^{-1}X_i>1\}=\nu_{\bm{X}}(\{\bm{x}\in \mathbb{E}^{p}:x_i>1\})=1,\\
            \end{aligned}
        \end{equation*}
    and similarly,
        \begin{equation*}
            \begin{aligned}
                \nu_{\bar{\bm{X}}}(\{\bm{x}\in \mathbb{E}^{2p}_+:x_{i+p}>1\})&=\lim_{n\rightarrow\infty}\textcolor{customizedColor}{n}\mathbb{P}\{-b_n^{-1}\bar{X}_i>1\}=\lim_{n\rightarrow\infty}\textcolor{customizedColor}{n}\mathbb{P}\{-b_n^{-1}\min\{X_i,0\}>1\}\\
                &=\lim_{n\rightarrow\infty}\textcolor{customizedColor}{n}\mathbb{P}\{b_n^{-1}X_i<-1\}=\nu_{\bm{X}}(\{\bm{x}\in \mathbb{E}^{p}:x_i<-1\})=1.\\
            \end{aligned}
        \end{equation*}
        
        Therefore, for all $i\in 1\dots,2p$, we have that $\nu_{\bar{\bm{X}}}(\{\bm{x}\in \mathbb{E}^{2p}_+:x_i>1\})=1$. Note that
        \begin{equation*}
            \begin{aligned}
                \nu_{\bar{\bm{X}}}(\{\bm{x}\in \mathbb{E}^{2p}_+:x_i>1\})&=\int_{\mathbb{S}^{2p-1}_+}\int_{1/\omega_i}^{\infty}2r^{-3}{\rm d}r{\rm d}H_{\bar{\bm{X}}}(\bm{\omega})\\
                &=\int_{\mathbb{S}^{2p-1}_+}\omega_i^2{\rm d}H_{\bar{\bm{X}}}(\bm{\omega})=1,
                 \end{aligned}
        \end{equation*}
        for all $i=1,\dots,2p$. Now recall that ${\mathbb{S}^{2p-1}_+}={\{\bm{\omega}\in \mathbb{R}^{2p}: ||\bm{\omega}||_2=1\}}$ and $\sum_{i=1}^{2p}\omega_i^2=1$ for ${\bm{\omega}=(\omega_1,\dots\omega_{2p})^\top\in \mathbb{S}^{2p-1}_+}$. Therefore, \[H_{\bar{\bm{X}}}(\mathbb{S}^{2p-1}_+)=\int_{\mathbb{S}^{2p-1}_+} 1 {\rm d}H_{\bar{\bm{X}}}(\bm{\omega})= \int_{\mathbb{S}^{2p-1}_+}\sum_{i=1}^{2p}\omega_i^2{\rm d}H_{\bar{\bm{X}}}(\bm{\omega})=\sum_{i=1}^{2p} \int_{\mathbb{S}^{2p-1}_+}\omega_i^2{\rm d}H_{\bar{\bm{X}}}(\bm{\omega})=2p,\]
        as needed.

Finally, we show that $\sigma(X_i^+,X_i^-)=0$ for all $i=1,\dots,p$. This follows as, for all $i=1,\dots,p$, we have that 
\begin{align*}
    \sigma(X_i^+,X_i^-)&=\lim_{r\rightarrow\infty} \mathbb{E}\left[\frac{X_i^+}{R_i}\frac{X_i^-}{R_i}|R_i>r\right]=\lim_{r\rightarrow\infty}\mathbb{E}\left[\frac{-\max\{X_i,0\}\min\{X_i,0\}}{R_i^2}|R_i>r\right]=0,
\end{align*}
where $R_i=||(X_i^+,X_i^-)||_2$.
\end{proof}

\subsection{Proof of Proposition~\textcolor{blue}{3.3}}
\label{sec:proof:tailInequality}

\begin{manualprop2}{3.3}{Extremal ball constraint}
    For $(X, Y)^\top\in {\rm BRV}^2(2)$ with $X^+,Y^+,Y^-\in \mathcal{V}^q_+$, the angles $\theta_{X^+,Y^+}$ and $\theta_{X^+,Y^-}$ satisfy the constraint $\frac{\pi}{2}-\theta_{X^+,Y^+} \leq \theta_{X^+,Y^-} \leq \frac{\pi}{2}$.
\end{manualprop2}

\begin{figure}[th!]
    \centering
    \includegraphics[width=\linewidth]{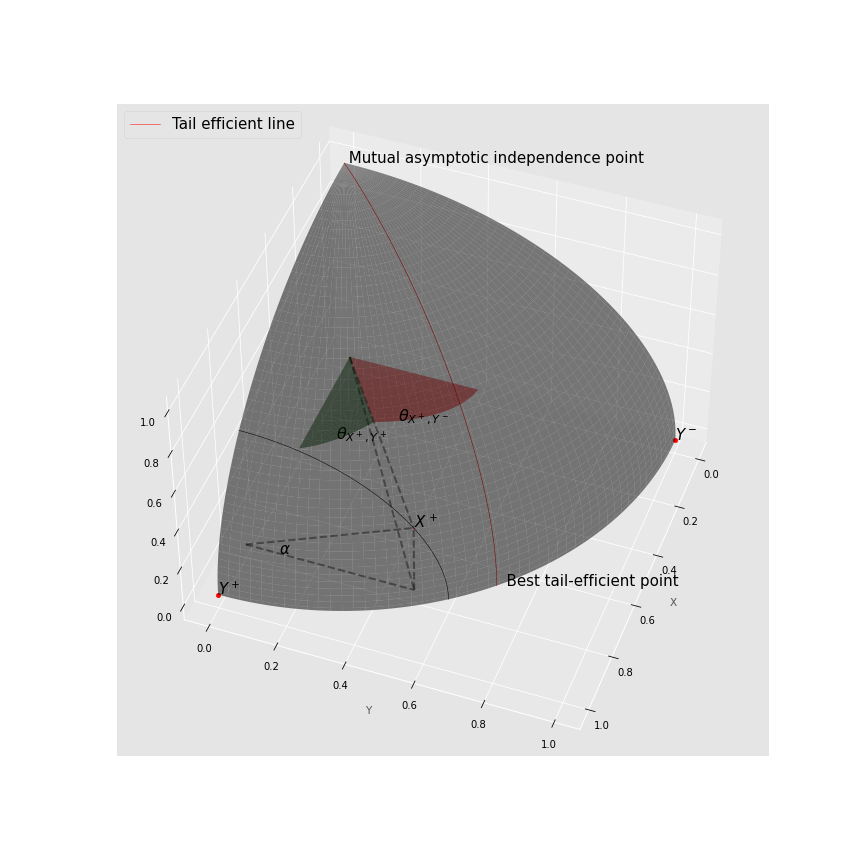}
    \caption{Illustration of the angle constraint given in Proposition~\textcolor{blue}{3.3}. Here $\theta_{X^+,Y^+}$ and $\theta_{X^+,Y^-}$ are the angles between $X^+$ and $Y^+$ and $X^+$ and $Y^-$, respectively; $\alpha$ is the angle between the line formed by the point \( X^+ \) and its perpendicular projection on the \( Y^+ \) axis and the base plane ($Y^+$--$Y^-$ plane); when $\alpha$ varies in $[0,{\pi \over 2}]$, the trajectory on the surface consists of $X^+$ with fixed $\theta_{X^+,Y^+}$ on the extremal ball. }
    \label{fig:illForinequaly}
\end{figure}

Figure \ref{fig:illForinequaly} illustrates the angle constraint given in Proposition~\textcolor{blue}{3.3}. We define $\alpha \in [0,\pi/2]$ as the angle between the line formed by the point \( X^+ \) and its perpendicular projection on the \( Y^+ \) axis and the plane connecting $Y^+$ and $Y^-$; see Figure~\ref{fig:illForinequaly}.

\begin{proof}[Proof of Proposition~\textcolor{blue}{3.3}]\label{proof:tailInequality}

    The trajectory of $X^+$ on the surface of the extremal ball, with the angle $\theta_{X^+,Y^+}$ fixed, is $$\left(\cos\theta_{X^+,Y^+},\cos\alpha\sin\theta_{X^+,Y^+},\sin\alpha\sin \theta_{X^+,Y^+}\right)$$ for $\alpha \in [0,\pi/2],$ as shown in Figure~\ref{fig:illForinequaly}. Then, we express $\theta_{X^+,Y^-}$ in terms of $\alpha$:
    \begin{equation*}
        \theta_{X^+,Y^-} =\cos^{-1}(\sin \theta_{X^+,Y^+} \cos \alpha), \quad \alpha\in [0,\pi/2],
    \end{equation*}
    is monotonically increasing as the partial derivative $\frac{\partial \theta_{X^+,Y^-}}{\partial \alpha} = {\sin \theta_{X^+,Y^+} \sin \alpha\over \sqrt{1 - (\sin \theta_{X^+,Y^+} \cos \alpha)^2}}\geq 0$ on $[0,\pi/2]$. The minimum of $\theta_{X^+,Y^-}$ on  $[0,\pi/2]$ is $\cos^{-1}(\sin \theta_{X^+,Y^+})=\frac{\pi}{2}-\theta_{X^+,Y^+}$. Therefore, $\frac{\pi}{2}-\theta_{X^+,Y^+} \leq \theta_{X^+,Y^-} \leq \frac{\pi}{2}$. 
    \end{proof}

\subsection{Proof of Proposition~\textcolor{blue}{3.4}}

\label{proof:sec:dtd}
\begin{manualprop2}{3.4}{\textcolor{customizedColor}{Properties of \tenameLower\ $\lambda$}}
    \textcolor{customizedColor}{The directional tail dependence $\lambda(X,Y)$ satisfies ${\lambda(X,Y)\in[-1,1]}$, where ${\lambda(X,Y)=-1}$ and ${\lambda(X,Y)=1}$ are achieved when ${\lim_{a\rightarrow \infty}\mathbb{P}[X>a \mid Y<-a]=1}$ and ${\lim_{a\rightarrow \infty}\mathbb{P}[X>a\mid Y>a]=1}$, respectively. Furthermore, $\lambda(X,Y)$ is an odd function of $Y$, i.e., $\lambda(X,-Y)=-\lambda(X,Y)$ for all $Y$ (this property does not hold for $X$).}
\end{manualprop2}

\begin{proof}[Proof of Proposition~\textcolor{blue}{3.4}]
We first show that $\lambda(X,Y)\in [-1,1]$, where $\lambda(X,Y)=-1$ and $\lambda(X,Y)=1$ are achieved when $\lim_{a\rightarrow \infty}\mathbb{P}[X>a \mid Y<-a]=1$ and $\lim_{a\rightarrow \infty}\mathbb{P}[X>a\mid Y>a]=1$, respectively. By Proposition~\textcolor{blue}{3.5}, we have that $\lambda(X,Y)=2\{\sigma(X^+,Y^+)-\sigma(X^+,Y^-)\}$. As $\sigma(X^+,Y^+) \in [0,{1 \over 2}]$ (and similarly for $\sigma(X^+,Y^-)$), we have that  $\lambda(X,Y)\in [-1,1]$. The boundary case $\lambda(X,Y)=1$ implies $\sigma(X^+,Y^+)=0.5$ and $\theta_{X^+,Y^+}=0$ which further implies $\sigma(X^+,Y^-)=0$ as $\theta_{X^+,Y^-}\in [\frac{\pi}{2}-\theta_{X^+,Y^+}, \frac{\pi}{2}]$ by Proposition~\textcolor{blue}{3.3}. The case $\sigma(X^+,Y^+)=0.5$ is achieved when $$\lim_{a\to\infty}\mathbb{P}[X^+>a\mid Y^+>a]=\lim_{a\to\infty}\mathbb{P}[X>a\mid Y>a]=1.$$
A similar argument holds for $\lambda(X,Y)=-1$.

To show that $\lambda(X,-Y)=-\lambda(X,Y)$, consider $$\lambda(X,-Y)=\int_{\{(\omega_x,\omega_{y})^\top\in \mathbb{S}^1:\omega_x\geq 0\}} \omega_x\omega_{y} {\rm d}H_{(X,-Y)}(\boldsymbol{\omega}),$$ where $\boldsymbol{\omega}=(\omega_x,\omega_{y})^\top$. Note that $\|(X,Y)^\top\|_2=\|(X,-Y)^\top\|_2,$ and so we have, for all $(\omega_x,\omega_{y})^\top \in \mathbb{S}^1$, that $H_{(X,-Y)}((\omega_x,\omega_y)^\top)=H_{(X,Y)}((\omega_x,-\omega_y)^\top)$. Thus, it follows that
 $$\lambda(X,-Y)=-\int_{\{(\omega_x,\omega_{y})^\top\in \mathbb{S}^1:\omega_x\geq 0\}} \omega_x\omega_{y} {\rm d}H_{(X,Y)}(\boldsymbol{\omega})=-\lambda(X,Y).$$ \end{proof}

\subsection{Proof of Proposition~\textcolor{blue}{3.5}}    \label{sec:proof:equivalentForm}

\begin{manualprop1}{3.5}
    \quad\vspace{-3mm}
    Consider the \tenameLower\ $\lambda(X,Y)$ as in Definition~\textcolor{blue}{3.4}. Then
    \begin{equation*}
        \lambda(X,Y)=3\int_{\{(\omega_x,\omega_y)^\top\in \mathbb{S}^1:\omega_x\geq 0\}}\omega_x\omega_y{\rm d}N_{(X^+,Y)}(\bm{\omega})=2\{\sigma(X^+,Y^+)-\sigma(X^+,Y^-)\},
        \tag{7}
    \end{equation*}
    where $N_{(X^+,Y)}(\cdot)$ is the normalized angular measure of $(X^+,Y)^\top \in {\rm BRV}^2(2)$.
\end{manualprop1}

To prove the equivalence of the two forms of $\lambda(X,Y)$, we first detail the following lemma which states that the angular measures of $H_{(X,Y)}(\cdot)$ and $H_{(X^+,Y^+)}(\cdot)$ are equal, except when evaluated at the axes.

    \begin{lemma}\label{lemma:MeasureFromWholeToQuadrant}
        Let $(X,Y)\in {\rm RV}^2(\alpha)$, $X^+=\max\{X,0\},$ and $X^-=-\min\{X,0\}$. The following two properties hold:
        \begin{itemize}
            \item For any $\Lambda \subseteq \mathring{\mathbb{S}}^1_+$ which is a Borel subset of ${\mathring{\mathbb{S}}^1_+:=\{x,y>0: x^2+y^2=1\}}$, we have that $H_{(X,Y)}(\Lambda)=H_{(X^+,Y^+)}(\Lambda)$;
            \item For any $\Lambda \subseteq \mathring{\mathbb{S}}^1_+$, we have $H_{(X,Y)}(\Lambda')=H_{(X^+,Y^-)}(\Lambda)$  where $\Lambda'=\{(x,-y):(x,y)\in \Lambda\}$.
        \end{itemize}
        \end{lemma}

    \begin{remark}
        When consider the truncated variables $X^+=\max\{X,0\}$ or $X^-=-\min\{X,0\}$, the angular measures of $(X^+,Y^+)^\top$ and $(X,Y)^\top$ differ at the boundaries, that is, where one angle is zero, e.g., $H_{(X^+,Y^+)}(\Lambda)\neq H_{(X,Y)}(\Lambda)$ for  $\Lambda=\{(0,1),(0,1)\}$. However, this change will not affect the value of the EDM, $\sigma(\cdot,\cdot)$, or DTD, $\lambda(\cdot,\cdot)$, as the product in their integrands evaluates to zero whenever at least one angle is zero. Lemma~\ref{lemma:MeasureFromWholeToQuadrant} highlights that the angular measures of   $(X^+,Y^+)^\top$ and $(X,Y)^\top$ do not differ elsewhere, and we can exploit this property to prove Proposition~\textcolor{blue}{3.5}.
    \end{remark}
        
\begin{proof}[Proof of Lemma \ref{lemma:MeasureFromWholeToQuadrant}]
For all $\Lambda \subseteq\mathring{\mathbb{S}}^1_+$, we have that
\begin{equation*}
    \begin{aligned}
    H_{(X,Y)}(\Lambda)&=\nu_{(X,Y)}\left(\left\{\sqrt{x^2+y^2}>1,\frac{(x,y)^\top}{||(x,y)^\top||_2}\in \Lambda\right\}\right) \\
    &=\lim_{n\rightarrow\infty}n\mathbb{P}\left\{b_n^{-1}\sqrt{X^2+Y^2}>1,\frac{(X,Y)^\top}{||(X,Y)^\top||_2}\in\Lambda\right\}\\
   &=\lim_{n\rightarrow\infty}n\mathbb{P}\left\{b_n^{-1}\sqrt{(X^+)^2+(Y^+)^2}>1,\frac{(X^+Y^+)^\top}{||(X^+,Y^+)^\top||_2}\in\Lambda\right\}\\
    &=\nu_{(X^+,Y^+)}\left(\left\{\sqrt{x^2+y^2}>1,\frac{(x,y)^\top}{||(x,y)^\top||_2}\in \Lambda\right\}\right)  \\
    &=H_{(X^+,Y^+)}(\Lambda).
    \end{aligned}
\end{equation*}
The third line follows as $X^+=X$ and $Y^+ = Y$ for any $(X,Y)^\top$ such that ${{(X,Y)^\top \over {||(X,Y)^\top||_2}}\in\mathring{\mathbb{S}}^1_+}$.
A similar argument can be used to show that, for $\Lambda\subseteq \mathring{\mathbb{S}}^1_+$, we have $H_{(X,Y)}(\Lambda')=H_{X^+,Y^-}(\Lambda)$. 
\end{proof}


\begin{proof}[Proof of Proposition~\textcolor{blue}{3.5}]
    We first show that $\lambda(X,Y)$ can be written as the sum of two integral parts: $m_1\int_{\mathbb{S}^1_+}\omega_x\omega_y{\rm d}N_{X^+,Y^+}(\bm{\omega})$ and $m_2\int_{\mathbb{S}^1_+}\omega_x\omega_y{\rm d}N_{X^+,Y^-}(\bm{\omega})$, which correspond to $m_1\sigma(X^+,Y^+)$ and $m_2\sigma(X^+,Y^-)$, respectively. We then illustrate that $m_1=m_2=2$. First, consider
    \begin{equation*}
        \begin{aligned}
            \lambda(X,Y)&=\int_{\{(\omega_x,\omega_y)^\top\in \mathbb{S}^1:\omega_x\geq 0\}}\omega_x\omega_y{\rm d}H_{(X,Y)}(\bm{\omega}) \\
            &=\int_{\{(\omega_x,\omega_y)^\top\in \mathbb{S}^1:\omega_x,\omega_y\geq 0\}}\omega_x\omega_y{\rm d}H_{(X,Y)}(\bm{\omega})+\int_{\{(\omega_x,\omega_y)^\top\in \mathbb{S}^1:\omega_x\geq 0,\omega_y\leq 0\}}\omega_x\omega_y{\rm d}H_{(X,Y)}(\bm{\omega})\\
            &=\int_{\{(\omega_x,\omega_y)^\top\in \mathbb{S}^1:\omega_x,\omega_y\geq 0\}}\omega_x\omega_y{\rm d}H_{(X,Y)}(\bm{\omega})-\int_{\{(\omega_x,\omega_y)^\top\in \mathbb{S}^1:\omega_x,\omega_y\geq 0\}}\omega_x\omega_y{\rm d}H_{(X,-Y)}(\bm{\omega})\\
            &=\int_{\mathring{\mathbb{S}}^1_+}\omega_x\omega_y{\rm d}H_{(X,Y)}(\bm{\omega})-\int_{\mathring{\mathbb{S}}^1_+}\omega_x\omega_y{\rm d}H_{(X,-Y)}(\bm{\omega})\\
            &=\int_{\mathring{\mathbb{S}}^1_+}\omega_x\omega_y{\rm d}H_{(X^+,Y^+)}(\bm{\omega})-\int_{\mathring{\mathbb{S}}^1_+}\omega_x\omega_y{\rm d}H_{(X^+,Y^-)}(\bm{\omega})\\
            &=\int_{\mathbb{S}^1_+}\omega_x\omega_y{\rm d}H_{(X^+,Y^+)}(\bm{\omega})-\int_{\mathbb{S}^1_+}\omega_x\omega_y{\rm d}H_{(X^+,Y^-)}(\bm{\omega})\\
            &=m_1\int_{\mathbb{S}^1_+}\omega_x\omega_y{\rm d}N_{(X^+,Y^+)}(\bm{\omega})-m_2\int_{\mathbb{S}^1_+}\omega_x\omega_y{\rm d}N_{(X^+,Y^-)}(\bm{\omega}),
        \end{aligned}
    \end{equation*}
where $\bm{\omega}=(\omega_x,\omega_y)^\top$ and $\mathring{\mathbb{S}}^1_+$ denotes the first quadrant of the unit circle with the boundaries excluded (see Lemma~\ref{lemma:MeasureFromWholeToQuadrant}).
Note that the third line of the above follows from the discussion in Section~\ref{proof:sec:dtd}, and the fifth line follows from Lemma~\ref{lemma:MeasureFromWholeToQuadrant}. Moreover, the domain of the integrals changes between lines; this follows as the integrand $\omega_x\omega_y$ is always equal to zero when at least one of $\omega_x=0$ or $\omega_y=0$ holds. 

Recall that $H_{(X^+,Y^+)}(\cdot)=m_1N_{(X^+,Y^+)}(\cdot)$ and $H_{(X^+,Y^-)}(\cdot)=m_2N_{(X^+,Y^-)}(\cdot)$. As we have that $N_{(X^+,Y^+)}(\mathbb{S}^1_+)=N_{(X^+,Y^-)}(\mathbb{S}^1_+)=1$ and $ \omega^2_x+\omega_y^2=1$ for all $(\omega_x,\omega_y)^\top \in \mathbb{S}^1_+$, it follows that
    \begin{equation*}
        \begin{aligned}
            m_1&=m_1N_{(X^+,Y^+)}(\mathbb{S}^1_+)=H_{(X^+,Y^+)}(\mathbb{S}^1_+)=\int_{\mathbb{S}^1_+}{\rm d}H_{(X^+,Y^+)}(\bm{\omega})\\
            &=\int_{\mathbb{S}^1_+}(\omega^2_x+\omega^2_y){\rm d}H_{(X^+,Y^+)}(\bm{\omega})=\int_{\mathbb{S}^1_+}\omega^2_x{\rm d}H_{(X^+,Y^+)}(\bm{\omega})+\int_{\mathbb{S}^1_+}\omega^2_y{\rm d}H_{(X^+,Y^+)}(\bm{\omega}).
      \end{aligned}
    \end{equation*}
    Recall that for $(X^+,Y^+)^\top\in {\rm RV}^2_+(2)$, $$\nu_{(X^+,Y^+)}(\{(x,y)^\top\in \mathbb{E}^2_+:||(x,y)||_2>r,(x,y)/||(x,y)||_2\in B_H)\}=r^{-2}H_{(X^+,Y^+)}(B_H),$$ where $B_H\subset \mathbb{S}^2_+$ is a Borel subset. Then
    \begin{align*}
        \nu_{(X^+,Y^+)}(\{(x,y)^\top\in\mathbb{E}^2_+&:x>1\})=\nu_{(X^+,Y^+)}(\{(x,y)^\top\in\mathbb{E}^2_+:\|(x,y)\|_2 >  1/\omega_x,\})\\
         &=\int_{\mathbb{S}^1_+}\int_{\frac{1}{\omega_x}}^{\infty} 2r^{-3}{\rm d}r{\rm d}H_{(X^+,Y^+)}(\bm{\omega})=\int_{\mathbb{S}^1_+}\omega^2_x{\rm d}H_{(X^+,Y^+)}(\bm{\omega}),
    \end{align*}
     where $\bm{\omega}= (\omega_x,\omega_y)^\top = {(x,y)^\top \over |(x,y)\|_2}$. 
     
     Similarly, ${\int_{\mathbb{S}^1_+}\omega^2_y{\rm d}H_{(X^+,Y^+)}(\bm{\omega})}={\nu_{(X^+,Y^+)}(\{(x,y)^\top\in\mathbb{E}^2_+:y>1\})}$. Then we have $$\nu_{(X^+,Y^+)}(\{(x,y)^\top\in\mathbb{E}^2_+:x>1\})=\nu_{(X^+,Y^+)}(\{(x,y)^\top\in\mathbb{E}^2_+:y>1\})=1$$ for ${(X,Y)^\top \in {\rm BRV^2(2)}}$.  Therefore, $m_1=2$. A similar argument can be used to show $m_2=2$.

To show that  $$\lambda(X,Y)=\int_{\{(\omega_x,\omega_{y})^\top\in \mathbb{S}^1:\omega_x\geq 0\}} \omega_x\omega_{y} {\rm d}H_{(X,Y)}(\boldsymbol{\omega})=3\int_{\{(\omega_x,\omega_y)^\top\in \mathbb{S}^1:\omega_x\geq 0\}}\omega_x\omega_y{\rm d}N_{(X^+,Y)}(\bm{\omega}),$$ we first note that $(X^+,Y)^\top$ is regular varying on the right half plane, with angular measure $H_{(X^+,Y)}(\cdot)$ and limit measure $\nu_{(X^+,Y)}(\cdot)$. From a similar argument to the proof of Lemma~\ref{lemma:MeasureFromWholeToQuadrant}, we have, for $\Lambda\subseteq  \{\mathbb{S}^1:x > 0\}$, that $H_{(X^+,Y)}(\Lambda)=H_{(X,Y)}(\Lambda)$ and so we can write $\lambda(X,Y)$ as $$\int_{\{(\omega_x,\omega_y)^\top\in \mathbb{S}^1:\omega_x\geq 0\}}\omega_x\omega_y{\rm d}H_{(X,Y)}(\bm{\omega})=\int_{\{(\omega_x,\omega_y)^\top\in \mathbb{S}^1:\omega_x\geq 0\}}\omega_x\omega_y{\rm d}H_{(X^+,Y)}(\bm{\omega}).$$ We then have $H_{(X^+,Y)}(\cdot)= m_3N_{(X^+,Y)}(\cdot)$ with $N_{(X^+,Y)}(\{(\omega_x,\omega_y)^\top\in \mathbb{S}^1:\omega_x\geq 0\})=1$ and so 
    \begin{equation*}
        \begin{aligned}
            m_3=H_{(X^+,Y)}(\{(\omega_x,\omega_y)^\top\in \mathbb{S}^1:\omega_x\geq 0\})&=\int_{\{(\omega_x,\omega_y)^\top\in \mathbb{S}^1:\omega_x\geq 0\}}(\omega^2_x+\omega^2_y){\rm d}H_{(X^+,Y)}(\bm{\omega}),
           \end{aligned}
    \end{equation*}
    where
      \begin{equation*}
        \begin{aligned}
            \int_{\{(\omega_x,\omega_y)^\top\in \mathbb{S}^1:\omega_x\geq 0\}}\omega^2_x{\rm d}H_{(X^+,Y)}(\bm{\omega})&=\int_{\{(\omega_x,\omega_y)^\top\in \mathbb{S}^1:\omega_x\geq 0\}}\int_{\frac{1}{\omega_x}}^{\infty} 2r^{-3}{\rm d}r{\rm d}H_{(X^+,Y)}(\bm{\omega})\\
            &=\nu_{(X^+,Y)}(\{(x,y)^\top\in\mathbb{E}^2_+:x>1\}),
                    \end{aligned}
    \end{equation*}
    and
       \begin{equation*}
        \begin{aligned}
            \int_{\{(\omega_x,\omega_y)^\top\in \mathbb{S}^1:\omega_x\geq 0\}}\omega^2_y{\rm d}&H_{(X^+,Y)}(\bm{\omega})=\int_{\{(\omega_x,\omega_y)^\top\in \mathbb{S}^1:\omega_x\geq 0\}}\int_{\frac{1}{\omega_y}}^{\infty} 2r^{-3}{\rm d}r{\rm d}H_{(X^+,Y)}(\bm{\omega})\\
            &=\nu_{(X^+,Y)}(\{(x,y)^\top\in\mathbb{E}^2_+:|y|>1\})\\
            &=\nu_{(X^+,Y)}(\{(x,y)^\top\in\mathbb{E}^2_+:y>1\}\cup \{(x,y)^\top\in\mathbb{E}^2_+:y<-1\} ).
        \end{aligned}
    \end{equation*}
    Thus, as $(X,Y)^\top\in {\rm BRV}^2(2)$, each of those three parts ${\nu_{(X^+,Y)}(\{(x,y)^\top\in\mathbb{E}^2_+:x>1\})}$, ${\nu_{(X^+,Y)}(\{(x,y)^\top\in\mathbb{E}^2_+:y>1\})}$ and 
 ${\nu_{(X^+,Y)}(\{(x,y)^\top\in\mathbb{E}^2_+:y<-1\})}$ all equals to $1$ and ${m_3=3}$ as needed.
\end{proof}    

\subsection{Proof of Proposition~\textcolor{blue}{3.6}}
\label{sec:proof:asymptoticNormality}

The Proof of Proposition~\textcolor{blue}{3.6} follows the same strategy as in \cite{larsson2012extremal}. The major difference comes from the estimator's definition of an ``exceedance''. In our case, we use $\mathbbm{1}\{r_{i} \geq r_0\}$ where $r_0=a(n / k)$ (same setting as used in \cite{resnick2004extremal}), while in \cite{larsson2012extremal}, they use $\mathbbm{1}\{r_{i} \geq r_{(k)}\}$ where $r_{(k)}$ is the $k$-th order statistic of $\left\{r_{i}\right\}_{i=1}^{n}$. 

\begin{manualprop2}{3.6}{Asymptotic normality of $\hat{\lambda}^1_n$}
        Let $(X,Y)^\top\in {\rm BRV}^2(2)$ with ${\omega^+_{x}=\frac{X^+}{R}}$, ${\omega_{y}=\frac{Y}{R}}$, where $X^+=\max\{X,0\}$ and $R=||(X^+,Y)^\top||_2$. Let $a(n)$ be a function satisfying $a(n)\rightarrow \infty$ and $n\mathbb{P}\{R>a(n)\}\rightarrow 1$ as $n\to \infty$. Define the ``bias'' process by
        \[B_n(t)=\frac{n}{\sigma_n\sqrt{k}}(\mathbb{E}\left[\omega^+_{x}\omega_{y}\mathbbm{1}\{a(n/k)^{-1}R\geq t\}\right]-\frac{\lambda}{3}\mathbb{P}[a(n/k)^{-1}R\geq t]),\]
    
        where $\sigma^2_{n}={\rm Var}[\omega^+_{x}\omega_{y}\mid a(n/k)^{-1}R\geq t]$. Assume that $n,k\to\infty$, $n/k\to \infty$, and that ${\lim_{n\to \infty}B_n(1)\xrightarrow{P} 0}$. Then,
        \begin{equation*}
            \frac{\sqrt{k}}{3}(\hat{\lambda}^1_n-\lambda)\xrightarrow{d}N(0,\sigma^2),
            \tag{8}
        \end{equation*} 
        where $\sigma^2={\rm Var}[\tilde{\omega}^+_{x}\tilde{\omega}_{y}]$, $\lambda=3\mathbb{E}[\tilde{\omega}^+_{x}\tilde{\omega}_{y}]$, and $(\tilde{\omega}^+_{x},\tilde{\omega}_{y})^\top$ has distribution $N_{(X^{+},Y)}(\cdot)$.
\end{manualprop2}

\begin{proof}[Proof of Proposition~\textcolor{blue}{3.6}]\label{proof:asymptoticNormality}
    Let $N_n(t)=\sum_{i=1}^{n}\mathbbm{1}\{a^{-1}(n/k)r_i \geq t^{-\frac{1}{2}}\}$, $t>0$. The expected number of exceedance over $r_0=a(n / k)$ is ${\mathbb{E}[N_{n}(1)]=k}$. Let $\{i_{(j,n)}(t),j\geq 1\}$ be the $j^{\textrm{th}}$ index in $i=1,2,\dots$ that for $t>0$, it satisfies ${a(n/k)^{-1}R_{i_{(j,n)}(s,t)}\geq t^{-\frac{1}{2}}}$. When is understood, we write $i(j,n)$ and $N_n$ instead of $i_{(j,n)}(t)$ and $N_n(t)$, respectively.
 
    Let $W_n(t)=\frac{1}{\sigma_n\sqrt{k}}\sum_{i=1}^n\left(\omega^+_{x,i}\omega_y-\frac{\lambda}{3}\right)\mathbbm{1}\{a^{-1}(n/k)r_i \geq t^{-\frac{1}{2}}\}$ where $t>0$. The estimator $\hat{\lambda}^1_n$ can be linked with $W_n(1)$ as $\frac{\sqrt{k}}{3}(\hat{\lambda}^1_n-\lambda)=\frac{\sigma_n k}{N_n(1)}W_n(1)$. If  $W_n(1) \Rightarrow Z$ holds, since ${\sigma_n\xrightarrow{P}\sigma}$, ${\frac{N_n(1)}{k}\xrightarrow{P}1}$, then $\lim_{n\to\infty}\frac{\sqrt{k}}{3}(\hat{\lambda}^1_n-\lambda)\xrightarrow{d} \sigma Z$. Now,
    \begin{equation*}
        \begin{aligned}
            W_n(t)=&\frac{1}{\sigma_n\sqrt{k}}\sum_{j=1}^{N_n}\left(\omega^+_{x,i(j,n)}\omega_{y,i(j,n)}-\frac{\lambda}{3}\right)\\
            =\frac{1}{\sigma_n\sqrt{k}}&\sum_{j=1}^{N_n}\left(\omega^+_{x,i(j,n)}\omega_{y,i(j,n)}-\mathbb{E}[\omega^+_{x,i(j,n)}\omega_{y,i(j,n)}]\right)+\frac{1}{\sigma_n\sqrt{k}}\sum_{j=1}^{N_n}\left(\mathbb{E}[\omega^+_{x,i(j,n)}\omega_{y,i(j,n)}]-\frac{\lambda}{3}\right) \\
            &:= C_n(t) + D_n(t).
        \end{aligned}
    \end{equation*}
First, consider $D_n$. As $\{(\omega^+_{x,i(j,n)},\omega_{y,i(j,n)})^\top,j\geq 1\}$ are independent and identically distributed, then
    \begin{equation*}
        \begin{aligned}
D_n(t)&=\frac{N_n}{\sigma_n\sqrt{k}}\left(\mathbb{E}[\omega^+_x\omega_y\mid a^{-1}(n/k)R\geq t^{-\frac{1}{2}}]-\frac{\lambda}{3}\right)\\
            =&\frac{N_n/k}{ \frac{n}{k}\mathbb{P}[a^{-1}(n/k)R \geq t^{-\frac{1}{2}}]}\\
            &\times \frac{n}{\sigma_n\sqrt{k}}\left(\mathbb{E}[\omega^+_{x}\omega_{y}\mathbbm{1}\{a^{-1}(n/k)R\geq t^{-\frac{1}{2}}\}]-\frac{\lambda}{3}\mathbb{P}[a^{-1}(n/k)R\geq t^{-\frac{1}{2}}]\right)\\
            =&\frac{N_n/k}{ \frac{n}{k}\mathbb{P}[a^{-1}(n/k)R\geq t^{-\frac{1}{2}}]}B_n(t),
        \end{aligned}
    \end{equation*}

where $N_n(t)/k\xrightarrow{P}t$ and $\frac{n}{k}\mathbb{P}[a^{-1}(n/k)R\geq t^{-\frac{1}{2}}]\rightarrow t$. Also by assuming $B_n(1)\xrightarrow{P}0$, then $D_n(1)=\frac{N_n(1)/k}{ \frac{n}{k}\mathbb{P}[a^{-1}(n/k)R\geq 1]}B_n(1)\xrightarrow{P} 0$.

Next, consider $C_n$. By \cite{larsson2012extremal}, Theorem 1, Equation (\textcolor{blue}{20}), $C_n(1)\Rightarrow Z$. Therefore, $W_n(1)=C_n(1)+D_n(1)$ where $C_n(1)\Rightarrow Z$ and $D_n(1)\xrightarrow{P}0$. By Slutzky's theorem, $W_n(1)\Rightarrow Z$.
\end{proof}

\section{High-frequency China's derivatives dataset}\label{sec:dataset}
To encourage research on this topic, we open-source our high-resolution dataset on China's derivatives market. The dataset can also be used in other research fields such as time series analysis, market microstructure, and macroeconomic analysis.

\subsection{Overview}

The subsequent sections are structured as followed. Section~\ref{sec:dataset:col} provides details on the collection of raw data from the exchanges and Section~\ref{sec:dataset:meta} details other meta information. Section~\ref{sec:dataset:topic} deliberates on the possible research avenues which these data enable. Section~\ref{sec:dataset:standard} concludes with details of a standardized rendition of our dataset, which is designed for easy handling by researchers and, hence, we choose to consider in our application (Section~\textcolor{blue}{5}).

\subsection{Raw data collection and description}
\label{sec:dataset:col}
Mainland China is home to six major exchanges, which each have their own specialisation:
\begin{itemize}
\item \textbf{Shanghai Futures Exchange (SHFE)}: Metal commodities.
\item \textbf{Shanghai International Energy Exchange (INE}): Energy commodities, with the notable introduction of the Shanghai Crude Oil in 2018.
\item \textbf{China Financial Futures Exchange (CFFEX)}: Financial contracts, including futures of national bonds and stock indices.
\item \textbf{Zhengzhou Commodity Exchange (CZCE)}: Agricultural products. 
\item \textbf{Dalian Commodity Exchange (DCE}): Agricultural products. 
\item \textbf{Guangzhou Futures Exchange (GFEX)}: this newest exchange not only focuses on carbon emissions rights and electricity but also exemplifies China's commitment to introducing pioneering financial products. According to its official website, GFEX has received approval from the China Securities Regulatory Commission (CSRC) to develop and list 16 futures products. 
\end{itemize}


All six exchanges share similar data disclosure policies. Data are provided in an aggregated form over regular time intervals. For public users, exchanges relay data to public channels every 500 milliseconds unless no market update has occurred, i.e., there has been neither a transaction nor an alteration to the order book in the last 500 millisecond; in this case, updates are not provided to the public channels. Table~\ref{tab:dataRecord} provides the fields of data released by the exchanges. The ``update time'' field is the timestamp provided by the exchanges, which also functions as a unique identifier for a record. Fields ``ask prices $1$ to $5$'' and ``bid prices $1$ to $5$''  denote the topmost prices on the ask and bid sides, respectively, while the ``ask volumes $1$ to $5$'' and ``bid volumes $1$ to $5$'' denote the available quantities at the associated volume prices on the order book. The ``volume'' reflects the total trading volume of the derivative over the preceding 500 milliseconds, and the ``turnover'' measures the trading value in CNY (currency) for the same duration. ``open interest'' represents the aggregate count of unsettled derivative contracts, while ``last price'' indicates the most recent transaction price.

\begin{table}[t!]
    \centering
    \caption{Fields of data types released by the six exchanges in mainland China.}
    \begin{tabular}{@{}llr@{}} \toprule
        Field    & Description \\ \midrule
        update time & the update timestamp \\
        ask price 1-5  & the top five prices on the ask side     \\
        bid price 1-5		& the top five prices on the bid side         \\
        ask volume 1-5 & the volume for the associated top five prices on the ask side     \\
        bid volume 1-5 & the volume for the associated top five prices on the bid side      \\ 
        last price       & the last executed price      \\
        volume       & the number of total traded volume in the preceding 500 milliseconds     \\
        open interest       &   total number of outstanding derivative contracts    \\
        turnover& the amount of traded volume in CNY currency  \\ \bottomrule
    \end{tabular}
    \label{tab:dataRecord}
    \end{table}

\begin{figure}[t!]
    \centering
    \begin{subfigure}[b]{0.8\textwidth}
        \includegraphics[width=\textwidth]{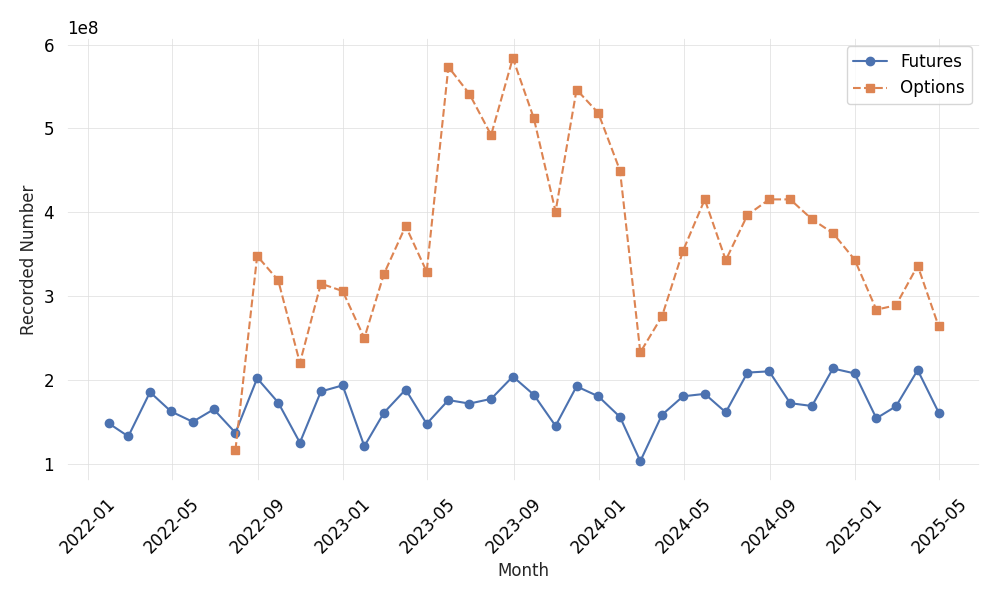}
    \end{subfigure}
    \caption{Monthly recorded numbers for futures and options.}
    \label{fig:weekly}
\end{figure}

In order to extract data from across China's derivatives market, we implement an online data collection program that utilizes application programming interfaces (APIs) to connect with brokers. This program has been operational and amassing raw data since August 2022. The derivatives market in China operates on a membership basis. Individuals aiming to trade in the market must do so through brokers. These brokers, being members of the exchange, shoulder the responsibility of risk management for traders and clearing of trades. Exchanges engage solely with brokers and both market and trading data are conveyed through these brokers. For algorithmic traders, brokers offer APIs to facilitate data access. Figure~\ref{fig:weekly} gives the number of monthly records accumulated through our data collection procedure. On average, daily records encompass 8 million entries for 700 futures instruments and 16 million entries for 10,000 option instruments. For futures, various maturities correspond to different instruments. In the case of options, combinations of differing maturities, strike prices, and types (call/put) lead to an array of instruments. Consequently, the instrument count for option data significantly surpasses that of futures data. We note that the options markets in China, while not as brisk as the futures market, demonstrates an upward trajectory in activity.

\subsection{Other meta data}
\label{sec:dataset:meta}
In addition to the raw data, we also provide other metadata that are useful for researchers. The metadata includes basic information about trading instruments. Specifically, The metadata includes the following information: (1) Commission fee --- the exchanges charge traders a fee for each transaction and use it as a tool to control liquidity. The commission fee changes occasionally. For high-frequency traders, the commission fee is an important factor to consider; (2) Instrument meta information --- this includes meta-information about a trading product such as the maturity date, underlying assets, tick price (the unit of minimum price change), volume multiplier (the volume of underlying assets per contract), etc.; (3) Product meta information --- the product table contains meta information for all underlying assets, including the exchange where the product is traded and the trading time window; (4) Trading day --- this indicates whether there is a day session and a night session for a trading day. On a normal working day (Monday to Friday), both night and day sessions are available. For holidays, there is no night session on the preceding day, and no neither night session nor day sessions within holidays.

\subsection{Potential research topics}\label{sec:dataset:topic}
With this dataset, a plethora of empirical research topics can be explored. Following a review of recent relevant literature on China's derivatives market, we have summarized several potential research avenues: (1) \textbf{Time-series dependence:} \cite{chen2021interdependence} investigated the interdependence between different sectors in China's derivatives market using time-varying vector autoregressive (TV-VAR) and graphical models. Their study encompassed six sectors: petrochemicals, energy, softs, non-ferrous metals, oil and fats, and grains. Between 2004 and 2019, they identified that the energy and petrochemical sectors primarily drove spillover effects. \cite{jia2016correlation} delved into the temporal dependence between China's agricultural futures markets and the US counterpart by examining lead-lag effects. \cite{yang2021global} analyzed the interplay between volatility in developed financial markets and the returns on China's crude oil, by employing both conventional causality analysis and causality in quantiles; (2) \textbf{Risk-contagions:} \cite{yang2021extreme} investigated the risk spillover between the Chinese and global crude oil futures markets. By constructing upside and downside Value-at-Risk (VaR) connectedness networks, they observed a marked surge in risk spillover post the onset of the COVID-19 pandemic; (3) \textbf{Market efficiency:} \cite{bohl2018speculative} examined speculative activity against return volatility for crude oil products traded on DCE and CZCE. By defining speculative activity as the ratio of trading volume to open interest and executing Granger causality analysis, they discerned that heightened speculative activity tends to destabilize returns and amplify volatility. Conversely, \cite{yang2020pricing} assessed the efficiency disparity between the futures market of crude oil in the INE and the spot markets, predominantly contained within the Asia-Pacific region; (4) \textbf{Market reaction to news:} \cite{li2022novel} explored the sentiment of relevant news and its repercussions on agricultural futures, by harnessing text analysis techniques. Additionally, \cite{li2022time} scrutinized the volatility spillover effects on crude oil products before and after the COVID-19 outbreak in 2022; (5) \textbf{Trading strategy:} \cite{zhang2022tradebot} developed a trading bot based on the reinforcement learning algorithm to trade in the China's derivatives market. This paper utilizes high-frequency data. However, the authors did not disclose the data source. \ifnum\blind=0\cite{jiang2022market} \else \textcolor{blue}{Anonym et al. }(\textcolor{blue}{2022}) \fi first studied the market-making strategy combined with reinforcement learning in China's derivatives Market. \cite{wu2022momentum} study the momentum strategy in China's derivatives market; (6) \textbf{Market microstructure:} So far, not many studies on the microstructure of the Chinese futures market can be found. However, in the meantime, this is an active topic in other markets such as cryptocurrency market (\cite{almeida2024cryptocurrency}) US stock market (\cite{o2015high}). The high-frequency data we provide can be used to study the microstructure of China's derivatives market; (7) \textbf{Option pricing:} \cite{li2018option} studied the correlation between past returns and the implied volatility spread (difference in implied volatility between call and put options with identical maturity and strike prices) for SSE 50 ETF options in China's derivatives market. Contrary to the Black-Scholes Model's prediction of a zero spread, they found a positive relationship with past returns, attributable to the momentum factor. \cite{liu2023risk} studied the risk appetite and options-implied information.

Most research mentioned above study daily data, and studies of sub-daily exchange data are largely missing from the literature. Our high-frequency data will be able to facilitate such studies and potentially corroborate the findings of the above studies.

\subsection{Standardized data}
\label{sec:dataset:standard}
To create a standardized dataset conducive to accessibility, we have structured the data according to specific guidelines. These standards not only ensure uniformity across different instruments but also facilitate its ease of use by other researchers. The core attributes of our standardized dataset are as follows: (1) \textbf{Standard Format:} The dataset adheres to a consistent structure regardless of the instrument, making it seamless for other researchers to employ; (2) \textbf{Market Representativeness:} For this study, we constrain our focus to commodity future derivatives. The data spans a year, from \TrainstartDate\ to \TestendDate, incorporating all products from the six aforementioned exchanges, except CFFEX. For each product, we select the most actively traded instrument, as gauged by its trading volume; (3) \textbf{Trading Time Window:} Notably, while some products undergo trading during the nighttime session, we've truncated the dataset to encapsulate only the daytime trading hours: 9:00 to 15:00; (4) \textbf{Pricing Convention:} We employ the mid-price as the standard price metric for each instrument. This mid-price is deduced by averaging the best bid and best ask prices. Subsequently, return series on a per-minute are derived using the formula $R_t=\frac{P_{t+1}}{P_t}-1$. Returns that straddle across two trading days are flagged as missing values. In subsequent analyses, these can be supplanted by the mean return, which approximates to zero.

In summary, the dataset comprises 6,615,735 observations across 55 distinct assets, with their classification detailed in Table~\ref{tab:products}. The classification methodology is based on the research report available at \href{https://inst.citicsf.com/EN/researchReport/reportDetails?researchId=3030056}{CITIC Securities Research Report}. The dataset is publicly available on Zenodo at \href{https://doi.org/10.5281/zenodo.15726389}{https://doi.org/10.5281/zenodo.15726389}. Higher-resolution data can be provided upon reasonable request to the authors.

\begin{table}[ht!]
    \caption{Commodity products in the standardized China's derivatives market data.}
    \centering
    \resizebox{\textwidth}{!}{
    {\renewcommand{\arraystretch}{0.6} 
\begin{tabular}{llllllll}
    \toprule
    {\textbf{code}} &  Name &\textcolor{customizedColor}{Categories} & Exchange & {\textbf{code}} & Name  &   \textcolor{customizedColor}{Categories} & Exchange\\
    \midrule
    \textbf{a}  &   soybean type.1 &   agricultural products/oil crops &  DCE&  \textbf{ag} &      silver&metals/precious metals &     SHFE \\
    \textbf{al} &       aluminum &metals/nonferrous metals &     SHFE &\textbf{ao} &     aluminium oxide &metals&     SHFE \\
    \textbf{AP} &      apple &   agricultural products/economic crops  &     CZCE  & \textbf{au} &      gold&metals/precious metals &     SHFE \\
    \textbf{bb} &      rubber board &   agricultural products/rubber\&woods&      DCE &\textbf{bc} &     international copper&metals/nonferrous metals &      INE\\
    \textbf{br} &   butadiene rubber &   agricultural products/rubber\&woods&     SHFE &\textbf{bu} &      bitumen & energy\& chemicals/oil\& gas&     SHFE \\
    \textbf{c}  &      corn &   agricultural products/grains &      DCE &\textbf{CF} &      cotton type a &   agricultural products/economic crops &     CZCE \\
    \textbf{CJ} &      red dates &   agricultural products/economic crops  &     CZCE &\textbf{cs} &    corn starch&   agricultural products/grains  &      DCE \\
    \textbf{cu} &       copper &metals/nonferrous metals &     SHFE &\textbf{CY} &      cotton yarn &   agricultural products/economic crops  &     CZCE \\
    \textbf{eb} &     styrene & energy\& chemicals/olefins &      DCE &\textbf{eg} &     ethylene glycol & energy\& chemicals/alcohols&      DCE \\
    \textbf{fb} &      fiberboard &  agricultural products/rubber\&woods &      DCE &\textbf{FG} &      glass & energy\& chemicals/inorganics &     CZCE \\
    \textbf{fu} &     fuel oil & energy\& chemicals/oil\& gas&     SHFE &\textbf{hc} &      hot rolled coil &metals/ferrous metals &     SHFE \\
    \textbf{i}  &     iron ore &metals/ferrous metals &      DCE &\textbf{IC} &   CSI 500 & financial futures/Equity index&    CFFEX \\
    \textbf{IF} &   CSI 300 & financial futures/Equity index&    CFFEX &\textbf{IH} &    SSE 50 & financial futures/Equity index&    CFFEX \\
    \textbf{IM} &  CSI 1000 & financial futures/Equity index&    CFFEX &\textbf{j}  &      coke & energy\& chemicals/coals&      DCE \\
    \textbf{jd} &      eggs &   agricultural products/animals&      DCE &\textbf{jm} &      coking coal & energy\& chemicals/coals&      DCE \\
    \textbf{JR} &      japonica rice &   agricultural products/grains&     CZCE &\textbf{l}  &      plastics & energy\& chemicals/olefins&      DCE \\
    \textbf{lc} &     lithium carbonate & metals/novel materials &     GFEX &\textbf{lh} &      live hog &   agricultural products/animals &      DCE \\
    \textbf{LR} &     late indica rice &   agricultural products/grains&     CZCE &\textbf{lu} &   low sulfur fuel oil& energy\& chemicals/oil\& gas &      INE\\
    \textbf{m}  &      soybean meal &   agricultural products/oil crops&      DCE &\textbf{MA} &      methanol& energy\& chemicals/alcohols &     CZCE \\
    \textbf{ni} &       nickel &metals/nonferrous metals &     SHFE &\textbf{nr} &    rubber No.20  &   agricultural products/rubber\&woods  &   INE \\
    \textbf{OI} &      vegetable oil &   agricultural products/oil crops&     CZCE &\textbf{p}  &     palm oil &   agricultural products/oil crops&      DCE \\
    \textbf{pb} &       lead &metals/nonferrous metals &     SHFE &\textbf{PF} &      short fiber & energy\& chemicals/aromatics&     CZCE \\
    \textbf{pg} &   liquefied petroleum gas& energy\& chemicals/oil\& gas  &      DCE &\textbf{PK} &      peanut &   agricultural products/economic crops  &     CZCE \\
    \textbf{PM} &      general wheat &   agricultural products/grains&     CZCE &\textbf{pp} &      polypropylene& energy\& chemicals/olefins &      DCE \\
    \textbf{rb} &     rebar &metals/ferrous metals &     SHFE & \textbf{b} &   soybean type.2 &   agricultural products/oil crops &      DCE \\
    \textbf{RI} &     early indica rice&   agricultural products/grains &     CZCE &\textbf{RM} &      vegetable meal &   agricultural products/oil crops &     CZCE \\
    \textbf{rr} &      japonica rice &   agricultural products&      DCE &\textbf{RS} &     rapeseed &   agricultural products/oil crops&     CZCE \\
    \textbf{ru} &     natural rubber  &  agricultural products/rubber\&woods&   SHFE &\textbf{SA} &      soda ash & energy\& chemicals/inorganics &     CZCE \\
    \textbf{sc} &      crude oil & energy\& chemicals/oil\&gas &      INE &\textbf{SF} &      ferrosilicon &metals/ferrous metals  &     CZCE \\
    \textbf{si} &     industrial silicon & metals/novel materials &     GFEX &\textbf{SM} &      manganese silicon & metals/ferrous metals &     CZCE \\
    \textbf{sn} &       tin &metals/nonferrous metals &     SHFE &\textbf{sp} &      pulp &   agricultural products/rubber\&woods &     SHFE \\
    \textbf{SR} &      white sugar &   agricultural products/economic crops &     CZCE &\textbf{ss} &     stainless steel &metals/ferrous metals &     SHFE \\
    \textbf{T}  &   10-year government bond & financial futures/interest rates &    CFFEX &\textbf{TA} &     PTA & energy\& chemicals/aromatics&     CZCE \\
    \textbf{TF} &    5-year government bond & financial futures/interest rates &    CFFEX &\textbf{TL} &   30-year government bond& financial futures/interest rates &    CFFEX \\
    \textbf{TS} &    2-year government bond & financial futures/interest rates &   CFFEX &\textbf{UR} &      urea & energy\& chemicals/inorganics&     CZCE \\
    \textbf{v}  &     PVC& energy\& chemicals/olefins &      DCE &\textbf{WH} &      strong wheat &   agricultural products/grains&     CZCE \\
    \textbf{wr} &      wire rod &metals/ferrous metals &     SHFE &\textbf{y}  &      soybean oil &   agricultural products/oil crops&      DCE \\
    \textbf{ZC} &     thermal coal& energy\& chemicals/coals &     CZCE &\textbf{zn} &       zinc &metals/nonferrous metals&     SHFE \\
    \textbf{PX} &     paraxylene& energy\& chemicals/aromatics &     CZCE &  \textbf{SH} & sodium hydroxide & energy\& chemicals/inorganics &     CZCE \\
    \textbf{ec} & SCFIS(Europe) & indices/indices &     SHFE  & & &  & \\

    \bottomrule
    \end{tabular}
    }
    }
    \label{tab:products}
\end{table}

\clearpage
\section{Further simulation studies}
\label{sec:moreSim}

\begin{figure}[h!]
    \centering
    \begin{subfigure}[b]{0.4\textwidth}
        \includegraphics[width=\linewidth]{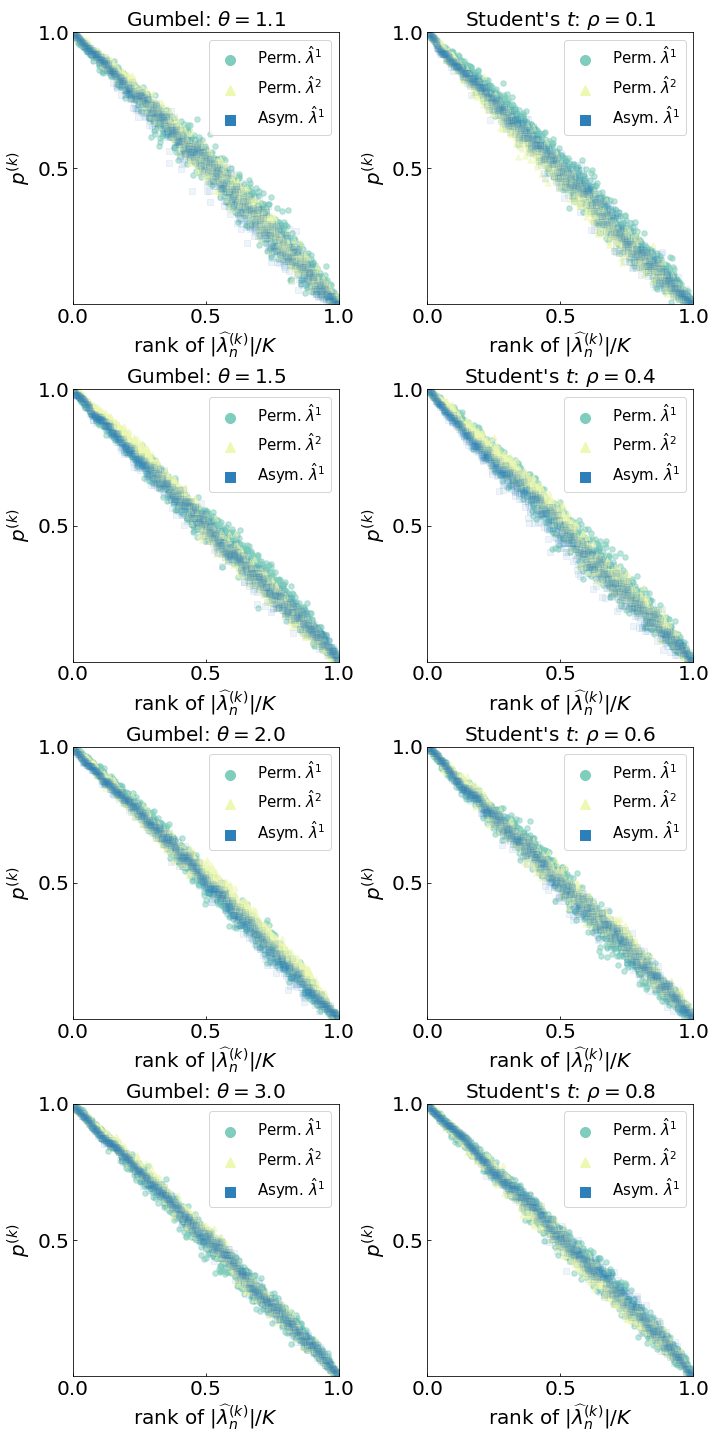}
    \end{subfigure}
    \begin{subfigure}[b]{0.4\textwidth}
        \includegraphics[width=\linewidth]{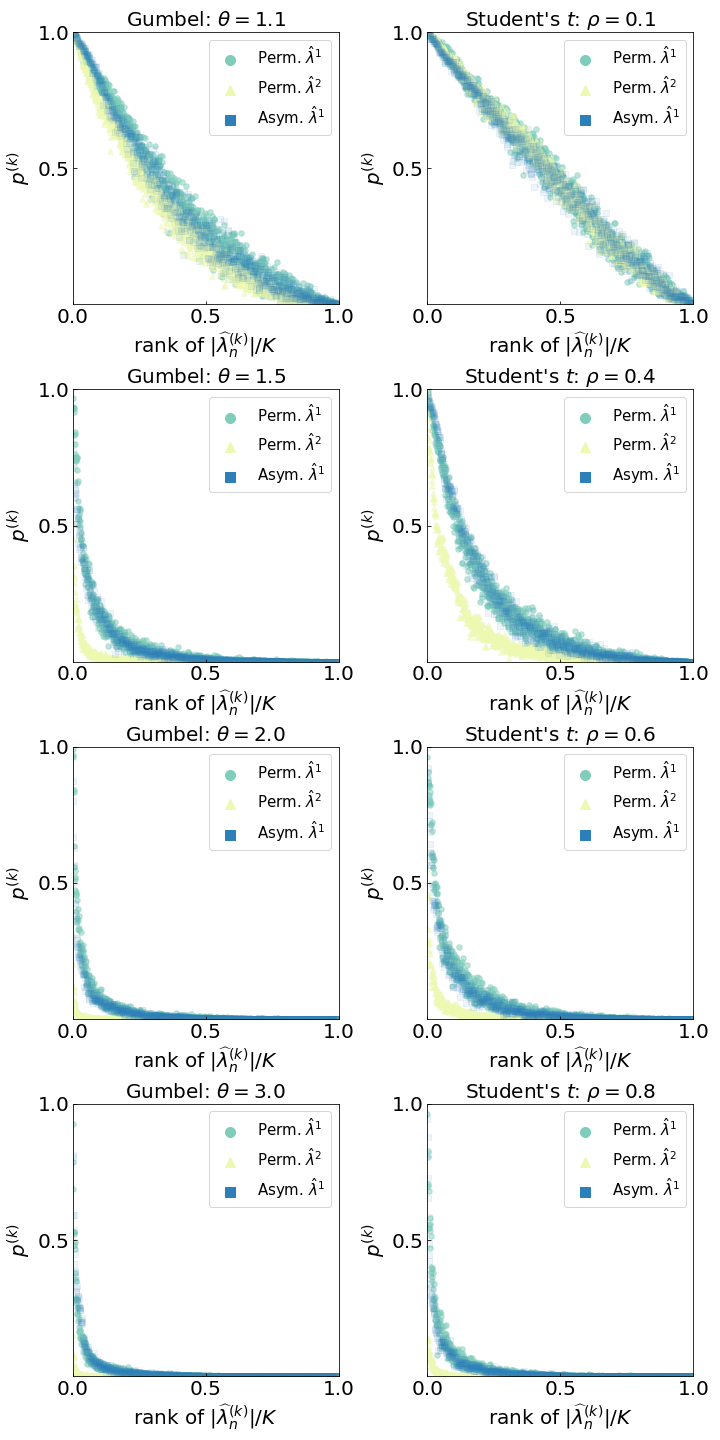}
    \end{subfigure}
\caption{Plot of p-values ($y$-axis) against empirical rank of test statistic $|\widehat{\lambda}_n^{(k)}(X,Y)|/K$ ($x$-axis); case $\lambda(X,Y)=0$ (Left two columns) and case $\lambda(X,Y)\neq0$ (Right two columns); Circular (green) and triangular (yellow) points correspond to permutation tests using $\widehat{\lambda}_n^1$ and $\widehat{\lambda}^2_n$, respectively; Square (blue) points correspond to $t$-tests using $\widehat{\lambda}_n^1$. The underlying copula differs with the columns: the first and third columns are results for Gumbel copula and the second and fourth columns are results for Student's $t$-copula. Rows correspond to different parameters for the copulas: for the Gumbel copula, $\theta\in\{1,1.5,2.5,3.0\}$ and for the Student's $t$-copula, $\rho\in\{0.1,0.4,0.6,0.8\}$ with degrees of freedom $ \nu$ fixed to be $4$.}
    \label{fig:sim:more}
\end{figure}

We conduct additional simulation studies to investigate the performance of our testing procedures with a lower sample size and lower exceedance threshold. Specifically, we repeat the simulation design from Section~\textcolor{blue}{4} of the main paper but with a smaller sample size of $n=1,000$ and a lower exceedance threshold at the $0.9$-quantile. All other settings remain the same as in Section~\textcolor{blue}{4} of the main paper. The results are shown in Figure~\ref{fig:sim:more}. The findings are consistent with those in Section~\textcolor{blue}{4} of the main paper, demonstrating that the permutation test with $\hat{\lambda}^2_n$ performs better than the $t$-test. The power to reject the null hypothesis when data are generated under alternative hypothesis are summarized in Table~\textcolor{blue}{1} of the main paper. In addition, the empirical rejection rates when data are simulated under the null hypothesis are reported in Table~\ref{tab:power:underNull} and Table~\ref{tab:power:underNullT}. As expected, the rejection rates generally align with the nominal significance levels. However, for $n = 1000$ and the $0.99$-quantile, only $1000 \times 0.01 = 10$ effective samples are available, and all three proposed tests tend to exhibit biased rejection frequencies, especially the permutation-based test using the first estimator and the asymptotic test. Among them, the permutation-based test using the second estimator, $\hat{\lambda}^2_n$, yields rejection rates closest to the nominal level. This can be attributed to its data efficiency as explained in Section~\textcolor{blue}{4.2}.

\begin{table}[t!]
    \centering
    \caption{Empirical size (i.e., percentage of null hypotheses rejected when data are generated under the null hypothesis at a given significance level $\alpha^*$) of the three proposed statistical tests, at $\alpha^* = 0.01$ and $\alpha^* = 0.05$, across different sample sizes, $q$-quantile and Gumbel copula settings. The number of permutation is fixed as $P=1000$. For the Gumbel copula, $\theta \in \{1, 1.5, 2.5, 3.0\}$. Entries with asterisk symbol denote significance levels that lie beyond three standard deviations from the mean.}
    \footnotesize
    \renewcommand{\arraystretch}{0.6}
    \begin{tabular}{c c c | ccc | ccc}
        \toprule
        \multirow{2}{*}{$\theta$} & \multirow{2}{*}{$n$} & \multirow{2}{*}{$q$-quantile}
        & \multicolumn{3}{c|}{$\alpha^* = 0.01$} 
        & \multicolumn{3}{c}{$\alpha^* = 0.05$} \\
        \cmidrule(lr){4-6} \cmidrule(lr){7-9}
        & & & Perm. $\hat{\lambda}_n^1$ & Perm. $\hat{\lambda}_n^2$ & Asym. $\hat{\lambda}_n^1$
            & Perm. $\hat{\lambda}_n^1$ & Perm. $\hat{\lambda}_n^2$ & Asym. $\hat{\lambda}_n^1$ \\
        \midrule
\multirow{8}{*}{$1.1$} 
    &\multirow{2}{*}{1000} & $q=0.90$ & $1.0\pm0.3\ $ & $0.7\pm0.3\ $ & $0.8\pm0.3\ $ & $5.3\pm0.7\ $ & $4.1\pm0.6\ $ & $4.2\pm0.6\ $ \\
             & & $q=0.99$ & $0.2\pm0.1^*$ & $1.5\pm0.4\ $ & $0.0\pm0.0^*$ & $3.9\pm0.6\ $ & $5.4\pm0.7\ $ & $1.9\pm0.4^*$ \\
            &\multirow{2}{*}{2500} & $q=0.90$ & $1.0\pm0.3\ $ & $0.6\pm0.2\ $ & $1.2\pm0.3\ $ & $5.6\pm0.7\ $ & $4.3\pm0.6\ $ & $5.8\pm0.7\ $ \\
             & & $q=0.99$ & $1.0\pm0.3\ $ & $1.0\pm0.3\ $ & $0.5\pm0.2\ $ & $5.4\pm0.7\ $ & $5.4\pm0.7\ $ & $5.1\pm0.7\ $ \\
            &\multirow{2}{*}{5000} & $q=0.90$ & $0.7\pm0.3\ $ & $1.0\pm0.3\ $ & $0.8\pm0.3\ $ & $3.5\pm0.6\ $ & $4.9\pm0.7\ $ & $6.2\pm0.8\ $ \\
             & & $q=0.99$ & $0.6\pm0.2\ $ & $0.6\pm0.2\ $ & $0.6\pm0.2\ $ & $5.4\pm0.7\ $ & $4.5\pm0.7\ $ & $4.6\pm0.7\ $ \\
            &\multirow{2}{*}{10000} & $q=0.90$ & $1.1\pm0.3\ $ & $1.1\pm0.3\ $ & $0.6\pm0.2\ $ & $5.0\pm0.7\ $ & $5.4\pm0.7\ $ & $4.6\pm0.7\ $ \\
             & & $q=0.99$ & $0.8\pm0.3\ $ & $0.9\pm0.3\ $ & $0.9\pm0.3\ $ & $5.4\pm0.7\ $ & $4.5\pm0.7\ $ & $5.3\pm0.7\ $ \\
            \bottomrule
    \multirow{8}{*}{$1.5$} 
    &\multirow{2}{*}{1000} & $q=0.90$ & $1.0\pm0.3\ $ & $1.5\pm0.4\ $ & $0.5\pm0.2\ $ & $5.0\pm0.7\ $ & $5.6\pm0.7\ $ & $3.9\pm0.6\ $ \\
             & & $q=0.99$ & $0.6\pm0.2\ $ & $1.2\pm0.3\ $ & $1.1\pm0.3\ $ & $3.2\pm0.6^*$ & $5.7\pm0.7\ $ & $6.3\pm0.8\ $ \\
            &\multirow{2}{*}{2500} & $q=0.90$ & $1.0\pm0.3\ $ & $1.0\pm0.3\ $ & $0.5\pm0.2\ $ & $5.1\pm0.7\ $ & $4.7\pm0.7\ $ & $3.6\pm0.6\ $ \\
             & & $q=0.99$ & $0.9\pm0.3\ $ & $1.0\pm0.3\ $ & $0.7\pm0.3\ $ & $4.4\pm0.6\ $ & $4.6\pm0.7\ $ & $5.7\pm0.7\ $ \\
            &\multirow{2}{*}{5000} & $q=0.90$ & $0.7\pm0.3\ $ & $0.9\pm0.3\ $ & $0.8\pm0.3\ $ & $4.6\pm0.7\ $ & $4.6\pm0.7\ $ & $5.0\pm0.7\ $ \\
             & & $q=0.99$ & $1.8\pm0.4\ $ & $1.3\pm0.4\ $ & $1.0\pm0.3\ $ & $5.6\pm0.7\ $ & $5.1\pm0.7\ $ & $5.8\pm0.7\ $ \\
            &\multirow{2}{*}{10000} & $q=0.90$ & $1.8\pm0.4\ $ & $1.2\pm0.3\ $ & $0.9\pm0.3\ $ & $5.2\pm0.7\ $ & $4.7\pm0.7\ $ & $5.2\pm0.7\ $ \\
             & & $q=0.99$ & $1.3\pm0.4\ $ & $1.0\pm0.3\ $ & $1.3\pm0.4\ $ & $5.1\pm0.7\ $ & $4.9\pm0.7\ $ & $5.9\pm0.7\ $ \\
            \bottomrule
    \multirow{8}{*}{$2.0$} 
    &\multirow{2}{*}{1000} & $q=0.90$ & $0.8\pm0.3\ $ & $0.7\pm0.3\ $ & $0.5\pm0.2\ $ & $3.8\pm0.6\ $ & $3.8\pm0.6\ $ & $3.1\pm0.5^*$ \\
             & & $q=0.99$ & $0.1\pm0.1^*$ & $0.9\pm0.3\ $ & $1.5\pm0.4\ $ & $2.8\pm0.5^*$ & $4.3\pm0.6\ $ & $5.3\pm0.7\ $ \\
            &\multirow{2}{*}{2500} & $q=0.90$ & $0.6\pm0.2\ $ & $0.9\pm0.3\ $ & $0.4\pm0.2^*$ & $4.7\pm0.7\ $ & $4.8\pm0.7\ $ & $3.2\pm0.6^*$ \\
             & & $q=0.99$ & $0.8\pm0.3\ $ & $0.8\pm0.3\ $ & $0.9\pm0.3\ $ & $4.1\pm0.6\ $ & $4.1\pm0.6\ $ & $4.7\pm0.7\ $ \\
            &\multirow{2}{*}{5000} & $q=0.90$ & $0.7\pm0.3\ $ & $0.8\pm0.3\ $ & $0.5\pm0.2\ $ & $5.5\pm0.7\ $ & $5.4\pm0.7\ $ & $4.0\pm0.6\ $ \\
             & & $q=0.99$ & $0.8\pm0.3\ $ & $1.1\pm0.3\ $ & $0.6\pm0.2\ $ & $4.9\pm0.7\ $ & $4.7\pm0.7\ $ & $3.6\pm0.6\ $ \\
            &\multirow{2}{*}{10000} & $q=0.90$ & $1.0\pm0.3\ $ & $0.9\pm0.3\ $ & $0.9\pm0.3\ $ & $4.7\pm0.7\ $ & $4.6\pm0.7\ $ & $4.9\pm0.7\ $ \\
             & & $q=0.99$ & $1.7\pm0.4\ $ & $0.9\pm0.3\ $ & $0.9\pm0.3\ $ & $5.2\pm0.7\ $ & $5.4\pm0.7\ $ & $5.2\pm0.7\ $ \\
            \bottomrule
    \multirow{8}{*}{$3.0$} 
    &\multirow{2}{*}{1000} & $q=0.90$ & $0.9\pm0.3\ $ & $0.4\pm0.2^*$ & $0.7\pm0.3\ $ & $4.5\pm0.7\ $ & $3.2\pm0.6^*$ & $4.4\pm0.6\ $ \\
             & & $q=0.99$ & $0.1\pm0.1^*$ & $1.6\pm0.4\ $ & $1.4\pm0.4\ $ & $3.2\pm0.6^*$ & $5.7\pm0.7\ $ & $4.9\pm0.7\ $ \\
            &\multirow{2}{*}{2500} & $q=0.90$ & $0.6\pm0.2\ $ & $0.6\pm0.2\ $ & $0.5\pm0.2\ $ & $3.4\pm0.6\ $ & $4.1\pm0.6\ $ & $3.5\pm0.6\ $ \\
             & & $q=0.99$ & $0.8\pm0.3\ $ & $1.5\pm0.4\ $ & $1.1\pm0.3\ $ & $4.9\pm0.7\ $ & $4.4\pm0.6\ $ & $5.3\pm0.7\ $ \\
            &\multirow{2}{*}{5000} & $q=0.90$ & $1.0\pm0.3\ $ & $0.4\pm0.2^*$ & $0.3\pm0.2^*$ & $3.3\pm0.6^*$ & $2.9\pm0.5^*$ & $3.7\pm0.6\ $ \\
             & & $q=0.99$ & $0.9\pm0.3\ $ & $1.8\pm0.4\ $ & $1.4\pm0.4\ $ & $4.9\pm0.7\ $ & $6.3\pm0.8\ $ & $5.3\pm0.7\ $ \\
            &\multirow{2}{*}{10000} & $q=0.90$ & $0.6\pm0.2\ $ & $0.3\pm0.2^*$ & $1.0\pm0.3\ $ & $4.5\pm0.7\ $ & $3.8\pm0.6\ $ & $4.1\pm0.6\ $ \\
             & & $q=0.99$ & $0.8\pm0.3\ $ & $1.6\pm0.4\ $ & $1.4\pm0.4\ $ & $5.7\pm0.7\ $ & $5.2\pm0.7\ $ & $6.2\pm0.8\ $ \\
            \bottomrule
    \end{tabular}
    \label{tab:power:underNull}
\end{table}

\begin{table}[t!]
    \centering
    \caption{Empirical size (i.e., percentage of null hypotheses rejected when data are generated under the null hypothesis at a given significance level $\alpha^*$) of the three proposed statistical tests, at $\alpha^* = 0.01$ and $\alpha^* = 0.05$, across different sample sizes, $q$-quantile and Student's $t$-copula settings. The number of permutation is fixed as $P=1000$. For the Student's $t$-copula, $\rho \in \{0.1, 0.4, 0.6, 0.8\}$ with degrees of freedom fixed at $\nu = 4$. Entries with asterisk symbol denote significance levels that lie beyond three standard deviations from the mean.}
    \footnotesize
    \renewcommand{\arraystretch}{0.6}
    \begin{tabular}{c c c | ccc | ccc}
        \toprule
        \multirow{2}{*}{$\rho$} & \multirow{2}{*}{$n$} & \multirow{2}{*}{$q$-quantile}
        & \multicolumn{3}{c|}{$\alpha^* = 0.01$} 
        & \multicolumn{3}{c}{$\alpha^* = 0.05$} \\
        \cmidrule(lr){4-6} \cmidrule(lr){7-9}
        & & & Perm. $\hat{\lambda}_n^1$ & Perm. $\hat{\lambda}_n^2$ & Asym. $\hat{\lambda}_n^1$
            & Perm. $\hat{\lambda}_n^1$ & Perm. $\hat{\lambda}_n^2$ & Asym. $\hat{\lambda}_n^1$ \\
        \midrule
\multirow{8}{*}{$0.1$} 
    &\multirow{2}{*}{1000} & $q=0.90$ & $1.0\pm0.3\ $ & $1.0\pm0.3\ $ & $0.5\pm0.2\ $ & $5.1\pm0.7\ $ & $5.5\pm0.7\ $ & $4.6\pm0.7\ $ \\
             & & $q=0.99$ & $0.4\pm0.2^*$ & $1.2\pm0.3\ $ & $0.2\pm0.1^*$ & $2.9\pm0.5^*$ & $3.6\pm0.6\ $ & $2.8\pm0.5^*$ \\
            &\multirow{2}{*}{2500} & $q=0.90$ & $1.4\pm0.4\ $ & $1.1\pm0.3\ $ & $0.6\pm0.2\ $ & $6.0\pm0.8\ $ & $5.7\pm0.7\ $ & $3.4\pm0.6\ $ \\
             & & $q=0.99$ & $1.1\pm0.3\ $ & $1.1\pm0.3\ $ & $1.3\pm0.4\ $ & $6.1\pm0.8\ $ & $4.3\pm0.6\ $ & $5.2\pm0.7\ $ \\
            &\multirow{2}{*}{5000} & $q=0.90$ & $0.6\pm0.2\ $ & $1.3\pm0.4\ $ & $1.3\pm0.4\ $ & $4.1\pm0.6\ $ & $4.6\pm0.7\ $ & $4.6\pm0.7\ $ \\
             & & $q=0.99$ & $1.3\pm0.4\ $ & $0.9\pm0.3\ $ & $0.8\pm0.3\ $ & $5.8\pm0.7\ $ & $4.4\pm0.6\ $ & $4.9\pm0.7\ $ \\
            &\multirow{2}{*}{10000} & $q=0.90$ & $0.7\pm0.3\ $ & $0.8\pm0.3\ $ & $1.3\pm0.4\ $ & $3.7\pm0.6\ $ & $5.0\pm0.7\ $ & $5.7\pm0.7\ $ \\
             & & $q=0.99$ & $1.2\pm0.3\ $ & $0.8\pm0.3\ $ & $0.8\pm0.3\ $ & $4.8\pm0.7\ $ & $5.0\pm0.7\ $ & $5.3\pm0.7\ $ \\
            \bottomrule
    \multirow{8}{*}{$0.4$} 
    &\multirow{2}{*}{1000} & $q=0.90$ & $1.0\pm0.3\ $ & $0.8\pm0.3\ $ & $1.3\pm0.4\ $ & $6.0\pm0.8\ $ & $5.6\pm0.7\ $ & $5.0\pm0.7\ $ \\
             & & $q=0.99$ & $0.0\pm0.0^*$ & $0.9\pm0.3\ $ & $0.7\pm0.3\ $ & $2.9\pm0.5^*$ & $4.1\pm0.6\ $ & $5.7\pm0.7\ $ \\
            &\multirow{2}{*}{2500} & $q=0.90$ & $1.4\pm0.4\ $ & $0.8\pm0.3\ $ & $0.4\pm0.2^*$ & $4.4\pm0.6\ $ & $4.4\pm0.6\ $ & $4.8\pm0.7\ $ \\
             & & $q=0.99$ & $1.1\pm0.3\ $ & $1.1\pm0.3\ $ & $1.2\pm0.3\ $ & $4.7\pm0.7\ $ & $5.2\pm0.7\ $ & $5.3\pm0.7\ $ \\
            &\multirow{2}{*}{5000} & $q=0.90$ & $1.4\pm0.4\ $ & $0.7\pm0.3\ $ & $0.8\pm0.3\ $ & $3.9\pm0.6\ $ & $4.2\pm0.6\ $ & $4.6\pm0.7\ $ \\
             & & $q=0.99$ & $0.8\pm0.3\ $ & $1.2\pm0.3\ $ & $1.4\pm0.4\ $ & $4.7\pm0.7\ $ & $5.6\pm0.7\ $ & $6.0\pm0.8\ $ \\
            &\multirow{2}{*}{10000} & $q=0.90$ & $1.7\pm0.4\ $ & $0.9\pm0.3\ $ & $0.7\pm0.3\ $ & $5.7\pm0.7\ $ & $4.9\pm0.7\ $ & $4.4\pm0.6\ $ \\
             & & $q=0.99$ & $1.5\pm0.4\ $ & $0.8\pm0.3\ $ & $1.6\pm0.4\ $ & $5.4\pm0.7\ $ & $4.8\pm0.7\ $ & $4.1\pm0.6\ $ \\
            \bottomrule
    \multirow{8}{*}{$0.6$} 
    &\multirow{2}{*}{1000} & $q=0.90$ & $0.6\pm0.2\ $ & $1.2\pm0.3\ $ & $0.9\pm0.3\ $ & $5.0\pm0.7\ $ & $4.0\pm0.6\ $ & $3.7\pm0.6\ $ \\
             & & $q=0.99$ & $0.1\pm0.1^*$ & $0.9\pm0.3\ $ & $0.3\pm0.2^*$ & $3.7\pm0.6\ $ & $4.5\pm0.7\ $ & $3.7\pm0.6\ $ \\
            &\multirow{2}{*}{2500} & $q=0.90$ & $1.3\pm0.4\ $ & $1.1\pm0.3\ $ & $0.5\pm0.2\ $ & $6.8\pm0.8\ $ & $5.0\pm0.7\ $ & $4.0\pm0.6\ $ \\
             & & $q=0.99$ & $1.5\pm0.4\ $ & $1.0\pm0.3\ $ & $0.9\pm0.3\ $ & $5.6\pm0.7\ $ & $4.6\pm0.7\ $ & $4.9\pm0.7\ $ \\
            &\multirow{2}{*}{5000} & $q=0.90$ & $0.4\pm0.2^*$ & $0.1\pm0.1^*$ & $0.8\pm0.3\ $ & $4.3\pm0.6\ $ & $3.2\pm0.6^*$ & $5.0\pm0.7\ $ \\
             & & $q=0.99$ & $0.7\pm0.3\ $ & $1.2\pm0.3\ $ & $1.8\pm0.4\ $ & $5.1\pm0.7\ $ & $4.9\pm0.7\ $ & $5.5\pm0.7\ $ \\
            &\multirow{2}{*}{10000} & $q=0.90$ & $0.8\pm0.3\ $ & $0.6\pm0.2\ $ & $0.7\pm0.3\ $ & $5.1\pm0.7\ $ & $3.6\pm0.6\ $ & $4.0\pm0.6\ $ \\
             & & $q=0.99$ & $1.2\pm0.3\ $ & $0.9\pm0.3\ $ & $1.1\pm0.3\ $ & $5.2\pm0.7\ $ & $5.6\pm0.7\ $ & $6.0\pm0.8\ $ \\
            \bottomrule
    \multirow{8}{*}{$0.8$} 
    &\multirow{2}{*}{1000} & $q=0.90$ & $1.0\pm0.3\ $ & $0.8\pm0.3\ $ & $1.2\pm0.3\ $ & $4.8\pm0.7\ $ & $4.2\pm0.6\ $ & $4.9\pm0.7\ $ \\
             & & $q=0.99$ & $0.2\pm0.1^*$ & $1.1\pm0.3\ $ & $1.8\pm0.4\ $ & $3.3\pm0.6^*$ & $5.0\pm0.7\ $ & $6.4\pm0.8\ $ \\
            &\multirow{2}{*}{2500} & $q=0.90$ & $0.7\pm0.3\ $ & $0.4\pm0.2^*$ & $0.8\pm0.3\ $ & $4.4\pm0.6\ $ & $3.9\pm0.6\ $ & $4.4\pm0.6\ $ \\
             & & $q=0.99$ & $1.2\pm0.3\ $ & $0.6\pm0.2\ $ & $1.6\pm0.4\ $ & $5.3\pm0.7\ $ & $4.8\pm0.7\ $ & $7.3\pm0.8\ $ \\
            &\multirow{2}{*}{5000} & $q=0.90$ & $0.6\pm0.2\ $ & $0.9\pm0.3\ $ & $0.6\pm0.2\ $ & $3.9\pm0.6\ $ & $4.3\pm0.6\ $ & $4.4\pm0.6\ $ \\
             & & $q=0.99$ & $1.1\pm0.3\ $ & $1.8\pm0.4\ $ & $0.8\pm0.3\ $ & $4.6\pm0.7\ $ & $5.6\pm0.7\ $ & $4.4\pm0.6\ $ \\
            &\multirow{2}{*}{10000} & $q=0.90$ & $0.8\pm0.3\ $ & $0.8\pm0.3\ $ & $1.1\pm0.3\ $ & $3.7\pm0.6\ $ & $3.1\pm0.5^*$ & $4.7\pm0.7\ $ \\
             & & $q=0.99$ & $1.2\pm0.3\ $ & $0.9\pm0.3\ $ & $0.5\pm0.2\ $ & $5.3\pm0.7\ $ & $4.6\pm0.7\ $ & $4.8\pm0.7\ $ \\
            \bottomrule
    \end{tabular}
    \label{tab:power:underNullT}
\end{table}

\clearpage

\section{Supplementary application details}
\label{sec:app:china}

This section supplements Section~\textcolor{blue}{5} of the main paper by providing additional analyses of China's futures market. In Section~\ref{sec:app:auto}, we estimate temporal auto-dependence. Section~\ref{sec:app:tailIndexForAsset} focuses on describing the heaviness of both the upper and lower tails of China's futures returns and, in Section~\ref{sec:app:cur}, we visualize the market-wide contemporaneous directional tail dependence structure in China's futures market using the extremal ball. In Section~\ref{sec:app:bootstrap}, we provide a bootstrap-based uncertainty assessment of the estimated extremal dependence structure.

\subsection{Temporal dependence analysis of China's derivative Markets}
\label{sec:app:auto}

\textcolor{customizedColor}{To investigate temporal dependence in the asset return time series, $\{R_{j,t}\}_{t=1}^T,j=1,\dots,55,$ we estimate the autocorrelation and extremogram~\citep{davis2012towards}, where the latter quantifies extremal temporal dependence. For tail dependence, we use the extremogram of the absolute value of the asset returns \( |R_{i,t}| \), which captures dependence in both the upper and lower tails. \cite{davis2012towards} provide the R package \texttt{extremogram}, with the \texttt{extremogram1} function for estimating temporal dependence and the \texttt{bootconf1} function for constructing bootstrap confidence intervals. For confidence intervals, we set a significance level of $0.05$ and, for the extremogram, an extra parameter, the threshold, is set to the empirical $0.99$-quantile.}

\begin{figure}[t!]
    \centering
    \includegraphics[width=\linewidth]{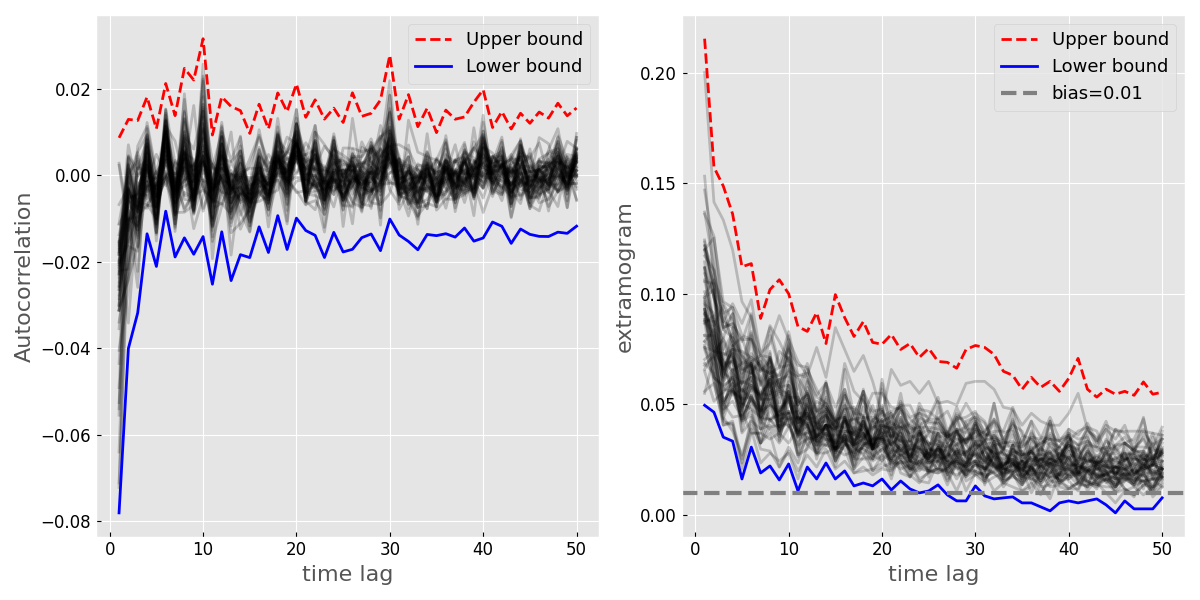}
    \caption{\textcolor{customizedColor}{Estimates of the autocorrelation (left) and extremogram (right) for each asset in China's futures market (black lines). The red and blue curves give the upper and lower confidence bounds for a significance level of $0.05$. The dashed line indicates the known bias in estimates of the extremogram; see \protect\cite{davis2012towards}}}
    \label{fig:autocorr_extremogram}
\end{figure}

\textcolor{customizedColor}{Figure~\ref{fig:autocorr_extremogram} presents the estimated autocorrelation and extremogram functions for each asset. The results indicate that (extremal) temporal dependence decays rapidly, falling below $0.05$ and $0.1$ after $10$ time lags for the autocorrelation and extremogram, respectively. Therefore, we conclude that temporal dependence in the data is quite weak, and samples $\bm{R}_{t_1}$ and $\bm{R}_{t_2}$ are approximately independent if $|t_1-t_2|>10$.}

\subsection{Tail index analysis of China's futures market data}
\label{sec:app:tailIndexForAsset}

\begin{figure}[t!]
    \centering
    \includegraphics[width=1\linewidth,center]{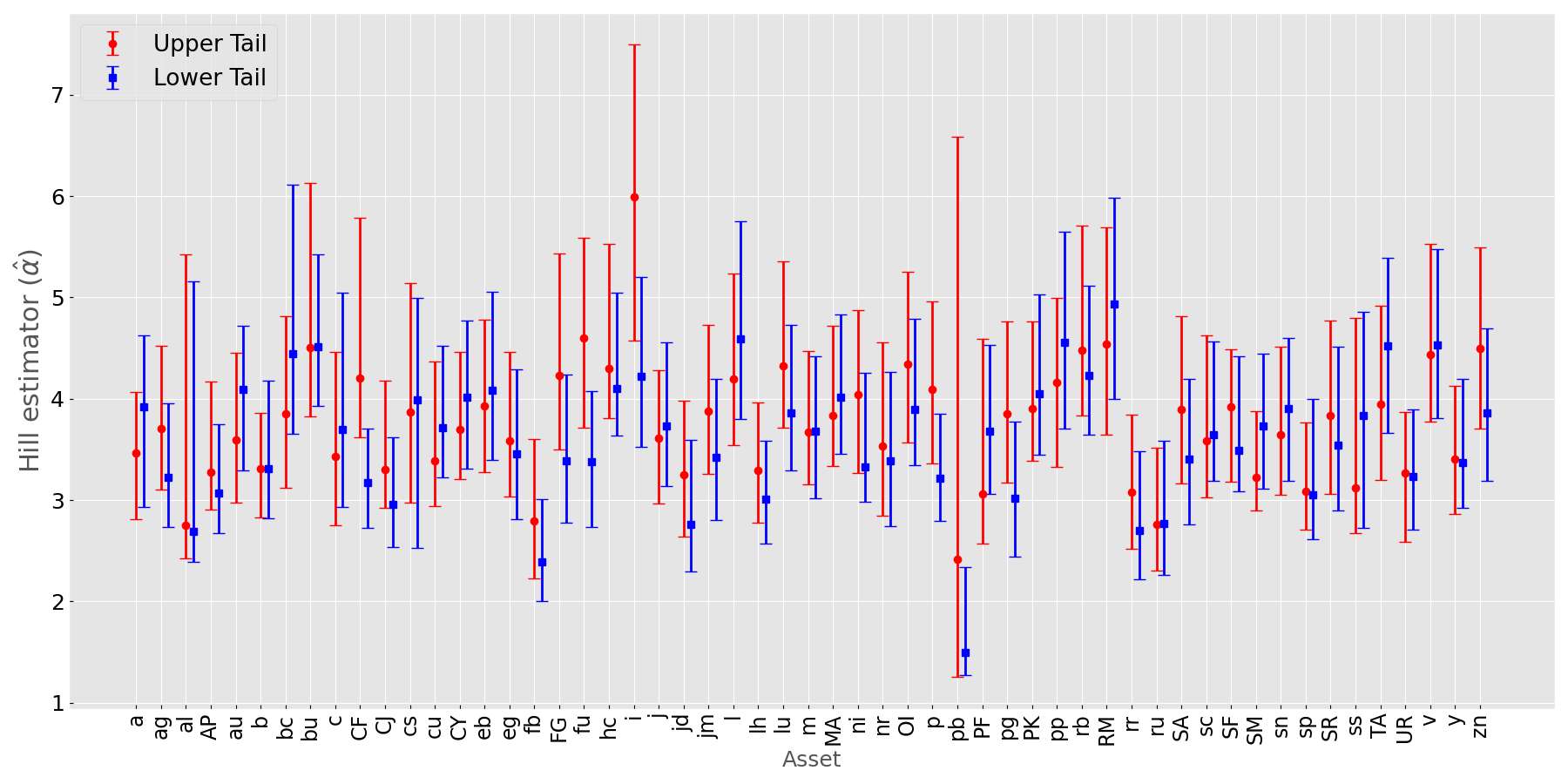}
    \caption{Estimates of the upper (red circle) and lower (blue square) tail indices, alongside $95\%$ confidence intervals, for each asset in China's futures market.}
    \label{fig:tailIndex}
\end{figure}
\begin{figure}[t!]
    \centering
    \includegraphics[width=1\linewidth,center]{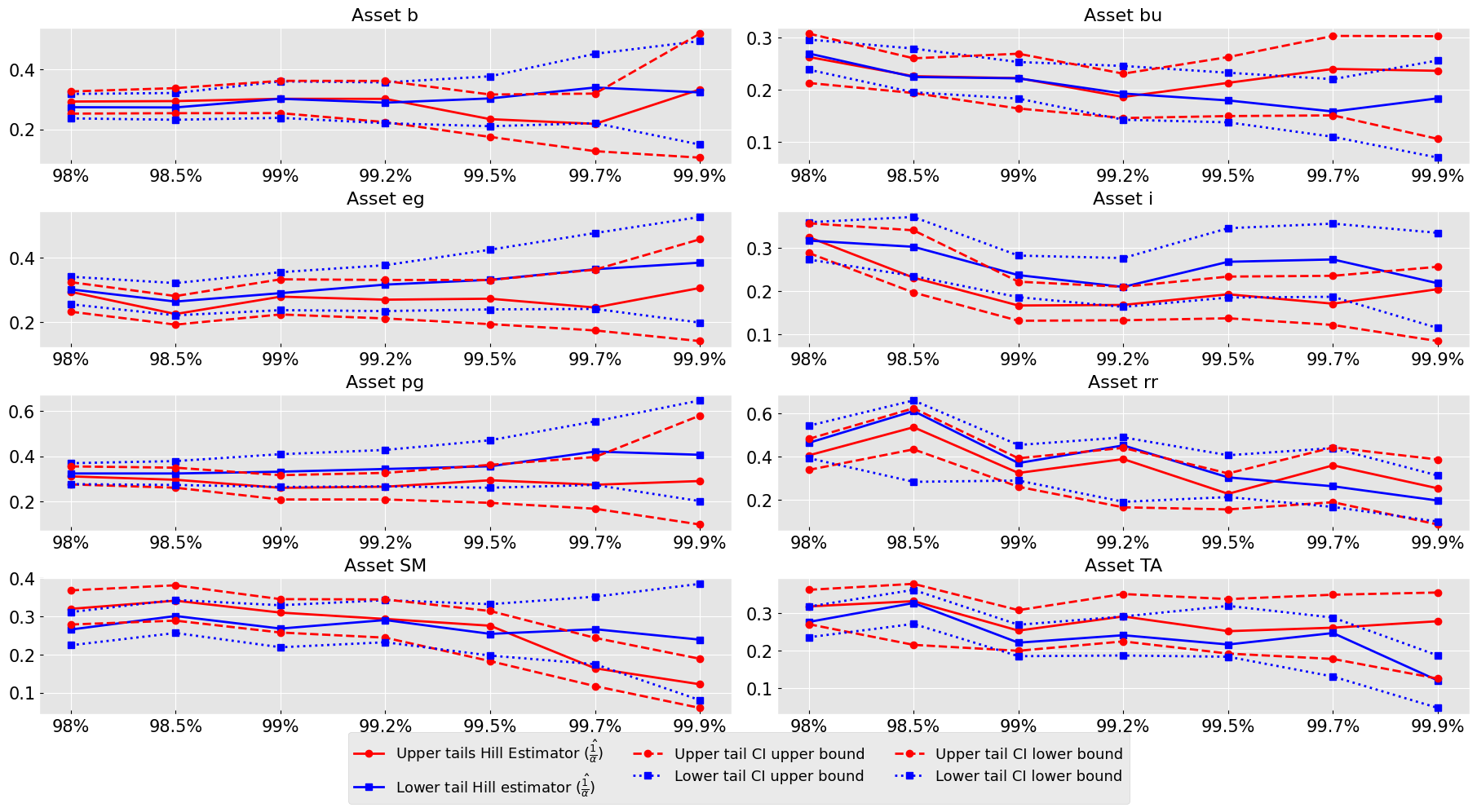}
    \caption{\textcolor{customizedColor}{Threshold stability plot for upper (red circle) and lower (blue square) tail index estimates for eight assets. The thresholds used in estimation are labeled on the $x$-axes, and range from the 0.98-quantile to the 0.999-quantile.}}
    \label{fig:stable_check_of_hill}
\end{figure}

\textcolor{customizedColor}{We use the~\cite{hill1975simple} estimator to estimate the tail index for the upper and lower tails of the asset returns. For a generic random variable $X$ with sample $X_1,\dots,X_n$, the Hill's estimate of its upper tail index is given by 
\begin{equation}\label{eq:hill}
\hat{\alpha}=\left({1 \over m-1}\sum_{i=1}^{m-1} \log \frac{X_{(i)}}{X_{(m)}}\right)^{-1},
\end{equation} 
where $X_{(1)}\geq X_{(2)}\geq \dots \geq X_{(m)}$ are the $m$-largest order statistics of the sample $X_1,\dots,X_n$; here we take $X_{(m)}$ to be  the $0.99$ sample quantile. To estimate the lower tail index, we can apply the same formula for negated samples $-X_1,\dots,-X_n$. Confidence intervals for tail index estimates are calculated using 1000 bootstrap replicates with a significance level of $0.05$. The average tail index across all \totalAsset\ assets is $3.7$ and $3.5$ for upper and lower tails, respectively, thus approximately equal. Figure~\ref{fig:tailIndex} presents estimates for each asset. We observe that both the upper and lower tails of all assets exhibit heavy-tailed behavior (as estimates of $\alpha$ are relatively low), which supports our use of the first marginal transformation in Section~\textcolor{blue}{3.2} of the main paper. We also select eight assets to check the threshold stability of the estimated tail index. The results are shown in Figure~\ref{fig:stable_check_of_hill}; we observe stability in these estimates, suggesting that our assumption that our data are regularly varying is well-founded.}

\subsection{Visualization of market-wide contemporaneous extremal dependence}
\label{sec:app:cur}
 
\begin{figure}[t!]
    \centering
    \includegraphics[width=0.65\linewidth]{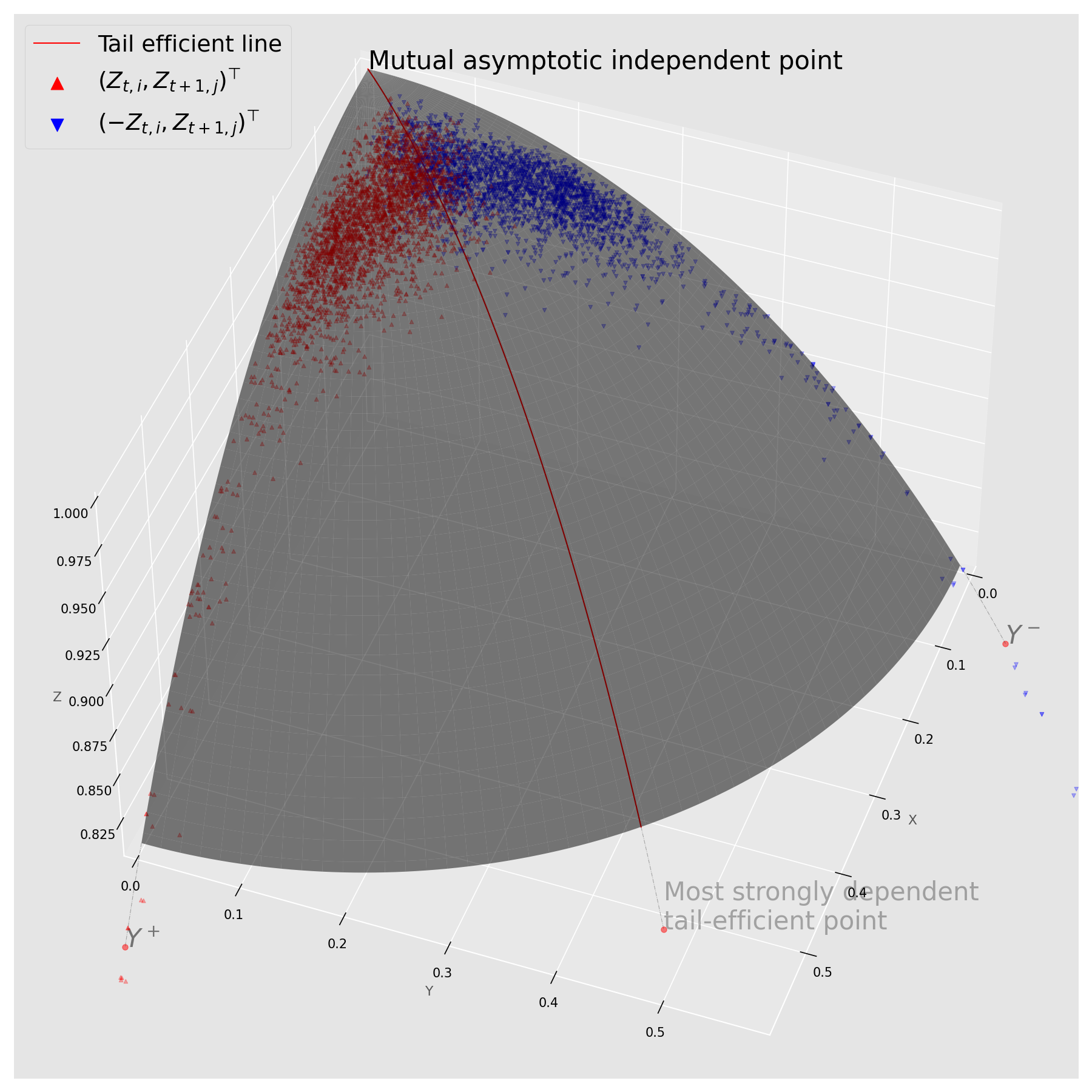}
    \caption{ Extremal ball for contemporaneous extremal dependencies in China's futures market. Red points denote pairs $(Z_{i,t},Z_{j,t})^\top$ and blue points denote $(-Z_{i,t},Z_{j,t})^\top$.}
    \label{fig:ExtremeBallCur}
\end{figure}

Figure~\ref{fig:ExtremeBallCur} visualizes directional tail asymmetry in the market-wide extremal dependence for all contemporaneous pairs $(-Z_{i,t},Z_{j,t})^\top$ and $(Z_{i,t},Z_{j,t})^\top$ where $i\neq j=1,\dots,55$. The former group corresponds to cases where the driving event is an extremal loss, while the latter corresponds to cases where the driving asset is an extremal gain. We have not conducted a detailed study of the directional tail asymmetry for contemporaneous relationships. Further investigation could reveal the pattern of concurrence of extremal events across assets. We defer this analysis to future research.


\subsection{Bootstrap analysis for uncertainty quantification}
\label{sec:app:bootstrap}

\textcolor{customizedColor}{To assess uncertainty in the results of our analysis that might arise from the marginal transformation, we perform the following non-parametric bootstrap study with $B=100$ replicates. For each bootstrap sample, we (i) perform the marginal tail-index-based transformation, (ii) re-estimate $\lambda$ for all pairs of assets, (iii) conduct a permutation test for the Efficient Tail Hypothesis, and (iv) backtest our investment strategy using tail-inefficient pairs (as described in Section~\textcolor{blue}{5.4} of the main paper) on the out-of-sample data.}

%

\textcolor{customizedColor}{We find that the ETH is rejected under significance level $\alpha^*=0.01$ for all $B=100$ bootstrap replicates. To quantify uncertainty in the estimated $\lambda$ values, we provide the bootstrap mean and standard deviation of $\lambda$ estimate for each pair of assets; see Table~\textcolor{blue}{1} of the main paper.}

\begin{figure}[t!]
    \centering
    \includegraphics[width=\textwidth]{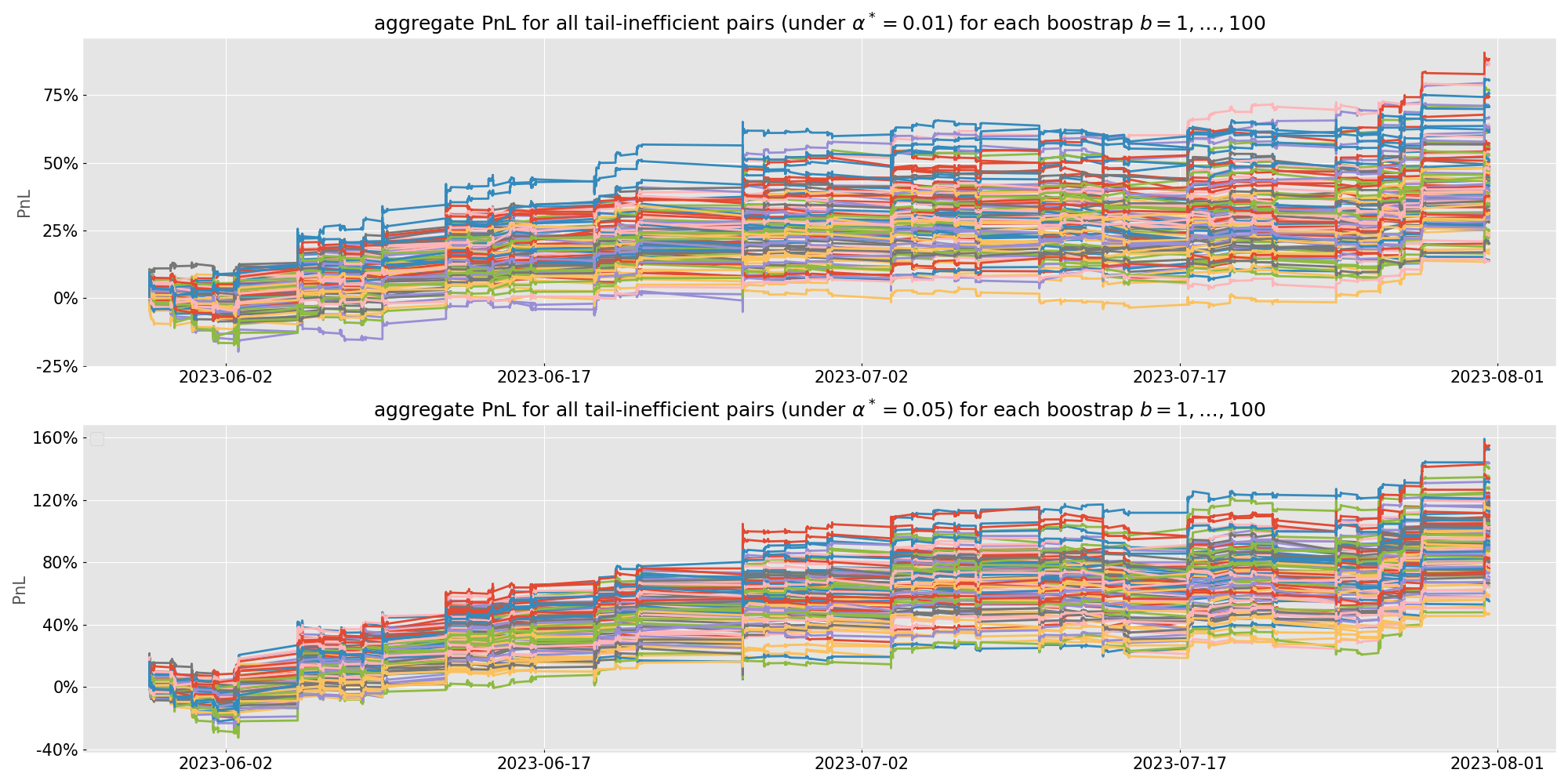} 
    \caption{\textcolor{customizedColor}{Aggregated Profit and Loss of a dynamic portfolio consisting of all tail-inefficient pairs with p-value under $\alpha^*=0.01$ (top) or under $\alpha^*=0.05$ (bottom). Different colored curves correspond to different bootstrap replicates ($B=100$). The $y$-axis is the cumulative Profit and Loss (\%) and the $x$-axis is the time index for the two-month test period.}}
    \label{fig:backtestBoostrap} 
\end{figure}

\textcolor{customizedColor}{For each bootstrap replicate, we select all tail-inefficient pairs with p-values smaller than $\alpha^*=0.01$ or $\alpha^*=0.05$ and construct an artificial dynamic portfolio, as described in Section~\textcolor{blue}{5.4} of the main paper. We calculate the time-varying aggregated Profit and Loss (PnL) across all tail-inefficient pairs, on out-of-sample data, and plot the $B=100$ PnL curves in Figure~\ref{fig:backtestBoostrap}. All bootstrap replicates show a positive aggregated PnL and similar dynamics. The results indicate that our dynamic portfolio is robust to uncertainty in the marginal transformation.}

\clearpage
\newpage

{\renewcommand{\baselinestretch}{0.80}\normalsize
\bibliographystyle{apalike}
\bibliography{refsx}
}

\end{document}